\documentclass[aps,pra,twocolumn,showpacs,groupedaddress]{revtex4}  

\usepackage[american]{babel}
\usepackage{graphicx}  
\usepackage{dcolumn}   
\usepackage{bm} 
\usepackage{amssymb}   
\usepackage{amsmath}
\usepackage{color} 

\def\op#1{{\Hat{\mathrm{#1}}}}
\def\vop#1{{\Hat{\mathbf{#1}}}}
\def\bra#1{\ensuremath{\langle{#1}\vert}}
\def\ket#1{\ensuremath{\vert{#1}\rangle}}
\def\bracket#1#2{\ensuremath{%
    \langle{#1}\mkern1.2mu\vert\mkern1.2mu{#2}\rangle}}
\def\expect#1{\ensuremath{\langle{#1}\rangle}}
\def\commutator#1#2{\mathinner{\mathopen[#1,#2\mathclose]}}

\def\abs#1{\mathinner{\lvert#1\rvert}}

\newcommand{\be}{\begin{equation}}
\newcommand{\ee}{\end{equation}}
\newcommand{\benn}{\begin{equation*}}
\newcommand{\eenn}{\end{equation*}}

\newcommand{\beq}{\begin{eqnarray}}
\newcommand{\eeq}{\end{eqnarray}}

\def\H1{\widehat{H}_1}

\newcommand{\lb}{\left[}
\newcommand{\rb}{\right]}
\newcommand{\lp}{\left(}
\newcommand{\rp}{\right)}

\let\oldsqrt\sqrt
\def\sqrt{\mathpalette\DHLhksqrt}
\def\DHLhksqrt#1#2{%
\setbox0=\hbox{$#1\oldsqrt{#2\,}$}\dimen0=\ht0
\advance\dimen0-0.2\ht0
\setbox2=\hbox{\vrule height\ht0 depth -\dimen0}%
{\box0\lower0.4pt\box2}}

\begin{document}

\title{Dynamics of the rotated Dicke model}
\author{Michael Tomka$^{\clubsuit}$ and Vladimir Gritsev$^{\clubsuit}$}
\affiliation{$^{\clubsuit}$Physics Department, University of
Fribourg, Chemin du Mus\'ee 3, 1700 Fribourg, Switzerland}

\date{\today}

\begin{abstract}
We study quantum dynamics of the rotationally driven Dicke model where
the collective spin is rotated around the $z$ axis with a finite
velocity.
In the absence of the rotating wave approximation we observe that for
several physically relevant initial states the position of the quantum
critical point is shifted by the amount given by the applied rotation
velocity. 
This allows us to probe the quantum criticality ``from a distance'' in
parameter space without actual crossing of the quantum critical
surface but instead by encircling it in the parameter space.
This may provide a useful experimental hint since the quantum state is
not destroyed by this protocol.
Moreover, for the coherent initial state we observe an interesting
non-equilibrium reentrant phenomenon of quantum critical behavior as a
function of the driving velocity and construct a non-equilibrium phase
diagram of the driven model.
\end{abstract}

\maketitle

\section{Introduction}

The Dicke model \cite{Dicke1954}, \cite{TC} was introduced 
to model the interaction of a collection of $N$ two-level
atoms with a single-mode radiation (bosonic) field through the dipole
coupling.
Under the condition that the ensemble of two-level atoms is confined
in a region of space which is much smaller than the
wavelength of the single-mode radiation field, the system of the
two-level emitters behaves as a collective large spin. 
The model became a paradigm for collective behavior of matter interacting with
light and is believed to have a phase transition at finite and zero
temperatures \cite{HL}, \cite{WH}, \cite{CGW}.
A super-radiant transition is characterized by the expectation value
of the photon number operator which distinguishes between normal phase
(where it is zero) and a super-radiant phase (where it is nonzero).
The issue of this phase transition has been debated \cite{RWZ} in
length in the literature by various authors using different methods.
The conclusion is that although is it difficult to realize the phase
transition in a real two-level system for which the model has been
introduced, it is possible to engineer it in some cavity QED setups.
For a recent review of these issues see, e.g.,
\cite{Keeling}, \cite{VD}. 

With a recent progress in quantum optics and cold atoms it became
possible to realize the Dicke model in a cavity QED systems and observe
a super-radiant phase transition and other collective behaviors
\cite{Esslinger}.
This advance became possible by achieving considerable degree of
tunebility of parameters of the system with high precision.
Moreover parameters can be controlled in real time.
This stimulates studies of non-equilibrium effects in this class of
systems. An interplay between collective and non-equilibrium effects
is pronounced close to the quantum phase transition. Several recent
theoretical works on driven non-equilibrium Dicke model, \cite{AGKP},
\cite{IT}, \cite{AH} and \cite{BERB} have focused on various effects
of time-depending driving of the interaction strength and/or
detuning.
Many of these non-equilibrium effects are related to the presence of a
quantum critical point in the equilibrium model. 

Here we propose to probe the quantum phase transition appearing in the
Dicke model in an indirect way.
We employ a certain degree of tunebility of the system to drive (rotate) the
collective spin by a certain time-dependent angle.
This makes it possible to probe the quantum criticality ``from a
distance'' in parameter space without crossing the equilibrium quantum
critical point.
Apart from this we observe new, reentrant quantum critical behavior in
such non-equilibrium protocol for certain initial state.
This behavior can be traced back to a mechanism of competition between
geometric and dynamical phases in the non-equilibrium dynamics~\cite{TPG}.  

\begin{figure}[h!]
  \begin{center}
    \includegraphics[scale=0.4]{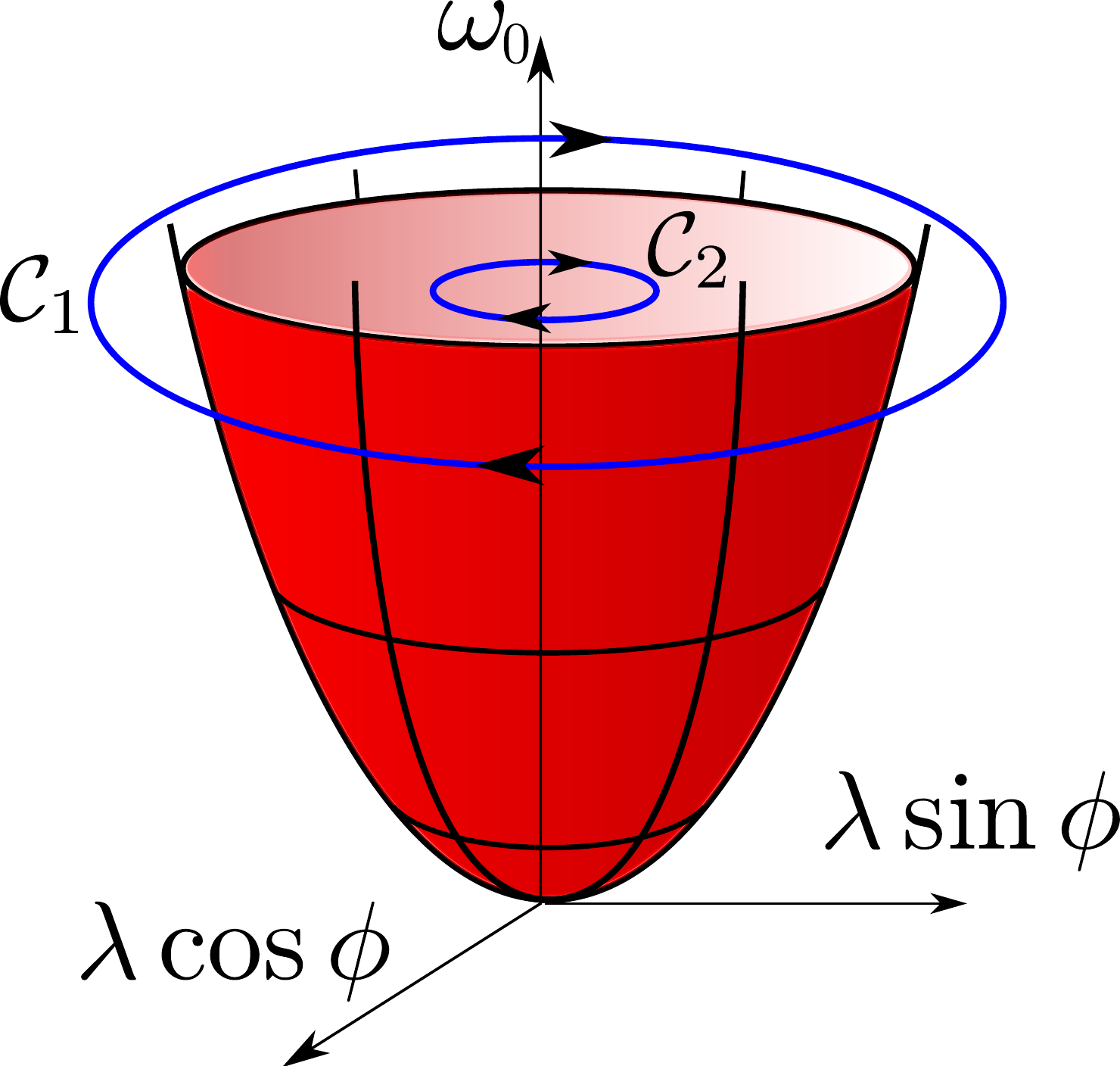}
    \caption{
      Equilibrium quantum phase diagram of the rotationally
      driven Dicke model as a function of atomic frequency
      $\omega_{0}$ and atom-photon coupling strength $\lambda$.
      Paths $\mathcal{C}_{1}$ and $\mathcal{C}_{2}$ are fixed by the
      time dependence of the driving parameter $\phi(t)$ (rotation
      angle around $z$-direction in spin space).
      Here the bosonic frequency is set to $\omega=1$.
      }
    \label{fig:dickemodeleqphdiag}
  \end{center}
\end{figure}

In addition to the previous studies, here we look into the quantum
dynamics of the rotationally driven Dicke model when the change of the
rotation angle is linear in time.
We distinguish between two types of protocols (see
Fig.~\ref{fig:dickemodeleqphdiag}): in one case we encircle the
critical surface in the parameter space from the outer region (contour
$\mathcal{C}_{1}$) while in the second case we drive the system in
such a way that the circle $\mathcal{C}_{2}$ stays inside the region
bounded by the critical surface.
These regions are distinguished by the presence (region
$\mathcal{C}_{1}$) or absence (region $\mathcal{C}_{2}$) of the
geometric phase contributions
to the total phase factor of each instantaneous eigenstate.
To characterize the non-equilibrium phase diagram we look into two
types of observables: (i) the mean photon number and (ii) a
topological number associated with the expectation value of the parity
operator.

Moreover we consider several different initial states: (i) atom-field coherent states (coherent state for
photons and pseudo-spin subsystem), (ii) Fock state and (iii) the
ground state of the system.
Our quasi-classical approach which is based on the coherent state
representation is supplemented by a numerical procedure. 
For initial coherent and ground states a non-equilibrium phase diagram
has an additional reentrant phase inside the super-radiant phase. 
For the aims of completeness and comparison with a driven system, we
also look into dynamics of the undriven Dicke model starting from
those initial states.
We study the system both for a finite number of two-level systems and
in the thermodynamic limit (TDL).
We found in particular that dynamics starting from the coherent state
coincides with the one for the initial ground state, both in the
driven and undriven cases.
Dynamics from the pure Fock state is featureless in the thermodynamic
limit: one need some fluctuations in the initial state to make it
evolve, at least in the quasi-classical approach.
This fluctuations are introduced by allowing small fluctuations in
the bosonic and the spin sectors; we call it ``nearly-Fock'' state. 
Our results for the non-equilibrium phase diagram of the driven Dicke model are summarized in Fig.~\ref{fig:adaneqpdscs}.

We note that here we study the Dicke model beyond the rotating wave
approximation which means that it is not integrable.
Dynamics of integrable variant of the Dicke model starting from
certain initial states (when the rotating wave approximation is made) has
been studied before in \cite{Yuzbashyan} and more recently in
\cite{TL}, \cite{Faribault}.
We however found some features (like e.g. formation of soliton-like
trains in the thermodynamic limit) which also hold in the
non-integrable case. 

First, we overview some basic facts about Dicke model (Section II),
while in the Section III we establish quasi-classical equations for
the dynamically driven model.
Section IV presents various results for several initial conditions and
comparison between dynamics of the driven and undriven models.

\section{The Dicke Model}

Here we consider the Dicke model without the rotating wave
approximation.
The quantum Hamiltonian of the Dicke model reads
\be
  \label{eq:dickeham}
  \op{H}_{\mathrm{D}} =
  \omega_{0} \op{J}_{z} +
  \omega \op{a}^{\dagger}\op{a} +
  \frac{\lambda}{\sqrt{N}}
  \lp \op{a}^{\dagger} + \op{a} \rp
  \lp \op{J}_{+} + \op{J}_{-} \rp,
\ee
where $\omega_{0}$ is the level-splitting of a two-level atom and the
collection of those $N$ identical two-level atoms is described by the
collective spin operators
$\op{J}_{\mu}=\sum_{i=1}^{N} \op{\sigma}_{i}^{\mu}/2$.
Here $\op{\sigma}_{i}^{\mu}$, $\mu=z,\pm$ are the usual Pauli matrices
which describe a two-level emitter.
The collective spin operators satisfy the $\mathrm{su}(2)$ Lie
algebra:
\be
  \commutator{\op{J}_{+}}{\op{J}_{-}} = 2 \op{J}_{z}, \quad
  \commutator{\op{J}_{z}}{\op{J}_{\pm}} = \pm \op{J}_{\pm}.
\ee
The Hilbert space of the $\mathrm{su}(2)$ Lie algebra is spanned by
the Dicke states $\ket{j,m}$, with $m= -j, -j+1, \ldots, j-1, j$ and
$j$ being integer or half integer.
The Dicke states $\ket{j,m}$ are the
eigenstates of $\op{J}^{2}$ and $\op{J}_{z}$ with eigenvalues $j(j+1)$
and $m$, respectively.
As in~\cite{Emary2003} we fix $j$ to its maximal value $j=N/2$.
Therefore the collection of two-level atoms can be interpreted as a
large ``spin'' (pseudo-spin) vector of magnitude $j=N/2$. 
The single-mode radiation field with frequency $\omega$ is described by the
bosonic creation and annihilation operators $\op{a}$, $\op{a}^{\dagger}$
satisfying the usual commutation relation
$\commutator{\op{a}}{\op{a}^{\dagger}} = \op{1}$.
The set of operators $\{ \op{a}^{\dagger}\op{a}, \op{a}^{\dagger},
\op{a}, \op{1} \}$ forms the Heisenberg-Weyl Lie algebra
$\mathrm{h}_{4}$.
The Hilbert space of this algebra is spanned by the Fock states
$\ket{n}$ with $n=0, 1, 2, 3, \ldots$, defined by
\be
  \op{a}^{\dagger}\ket{n}=\sqrt{n+1}\ket{n+1}, \quad
  \op{a}\ket{n}=\sqrt{n}\ket{n-1}.
\ee
Finally, the term
$\frac{\lambda}{\sqrt{N}}(\op{a}^{\dagger}+\op{a})(\op{J}_{+}+\op{J}_{-})$
describes the atom-field interaction within the dipole approximation
where the coupling strength is $\lambda$ and the factor
$\frac{1}{\sqrt{N}}$ stems from the fact that the original dipole
coupling strength is proportional to $\frac{1}{\sqrt{V}}$.

The symmetries of the Dicke model are found by applying a unitary
transformation of the form
\be
  \op{U}(\varphi_{0}) = \exp\lb i \varphi_{0} \op{N} \rb,
\ee
to the Dicke Hamiltonian $\op{H}_{\mathrm{D}}$.
Here $\op{N}=\op{a}^{\dagger}\op{a} + \op{J}_{z} + j$ is the
excitation number operator, the operator
which counts the number of excited quanta in the system and $\varphi_{0}$
is some rotation angle.
Applying this transformation we observe that
\be
  \op{U} \op{a} \op{U}^{\dagger} = e^{i\varphi_{0}} \op{a}, \quad
  \op{U} \op{J}_{+} \op{U}^{\dagger} = e^{i\varphi_{0}} \op{J}_{+},
\ee
and therefore, the transformed Dicke Hamiltonian reads
\begin{align}
  & \op{U} \op{H}_{\mathrm{D}} \op{U}^{\dagger} 
  =
  \omega_{0} \op{J}_{z} + \omega \op{a}^{\dagger}\op{a} \nonumber \\
  &+ \frac{\lambda}{\sqrt{2j}}
   \lp 
    \op{a}^{\dagger}\op{J}_{+}e^{2i\varphi_{0}}
   + \op{a}^{\dagger}\op{J}_{-}
   + \op{a}\op{J}_{+}
   + \op{a}\op{J}_{-} e^{-2i\varphi_{0}} 
   \rp.
\end{align}
The Dicke model remains invariant under this transformation only if
$\varphi_{0}=\pi$.
Hence, the associated symmetry of the Dicke model without the
rotating wave approximation is the parity, since
\be
  \op{U}(\varphi_{0}=\pi) = \exp\lb i \pi \op{N} \rb =: \op{\Pi}.
\ee
This has the interpretation as a $\mathbb{Z}_{2}$ symmetry because it
has the eigenvalues  $e^{i \pi (n+m+j)} = \pm 1$.
Thus the Hilbert space breaks up into two non-interacting subspaces,
one with an even excitation number $n+m+j$ and one with an odd one.
Therefore, the Dicke model without the rotating wave approximation
possesses a $\mathbb{Z}_{2}$ symmetry. 

The Dicke model~(\ref{eq:dickeham}) has a
quantum phase transition in the thermodynamic limit $N \to \infty$
($j \to \infty$) at the quantum critical point $\lambda_{c} =
\frac{\sqrt{\omega\omega_{0}}}{2}$.
Quantum critical points (criticalities) are points in the 
underlying parameter space $(\omega_{0}, \omega, \lambda)$ at
which the excitation energy vanishes.
Hence, the ground state of the system changes qualitatively from a normal
to a super-radiant one.
In the super-radiant phase the ground state is a macroscopically excited
collective state that possesses the potential to super-radiate.
A phase transition is associated to spontaneous breakdown of the
$\mathbb{Z}_{2}$ symmetry described above. 
The super-radiant emission is a collective effect involving all the
$N$ two-level atoms in the sample.
In this case the decay rate is $\Gamma \sim N^{2}\gamma_{0}$ instead
of $N \gamma_{0}$, the rate for independent atom emission.
Here $\gamma_{0}$ is the decay rate of an excited two-level atom.

In this paper we study a rotationally driven version of the Dicke
model~\cite{Plastina2006},~\cite{Chen2006}, which results from
applying a rotation of angle $\phi$ around the $z$ axis to each
two-level atom.
Explicitly it has the form of time-dependent unitary rotation applied to
$z$ axis, 
\be
  \op{R}_{z}(t) = \exp\lb -i\phi(t)\op{J}_{z} \rb.
\ee
The resulting time dependent Hamiltonian $\op{H}_{\mathrm{RD}}(t)=
\op{R}_{z}^{\dagger}(t) \op{H}_{\mathrm{D}}(\omega_{0}, \omega, \lambda) \op{R}_{z}(t)$ reads
\begin{align}
 \label{eq:hamrd}
 \op{H}_{\mathrm{RD}}(t) &=
 \omega_{0} \op{J}_{z} +
 \omega \op{a}^{\dagger}\op{a} + \nonumber \\
 &+
 \frac{\lambda}{\sqrt{2j}} \lp \op{a}^{\dagger} + \op{a} \rp \lp
 e^{i\phi(t)}\op{J}_{+} + e^{-i\phi(t)}\op{J}_{-} \rp.
\end{align}
Hence, the applied time dependent rotation $\op{R}_{z}^{\dagger}(t)$
can be interpreted as a rotation of the pseudo-spin vector $\vop{J}$ of
length $j=N/2$ around the $z$ axis by an angle $\phi(t)$.

Since the rotated Dicke Hamiltonian $\op{H}_{\mathrm{RD}}(t)$ is
obtained by applying a unitary 
transformation, namely by the rotation $\op{R}_{z}(t)$, 
the eigenvalues
(eigenenergies) of the rotated Hamiltonian remain the same.
Therefore, the quantum criticalities of the rotated Dicke Hamiltonian can be
obtained by applying a rotation around the $z$ axis to
the time independent Dicke Hamiltonian.
We use this fact and the results of Ref.~\cite{Emary2003} where the
time independent Dicke Hamiltonian is 
diagonalized in the thermodynamic limit.
In the ``normal'' phase (NP) the excitation energy reads
\be
  \epsilon_{-}^{(\mathrm{NP})} =
  \sqrt{
   \frac{1}{2}
    \lp
       \omega^{2}
     + \omega_{0}^{2}
     - \sqrt{(\omega_{0}^{2}-\omega^{2})^{2}+16\lambda^{2}\omega\omega_{0}}
    \rp
  },
\ee
in which the reality condition $\lambda \leq
\frac{\sqrt{\omega\omega_{0}}}{2}$ needs to be fulfilled. 
Similarly, in the ``super-radiant'' phase (SRP) the excitation energy is
\be
  \epsilon_{-}^{(\mathrm{SRP})} =
    \frac{1}{\sqrt{2}}
    \lp
       \frac{16\lambda^{4}}{\omega^{2}}
     + \omega^{2}
     - \sqrt{f(\omega,\lambda)}
    \rp^{\frac{1}{2}},
\ee
which is real under the condition that $\lambda \geq
\frac{\sqrt{\omega\omega_{0}}}{2}$, where
$f(\omega,\lambda)=\lp\frac{16\lambda^{4}}{\omega^{2}}-\omega^{2}\rp^{2}+4\omega^{2}\omega_{0}^{2}$.
Hence, the critical coupling at which the quantum phase transition occurs is given by
$\lambda_{c} = \frac{\sqrt{\omega\omega_{0}}}{2}$.
Using this relation we plot the equilibrium phase diagram for $\omega=1$.
We parametrize it by $\lambda \cos\phi$ for the $x$ axis, and
$\lambda\sin\phi$ for the $y$ axis and $\omega_{0}$ on the $z$ axis.
This is shown in Fig.~\ref{fig:dickemodeleqphdiag} for the purpose of
illustration of our setup.

\section{Coherent state representation - Classical Analogue - Time
  dependent mean field theory}

For concreteness of our protocol we will study quantum dynamics governed by the
Hamiltonian~(\ref{eq:hamrd}) when $\phi(t)$ is linear in time
\be
  \phi(t) = \delta_{\phi} \, t,
\ee
where $\delta_{\phi}$ is the driving velocity and the time interval is $0<t<t_{f}$.
This driving protocol generates circular paths in parameter space (see
Fig.~\ref{fig:dickemodeleqphdiag}).
For $t_{f}=2\pi/\delta_{\phi}$ the path corresponds to one
closed circle.
First we will be interested in the mean photon number
$\expect{\op{a}^{\dagger}\op{a}}$ for the closed circular
paths $\mathcal{C}_{1}$ and $\mathcal{C}_{2}$
\be
  \expect{\op{a}^{\dagger}\op{a}}
  =
  \bra{\psi(t_{f})} \op{a}^{\dagger}\op{a} \ket{\psi(t_{f})},
\ee
and for different initial quantum states $\ket{\psi(0)}$.
Time evolution is governed by 
the time dependent Schr\"odinger equation
$i\partial_{t}\ket{\psi(t)} = \op{H}_{\mathrm{RD}}(t) \ket{\psi(t)}$.
Unfortunately, the Hamiltonian~(\ref{eq:hamrd})
$\op{H}_{\mathrm{RD}}(t)$ is not a linear function of the generators
of the dynamical group, here $\mathrm{SU}(2)$ and $\mathrm{H}_{4}$.
Therefore, an exact solution of the time dependent Schr\"odinger
equation is difficult to obtain and we have to rely on some approximate methods
which may become exact in certain limits (e.g. in the thermodynamic
limit of $j\rightarrow\infty$).
For this reason we will be interested in the dynamics in the large
spin limit $(j \to \infty)$ and also since the actual quantum phase
transition only occurs in the thermodynamic limit.
In this respect the use of the basis of coherent states is optimal
because of the semi-classical nature of this basis.
Further, the use of the coherent states allows us to write some time
dependent mean field approximation for the dynamics which in the
thermodynamic limit (semi-classical limit) becomes exact.
In particular, the spin coherent states~\cite{Zhang1990}
($\mathrm{SU}(2)$ coherent states) are defined as 
\be
  \ket{\zeta(t)} =
  \frac{1}{\lp 1 + \zeta(t) \zeta(t)^{\ast} \rp^{j}}
  \exp\lp \zeta(t) \op{J}_{+} \rp
  \ket{j,-j}
\ee
where $\ket{j,-j}$ is the lowest eigenstate of $\op{J}_{z}$,
$\op{J}_{z}\ket{j,-j}=-j\ket{j,-j}$.
The parameter $\zeta(t) \in \mathbb{C}$ results from the
parametrization of the two-dimensional sphere $S^{2}$ by the
stereographic projection
\be
  \zeta = \tan\frac{\theta}{2}e^{i\varphi}.
\ee
Matrix elements of the spin operators in this basis are given by 
\begin{align}
  \bra{\zeta} \op{J}_{+} \ket{\zeta}
  &= 2j \frac{\zeta^{\ast}}{1+\zeta\zeta^{\ast}}, \\
  \bra{\zeta} \op{J}_{-} \ket{\zeta}
  &= \lp \bra{\zeta} \op{J}_{+} \ket{\zeta} \rp^{\ast}, \\
  \bra{\zeta} \op{J}_{z} \ket{\zeta} 
  &= \zeta \bra{\zeta} \op{J}_{+} \ket{\zeta} - j
  = - j\frac{1-\zeta\zeta^{\ast}}{1+\zeta\zeta^{\ast}}.
\end{align}
Similarly, for the bosonic part of the system we have \cite{Zhang1990} 
\be
  \ket{\alpha(t)} = 
  \exp\lp - \frac{1}{2} \alpha(t) \alpha^{\ast}(t) \rp
  \exp\lp \alpha(t) \op{a}^{\dagger} \rp
  \ket{0},
\ee
where $\ket{0}$ is the field vacuum state $\op{a}\ket{0}=0$ and 
$\alpha(t) \in \mathbb{C}$ a complex number.
It is also convenient to introduce the following parametrization
\beq
  \label{eq:paramalpazeta}
  \zeta(t) &=&
  \frac{q_{1}(t) + i p_{1}(t)}{\sqrt{4j-\lp q_{1}^{2}(t) +
      p_{1}^{2}(t) \rp}}, \\
  \label{eq:paramalpazetaalpha}
  \alpha(t) &=& \frac{1}{\sqrt{2}} \lp q_{2}(t) + i p_{2}(t) \rp, 
\eeq
where $(q_{1},p_{1},q_{2},p_{2})$ describe the phase space of the
system under consideration and the indices 1 and 2 label the spin
and the field subsystem, respectively.

Following the standard procedure
of~\cite{Zhang1990},~\cite{Gilmore1993} we can obtain time dependent
mean field equations of motions.
Namely, defining the classical Hamiltonian as 
\begin{align}
& \bra{\alpha}\bra{\zeta} \op{H}_{\mathrm{RD}}(t) \ket{\zeta}\ket{\alpha}
  \equiv
  \mathcal{H}_{\mathrm{cl}}(t) \nonumber \\
& =
  \frac{1}{2} \omega_{0} \lp q_{1}^{2} + p_{1}^{2} - 2j \rp
  +
  \frac{1}{2} \omega \lp q_{2}^{2} + p_{2}^{2} \rp
  + \nonumber \\
& +
  2 \lambda \sqrt{\frac{4j-(q_{1}^{2}+p_{1}^{2})}{4j}}
  \lp q_{1}\cos\phi(t) + p_{1}\sin\phi(t) \rp
  q_{2}.
\end{align}
we consider a minimization problem for the action integral given by
\be
  \mathcal{S} = \int_{t_{i}}^{t_{f}} \mathcal{L}(q_{1}, \dot{q}_{1}, q_{2}, \dot{q}_{2},t) dt,
\ee
where the corresponding Lagrangian is given by
\be
  \mathcal{L}(q_{1}, \dot{q}_{1}, q_{2}, \dot{q}_{2},t) =
  \bra{\alpha}\bra{\zeta} i \partial_{t} - \op{H}_{\mathrm{RD}} \ket{\zeta}\ket{\alpha}.
\ee
The minimization leads to the Euler-Lagrange equations of motions
\be
  \frac{d}{dt} \lp \frac{\partial \mathcal{L}}{\partial \dot{q}_{i}} \rp - \frac{\partial \mathcal{L}}{\partial q_{i}}
  = 0,
\ee
from which we obtain the Hamiltonian equations of motion
$\frac{\partial \mathcal{H}_{\mathrm{cl}}}{\partial q_{i}} = -
\dot{p}_{i}$ and $\frac{\partial \mathcal{H}_{cl}}{\partial p_{i}}
= \dot{q}_{i}$ taking the following form
\begin{align}
  \label{eq:tdmfeefirst}
  \dot{q}_{1} &=
  \omega_{0} p_{1}
  - 2 \lambda
    \frac{\lp \cos \phi(t) q_{1} + \sin\phi(t) p_{1} \rp  p_{1} q_{2}}
         {\sqrt{4j \lp 4j - \lp q_{1}^{2} + p_{1}^{2} \rp \rp}}  + \nonumber \\
  &+ 2 \lambda
     \sqrt{\frac{4j - \lp q_{1}^{2} + p_{1}^{2}\rp}{4j}} 
     \sin\phi(t) q_{2}, \\
  \dot{p}_{1} &=
  - \omega_{0} q_{1}
  + 2 \lambda
    \frac{q_{1} \lp \cos \phi(t) q_{1} + \sin\phi(t) p_{1} \rp q_{2}}
         {\sqrt{4j \lp 4j - \lp q_{1}^{2} + p_{1}^{2} \rp \rp}}  + \nonumber \\
  &- 2 \lambda
     \sqrt{\frac{4j - \lp q_{1}^{2} + p_{1}^{2}\rp}{4j}} 
     \cos\phi(t) q_{2}, \\
  \dot{q}_{2} &=
  \omega p_{2}, \\
  \label{eq:tdmfeelast}
  \dot{p}_{2} &=
  -\omega q_{2}
  - 2\lambda \sqrt{\frac{4j - \lp q_{1}^{2} + p_{1}^{2}\rp}{4j}}
    \lp \cos \phi(t) q_{1} + \sin\phi(t) p_{1} \rp.
\end{align}

\begin{figure}[h!]
  \begin{center}
    \includegraphics[scale=0.6]{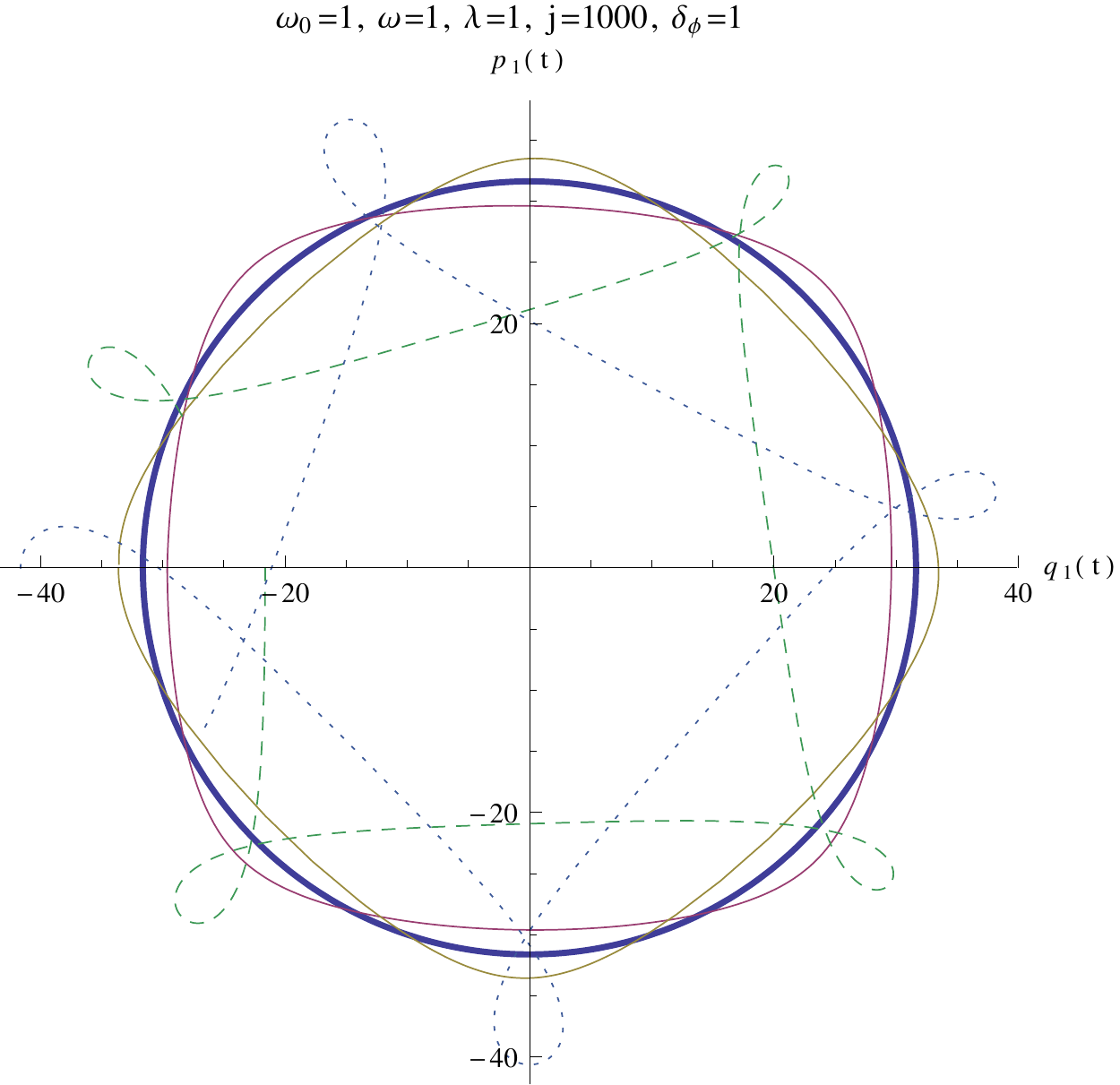}
    \caption{
      Plot of the phase portrait $(q_{1}(t),p_{1}(t))$ for different
      initial conditions.
      }
    \label{fig:phaseportrait}
  \end{center}
\end{figure}

First, let us determine the fixed points of these equations of motions
in the classical phase space $(q_{1}, p_{1}, q_{2}, p_{2})$.
Since, these equations are explicitly time dependent we need to apply a
canonical transformation $(q_{1}, p_{1}, q_{2}, p_{2}) \to (Q_{1},
P_{1}, q_{2}, p_{2})$ that removes the explicit time dependence and
then find out the fixed points.
The first and the simplest fixed point is
\be
  q_{1,c_{1}} = 0, p_{1,c_{1}} = 0, q_{2,c_{1}} = 0, p_{2,c_{1}} = 0. 
\ee
This fixed point is only stable if
\be
  \lambda < \frac{\sqrt{\omega(\omega_{0}+\delta_{\phi})}}{2}.
\ee
In addition, there are two fixed ``circles'' given by
\begin{align}
  q_{1,c_{2}} &= -\sqrt{2j \lp 1 - \frac{\omega(\omega_{0} + \delta_{\phi})}{4\lambda^{2}} \rp} \cos\phi(t), \\
  p_{1,c_{2}} &= -\sqrt{2j \lp 1 - \frac{\omega(\omega_{0} + \delta_{\phi})}{4\lambda^{2}} \rp} \sin\phi(t), \\ 
  q_{2,c_{2}} &= \frac{2\lambda}{\omega} \sqrt{j \lp 1 - \lp \frac{\omega(\omega_{0} + \delta_{\phi})}{4\lambda^{2}}\rp^{2} \rp}, \\
  p_{2,c_{2}} &= 0,
\end{align}
and
\begin{align}
  q_{1,c_{3}} &= \sqrt{2j \lp 1 - \frac{\omega(\omega_{0} + \delta_{\phi})}{4\lambda^{2}} \rp} \cos\phi(t), \\
  p_{1,c_{3}} &= \sqrt{2j \lp 1 - \frac{\omega(\omega_{0} + \delta_{\phi})}{4\lambda^{2}} \rp} \sin\phi(t), \\
  q_{2,c_{3}} &= -\frac{2\lambda}{\omega} \sqrt{j \lp 1 - \lp \frac{\omega(\omega_{0} + \delta_{\phi})}{4\lambda^{2}}\rp^{2} \rp}, \\
  p_{2,c_{3}} &= 0,
\end{align}
which are only real and stable if 
\be
  \lambda >
  \frac{\sqrt{\omega(\omega_{0}+\delta_{\phi})}}{2}.
\ee
We plot the phase portrait of the system in Fig.~\ref{fig:phaseportrait}.

It is instructive to note that the classical equations above can also
be derived following a different procedure.
Using the approach of Ref.~\cite{Emary2003} one can introduce a
Holstein-Primakoff representation for the spin variables as
\begin{align}
  \op{J}_{+}
  &= \op{b}^{\dagger} \sqrt{2j} \sqrt{1-\frac{1}{2j}\op{b}^{\dagger}\op{b}}, \\
  \op{J}_{-}
  &= \sqrt{2j} \sqrt{1-\frac{1}{2j}\op{b}^{\dagger}\op{b}} \, \op{b}, \\
  \op{J}_{z}
  &= \op{b}^{\dagger}\op{b} - j.
\end{align}
Next we apply a time-dependent displacement to both the bosonic field
operators $\op{a}$ and the Holstein-Primakoff bosons $\op{b}$, 
\begin{align}
  \op{D}_{\op{a}}(\alpha(t)) &= \exp\lp \alpha(t)\op{a}^{\dagger} - \alpha^{\ast}(t)\op{a} \rp, \\
  \op{D}_{\op{b}}(\beta(t)) &= \exp\lp \beta(t)\op{b}^{\dagger} - \beta^{\ast}(t)\op{b} \rp.
\end{align}
This transformation takes into account the fact that the bosonic field operators and the atomic
ensemble may acquire a macroscopic occupations.
In the thermodynamic limit $(j \to \infty)$ we find the following
equations for the displacement parameters $\alpha(t)$ and $\beta(t)$:
\begin{align}
& i \dot{\alpha}
 - \omega \alpha
 - \lambda \sqrt{\frac{2j-\beta\beta^{\ast}}{2j}}
   \lp \beta^{\ast}e^{i\phi(t)} + \beta e^{-i\phi(t)}\rp
 = 0, \\
& - i \dot{\alpha}^{\ast}
  - \omega \alpha^{\ast}
  - \lambda \sqrt{\frac{2j-\beta\beta^{\ast}}{2j}}
    \lp \beta^{\ast} e^{i\phi} + \beta e^{-i\phi} \rp
  = 0, \\
& i \dot{\beta} - \omega_{0}\beta
  -
  \lambda\sqrt{\frac{2j-\beta\beta^{\ast}}{2j}}\lp\alpha+\alpha^{\ast}\rp
  e^{i\phi} + \nonumber \\
& +
  \frac{1}{2}\lambda\frac{1}{\sqrt{2j}}\frac{1}{\sqrt{2j-\beta\beta^{\ast}}}
  \lp\alpha+\alpha^{\ast}\rp
  \lp\beta^{\ast}e^{i\phi}+\beta e^{-i\phi}\rp\beta = 0, \\
& -i \dot{\beta}^{\ast} - \omega_{0}\beta^{\ast}
  -\lambda\sqrt{\frac{2j-\beta\beta^{\ast}}{2j}}\lp\alpha+\alpha^{\ast}\rp
  e^{-i\phi} + \nonumber \\
& +
  \frac{1}{2}\lambda\frac{1}{\sqrt{2j}}\frac{1}{\sqrt{2j-\beta\beta^{\ast}}}
  \lp\alpha+\alpha^{\ast}\rp
  \lp\beta^{\ast}e^{i\phi}+\beta e^{-i\phi}\rp\beta^{\ast} = 0,
\end{align}
which are equivalent to the system of equations obtained by the mean field
approximation~(\ref{eq:tdmfeefirst})-(\ref{eq:tdmfeelast}) if we use
the following identification
\be
  \beta = \frac{1}{\sqrt{2}} \lp q_{1} + i p_{1} \rp, \quad
  \alpha = \frac{1}{\sqrt{2}} \lp q_{2} + i p_{2} \rp.
\ee

\section{Results for the mean photon number and the parity}

In this Section we study the time evolution of the scaled mean photon
number and the parity for the rotated Dicke model.
We explore several initial states and compare the results between
driven and undriven situations.
Namely, we consider several physically-relevant initial states: (i) coherent
states $\ket{\alpha}\ket{\zeta}$, (ii) Fock state
$\ket{n=0}\ket{j,m=-j}$ and (iii) the ground state
$\ket{\mathrm{GS}}$.
Before going into detailed studies we introduce important observables
and review the details of numerical procedure.

\subsection{General setup}

\subsubsection{Mean photon number}

The scaled mean photon number for the rotationally driven Dicke model is defined by
\begin{align}
  \frac{1}{j}\expect{\op{a}^{\dagger}\op{a}}
  &=
  \frac{1}{j}\bra{\psi(t_{f})} \op{a}^{\dagger}\op{a}
  \ket{\psi(t_{f})} \nonumber \\
  &= \frac{1}{j}
    \bra{\psi(t_{0})} \op{\mathcal{U}}^{\dagger}(t_{f})
    \, \op{a}^{\dagger} \op{a} \, \,
    \op{\mathcal{U}}(t_{f}) \ket{\psi(t_{0})},
\end{align}
where the evolution operator is the time ordered exponent
\be
 \op{\mathcal{U}}(t) =
 \mathcal{T} \exp\lp -i \int_{t_{0}}^{t}dt'\,\op{H}_{\mathrm{RD}}(t') \rp.
\ee
In particular we will focus on the dependence of
$\frac{1}{j}\expect{\op{a}^{\dagger}\op{a}}$ on the coupling
strength $\lambda$ and the rotation velocity $\delta_{\phi}$ after one
or many revolutions for different initial states.
Specifically, we are interested in the effect of the time-dependent
rotation on the mean photon number.
This gives us an insight into the interplay of the geometric phase and
non-equilibrium time evolution.

In the TDL ($j\to\infty$) the time evolution of the scaled mean photon
number can be derived from the time dependent mean field
equations~(\ref{eq:tdmfeefirst})-(\ref{eq:tdmfeelast}).
Indeed
\be
  \frac{1}{j} \expect{\op{a}^{\dagger}\op{a}}
  =
  \frac{1}{j} \abs{\alpha(t_{f})}^{2}
  =
  \frac{1}{2j} \lp q_{2}(t_{f})^{2} + p_{2}(t_{f})^{2} \rp
\ee
if the system is initially (at $t=t_{0}$) prepared in the product state
\be
  \ket{\psi(t_{0})} = \ket{\alpha(t_{0})} \ket{\zeta(t_{0})}
\ee
of the field coherent state $\ket{\alpha}$ and the spin coherent state
$\ket{\zeta}$,
\begin{align}
  \ket{\alpha(t)} &=
  e^{-\frac{1}{2}\alpha\alpha^{\ast}} \sum_{n=0}^{\infty}
  \frac{\alpha^{n}}{\sqrt{n!}} \ket{n}, \\
  \ket{\zeta(t)} &=
  \frac{1}{(1+\zeta\zeta^{\ast})^{j}}
  \sum_{m=-j}^{j} \zeta^{m+j} \sqrt{\binom{2j}{m+j}} \ket{j,m}.
\end{align}

\subsubsection{Parity operator}

An additional information on dynamical behavior is provided by the
observable, which has a meaning of discrete topological number, namely
the parity of a time-evolved state. 
The parity operator $\op{\Pi}$ of the Dicke model reads
\be
  \op{\Pi} = \exp\lp i \pi \op{N} \rp,
\ee
where
\be
 \op{N}=\op{a}^{\dagger}\op{a} + \op{J}_{z} + j
\ee
counts the number of excited quanta in the system.
Applying it to a basis state $\ket{n}\ket{j,m}$ yields
\be
  \op{\Pi} \ket{n}\ket{j,m} = e^{i\pi(n+m+j)} \ket{n}\ket{j,m},
\ee
since $\op{a}^{\dagger}\op{a}\ket{n}=n\ket{n}$, $\op{J}_{z}\ket{j,m}=m\ket{j,m}$
and $\vop{J}^{2} \ket{j,m}=j(j+1)\ket{j,m}$.
Thus the parity operator has the eigenstates $\ket{n}\ket{j,m}$ and
the corresponding eigenvalues $\pm 1$ (depending whether $n+m+j$ is
even or odd).

The time evolution of the parity operator is given by
\be
  \bra{\psi(t)} \op{\Pi} \ket{\psi(t)} = 
  \sum_{n=0}^{\infty} \sum_{m=-j}^{j}
  \abs{\kappa_{n,m}(t)}^{2} e^{i\pi(n+m+j)},
\ee
where $\kappa_{n,m}(t)$ is defined by the expansion of a state
$\ket{\psi(t)}$ in the basis $\ket{n}\ket{j,m}$, i.e., 
\be
  \ket{\psi(t)} = 
  \sum_{n=0}^{\infty} \sum_{m=-j}^{j}
  \kappa_{n,m}(t) \ket{n}\ket{j,m}.
\ee


In the TDL ($j \to \infty$) we use the solutions of the mean field
equations~(\ref{eq:tdmfeefirst})-(\ref{eq:tdmfeelast}) 
\be
  \alpha = \frac{1}{\sqrt{2}} \lp q_{2} + i p_{2}\rp, \qquad \zeta =
  \frac{q_{1} + i p_{1}}{\sqrt{4j-(q_{1}^{2}+p_{1}^{2})}}
\ee
to calculate the time evolution of the parity for the dynamics
starting from a coherent state
$\ket{\psi(t_{0})}=\ket{\alpha(t_{0})}\ket{\zeta(t_{0})}$.
The time evolution of the parity can be calculated by using
the following properties of the field coherent states
\begin{align}
  e^{i\pi\op{a}^{\dagger}\op{a}}\ket{\alpha} &= \ket{\beta},
  \quad \beta:=e^{i\pi}\alpha, \\
  \bracket{\alpha}{\beta} &=
  \mathrm{exp}\lb \alpha^{\ast}\beta - \frac{1}{2}(\alpha^{\ast}\alpha +\beta^{\ast}\beta ) \rb, \\
  \exp(\mu \op{a}^{\dagger}\op{a}) 
  &= \sum_{k=0}^{\infty} \frac{\lp
    e^{\mu}-1 \rp^{k}}{k!} (\op{a}^{\dagger})^{k}(\op{a})^{k}
\end{align}
and the spin coherent states
\begin{align}
  e^{i\pi \op{J}_{z}} \ket{\zeta} &= \ket{\eta} e^{i\pi(-j)}, \quad
  \eta := e^{i\pi} \zeta, \\
  \bracket{\zeta}{\eta} &= \frac{(1 + \eta \zeta^{\ast})^{2j}}{(1+\zeta\zeta^{\ast})^{j}(1+\eta\eta^{\ast})^{j}}.
\end{align} 
The time evolution of the parity in the coherent state
$\ket{\alpha}\ket{\zeta}$ therefore reads
\be
  \label{eq:parcs}
  \bra{\alpha}\bra{\zeta}\op{\Pi} \ket{\zeta}\ket{\alpha}
  =
  \mathrm{exp}\lp \alpha\alpha^{\ast} (e^{i\pi}-1) \rp
  \lp \frac{1-\zeta\zeta^{\ast}}{1+\zeta\zeta^{\ast}} \rp^{2j}.
\ee


\subsubsection{Numerical procedure}

For a finite number of two-level atoms $N=2j$ or pseudo-spin of finite
length $j=N/2$ the time evolution can not be calculated exactly.
In addition, to study the validity of the time dependent mean field
evolution equations we compare the mean field solution with exact
numerical solution of the Schr\"odinger equation.
For the numerical calculations we truncate the bosonic Hilbert space
up to $n_{\mathrm{M}}$ bosons but we always keep the full Hilbert space
of the pseudo-spin.
We kept $n_{\mathrm{M}}$ always high enough to assure that the error of the
numerical data is on the level of the machine precision.
For the numerical solution we use the Chebyshev scheme~\cite{Tal1984}
to calculate the time evolution.
First, we apply a transformation into a co-rotating basis 
$\ket{\psi}_{\mathrm{ROT}} = \op{R}_{z}(t) \ket{\psi}$: 
\begin{align}
  \expect{\op{a}^{\dagger}\op{a}} &= 
  \bra{\psi(t_{f})} \op{a}^{\dagger}\op{a} \ket{\psi(t_{f})}
  \nonumber \\
  &=
  {}_{\mathrm{ROT}}\bra{\psi(t_{0})} e^{i\op{H}_{\mathrm{ROT}}  t_{f}}
  \,
  \op{a}^{\dagger}\op{a}
  \,
  e^{-i\op{H}_{\mathrm{ROT}} t_{f}} \ket{\psi(t_{0})}_{\mathrm{ROT}}.
\end{align}
Then we compute the time propagator of the time independent
Hamiltonian
\be
  \op{H}_{\mathrm{ROT}}
  =
  (\omega_{0} + \delta_{\phi}) \op{J}_{z}
  +
  \omega \op{a}^{\dagger}\op{a}
  +
  \frac{\lambda}{\sqrt{2j}}
  \lp \op{a}^{\dagger} + \op{a} \rp \lp \op{J}_{+} + \op{J}_{-} \rp,
\ee
which is then expanded in a Chebyshev series:
\be
 e^{-i \op{H}_{\mathrm{ROT}} \Delta t} \approx \sum_{k=0}^{M} a_{k}
 T_{k}(\op{h}_{\mathrm{ROT}}),
\ee
where $T_{k}(\op{h}_{\mathrm{ROT}})$ are the Chebyshev polynomials of order
$k$ and $\op{h}_{\mathrm{ROT}}$ is a rescaled Hamiltonian
\be
  \op{h}_{\mathrm{ROT}} = \frac{2\op{H}_{\mathrm{ROT}} - (E_{\mathrm{Max}}+E_{\mathrm{Min}})\op{1}}{E_{\mathrm{Max}}-E_{\mathrm{Min}}}.
\ee
Here $E_{\mathrm{Max}}$ and $E_{\mathrm{Min}}$ are the largest and the smallest
eigenvalue of $\op{H}_{\mathrm{ROT}}$.
The expansion coefficients $a_{k}$ are determined by
\begin{align}
  a_{k}
  =&
  (-i)^{k} 
  \exp\lp - i \Delta t \frac{1}{2}
  (E_{\mathrm{Max}} + E_{\mathrm{Min}})
  \rp
  (2-\delta_{k,0}) \nonumber \\
  & J_{k}(\Delta t\frac{1}{2}(E_{\mathrm{Max}} - E_{\mathrm{Min}})),
\end{align}
where $\delta_{k,0}$ is the Kronecker delta and $J_{k}(x)$ are the
Bessel functions of the first kind.

In the same way we also compute the time evolution
of the parity operator for a system with a finite number of
two-level atoms.

\subsection{Stationary Dicke initial state: $\ket{\alpha_{c_{\mathrm{st}}}}\ket{\zeta_{c_{\mathrm{st}}}}$}

First, we study the mean photon number if the system is initially
(at $t_{0}=0$) prepared in the product state of the field and spin
coherent states
\be
  \ket{\psi(0)} = \ket{\alpha_{c_{\mathrm{st}}}} \otimes \ket{\zeta_{c_{\mathrm{st}}}}
\ee
with
\begin{align}
 &\alpha_{c_{\mathrm{st}}} =
   \begin{cases}
    \frac{2 \lambda}{\omega} \sqrt{\frac{j}{2} \lp 1 -
    \lp\frac{\omega\omega_{0}}{4\lambda^{2}}\rp^{2} \rp}, & \lambda \geq \frac{1}{2} \sqrt{\omega\omega_{0}} \\
    0, & \lambda < \frac{1}{2} \sqrt{\omega\omega_{0}}
   \end{cases}, 
   \\
  &\zeta_{c_{\mathrm{st}}} =
   \begin{cases}
    -
    \sqrt{
     \frac{4\lambda^{2} - \omega\omega_{0}}{4\lambda^{2} + \omega\omega_{0}}
    },
    & \lambda \geq \frac{1}{2} \sqrt{\omega\omega_{0}} \\
    0, & \lambda < \frac{1}{2} \sqrt{\omega\omega_{0}}
   \end{cases}.
\end{align}
Coherent state with these parameters corresponds to the
stationary state of the time
independent (unrotated) Dicke model
(equations~(\ref{eq:tdmfeefirst})-(\ref{eq:tdmfeelast}) for $\delta_{\phi}=0$). 
This can be seen by inverting the parametrization used
previously, see Eq.~(\ref{eq:paramalpazeta}) and Eq.~(\ref{eq:paramalpazetaalpha}).
Therefore, we obtain
\begin{align}
  q_{1}(0) &=
    \begin{cases}
     - 
     \sqrt{
      2j \lp
           1 - \frac{\omega\omega_{0}}{4\lambda^{2}}
         \rp
          }, & \lambda \geq \frac{1}{2} \sqrt{\omega\omega_{0}} \\
     0, & \lambda < \frac{1}{2} \sqrt{\omega\omega_{0}}
    \end{cases}
    ,
    \\
  p_{1}(0) &= 0, \\
  q_{2}(0) &=
   \begin{cases}
     \frac{2\lambda}{\omega}
     \sqrt{
           j \lp 1 - \lp
           \frac{\omega\omega_{0}}{4\lambda^{2}} \rp^{2} \rp
     }, & \lambda \geq \frac{1}{2} \sqrt{\omega\omega_{0}} \\
     0, & \lambda < \frac{1}{2} \sqrt{\omega\omega_{0}}
    \end{cases}
    ,
    \\
  p_{2}(0) &= 0.
\end{align}
Our driving protocol is the following: we start in the initial
state $\ket{\alpha_{c_{\mathrm{st}}}}\ket{\zeta_{c_{\mathrm{st}}}}$
and then we switch on the rotation at $t_{0}=0$ and let the system
evolve until $t_{f}=n_{\mathrm{R}} \, 2\pi/\delta_{\phi}$, where $n_{\mathrm{R}}$
indicates the number of rotations.
In order to understand the influence of the rotation we also computed
the time evolution where we do not switch on the rotation at $t_{0}=0$
but where we let the system evolve according to the usual
Dicke Hamiltonian $\op{H}_{\mathrm{D}}$.

\subsubsection{Time dependence of the mean photon number}

\begin{figure}[h!]
  \begin{center}
    \includegraphics[scale=0.43]{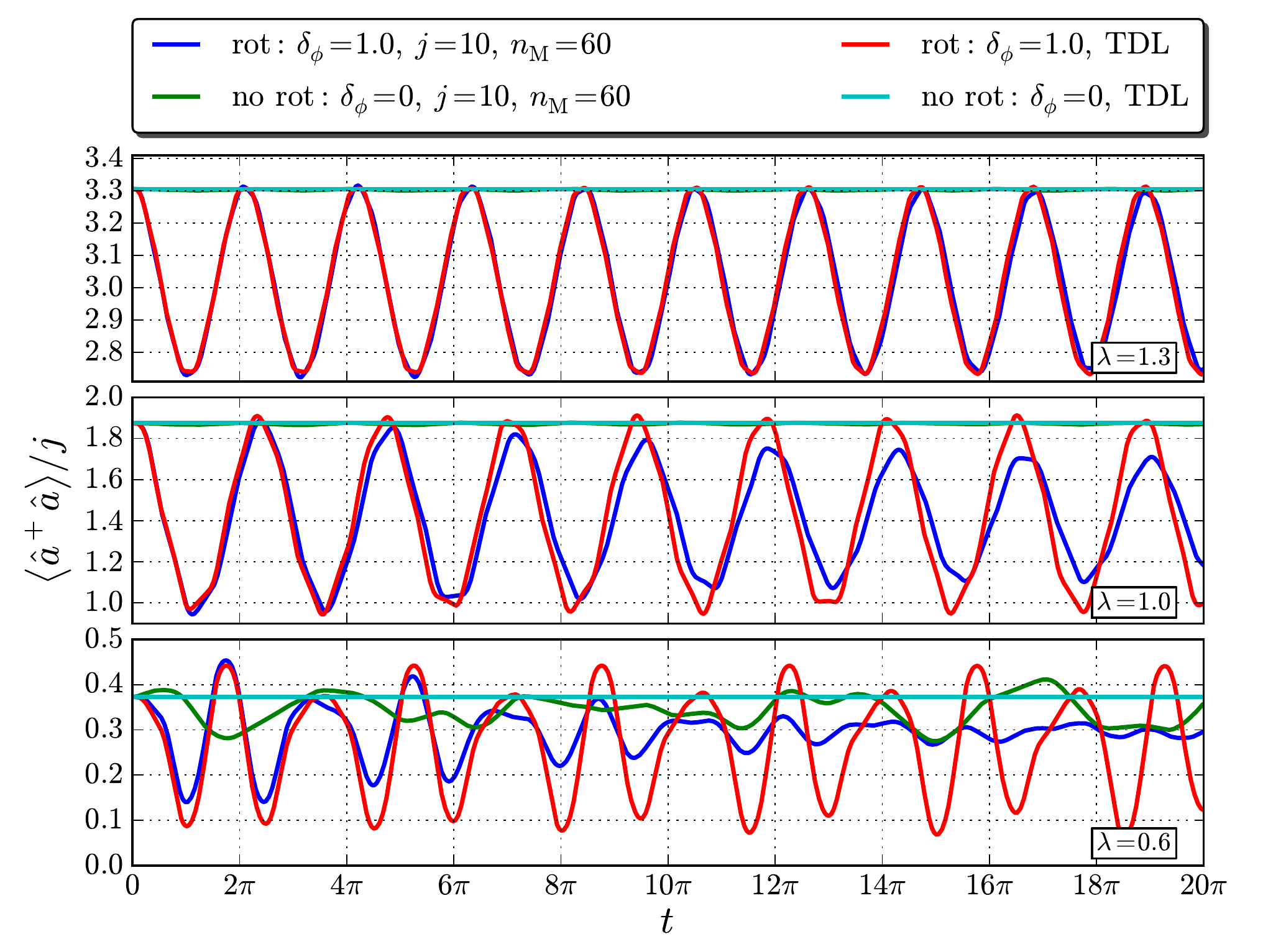}
    \caption{
      Time dependence of the scaled mean photon number on resonance
      $\omega = \omega_{0} = 1.0$ for different values of the coupling
      strength $\lambda=0.6, 1.0, 1.3$.
      The red and cyan curves are the time evolution obtained from the time
      dependent mean field equations and the blue and green from the
      numerical integration of the time dependent Schr\"odinger
      equation for a finite sized system.
      Here the red and blue lines correspond to
      rotationally driven system with a driving velocity of $\delta_{\phi}=1.0$ and the
      cyan and green curves correspond to the time evolution with no rotational
      driving.
      }
    \label{fig:adatimescs}
  \end{center}
\end{figure}
Fig.~\ref{fig:adatimescs} compares the time evolution of the
scaled mean photon number, $\expect{\op{a}^{\dagger}\op{a}}(t)/j$ for
the rotationally driven Dicke model and the time evolution of the
undriven Dicke model.
Part of the data refer to the TDL at $j \to \infty$, using the mean field solution (red
and cyan), while the other part corresponds to the system with a finite number of two level
atoms (blue and green).
We observe that the mean field solution works very well in the strong
coupling regime and it becomes more and more accurate with increasing
$j$, as expected.

\subsubsection{Dependence of the mean photon number on the atom-field
  coupling strength $\lambda$}

\begin{figure}[h!]
  \begin{center}
    \includegraphics[scale=0.45]{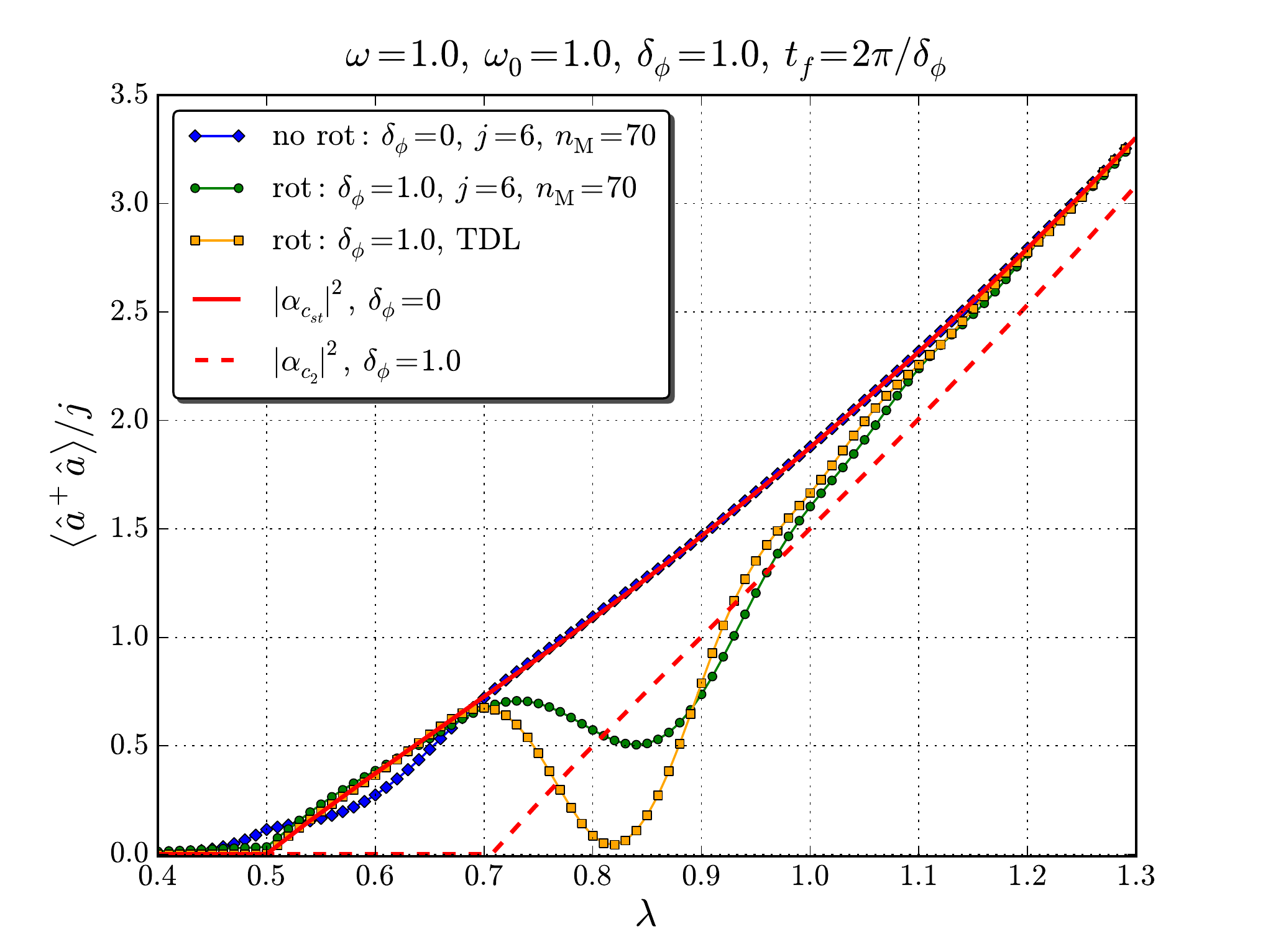}
    \caption{
      The scaled mean photon number as a function of the
      coupling strength. Comparing the mean field solution
      ($j\to\infty$) with the numerical solution for a system with a
      finite number of atoms. We also compare the rotationally
      driven Dicke model with the undriven Dicke model.
      }
    \label{fig:adalamscs}
  \end{center}
\end{figure}

The dependence of the scaled mean photon number on the coupling
strength $\lambda$ after time evolution of the system corresponding to a single circle,
$t_{f}=2\pi/\delta_{\phi}$, is shown in
Fig.~\ref{fig:adalamscs} for a fixed driving velocity
$\delta_{\phi}=1.0$ and on resonance $\omega=\omega_{0}=1.0$.  
Again, the mean photon number is computed using the time dependent
mean field equations which should be valid in the TDL ($j\to\infty$,
see orange curve) and also by solving the Schr\"odinger equation numerically (here
$j=6$, see green curve).
We also compare the results with the evolution of the Dicke
model in the absence of rotations (blue curve).

A clear feature of the rotational driving is the appearance of a
minimum in the super radiant phase (in the case of the Fig.~\ref{fig:adalamscs}
around $\lambda \approx 0.82$).
By looking into the time evolution of the scaled mean photon number we
observe from Fig.~\ref{fig:adatimescs} that the time evolution is not
periodic in the region of the minimum which may be attributed to the
onset of chaotic behavior noticed in Ref.~\cite{Emary2003}.
The new minimum is an indication of the competition between two
stationary states, corresponding to driven and unrotated Dicke models.

\begin{figure}[h!]
  \begin{center}
    \includegraphics[scale=0.43]{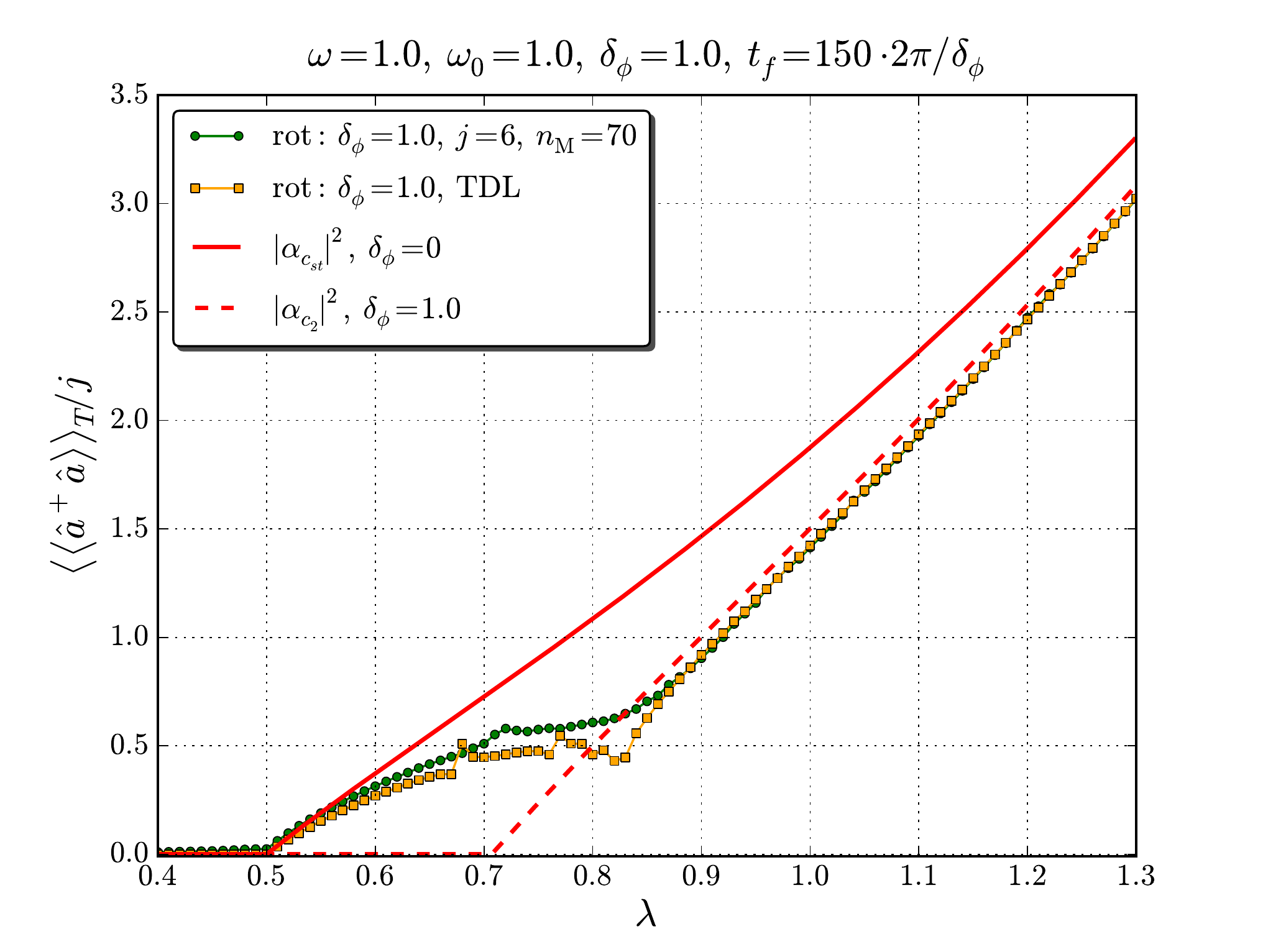}
    \caption{
      Time averaged mean photon number as a function of $\lambda$. 
      The presence of a new minimum in the super radiant phase
      indicates a reentrant behavior as a consequence of a competition
      between geometric and dynamical phases~\cite{TPG}.
      }
    \label{fig:adalamtascs}
  \end{center}
\end{figure}

Now let us consider an interesting limit of multiple rotations $\phi_{f}
\gg 1$.
This limit is interesting because of a competition between dynamical
and geometric phases \cite{TPG}.
In this limit it is natural to look into the time averaged scaled mean
photon number
\be
  \langle\expect{\op{a}^{\dagger}\op{a}}\rangle_{T}/j :=
  \frac{1}{t_{f}-t_{0}} \int_{t_{0}}^{t_{f}} dt \, \expect{\op{a}^{\dagger}\op{a}}/j.
  \label{eq:timeavesmph}
\ee
In Fig.~\ref{fig:adalamtascs} we illustrate the time averaged scaled mean
photon number obtained for a fixed rotation velocity $\delta_{\phi}=1.0$
by solving the mean field equations (orange curve) and by
solving the Schr\"odinger equation for a system with $j=6$ (green
curve). 
We conclude therefore that the averaging over many rotation reveals a
meta-stable phase in the 
super-radiant phase $\lambda>\lambda_{c}$. 
This phase is determined by a dynamical critical atom-field
coupling strength $\lambda_{c}^{(\mathrm{dyn})}$.
In the case of $\omega=\omega_{0}=1.0$ and $\delta_{\phi}=1.0$ it is
given by $\lambda_{c}^{(\mathrm{dyn})}\approx0.82$.

\subsubsection{Dependence of the mean photon number on the rotation
  velocity $\delta_{\phi}$}

\begin{figure}[h!]
  \begin{center}
    \includegraphics[scale=0.43]{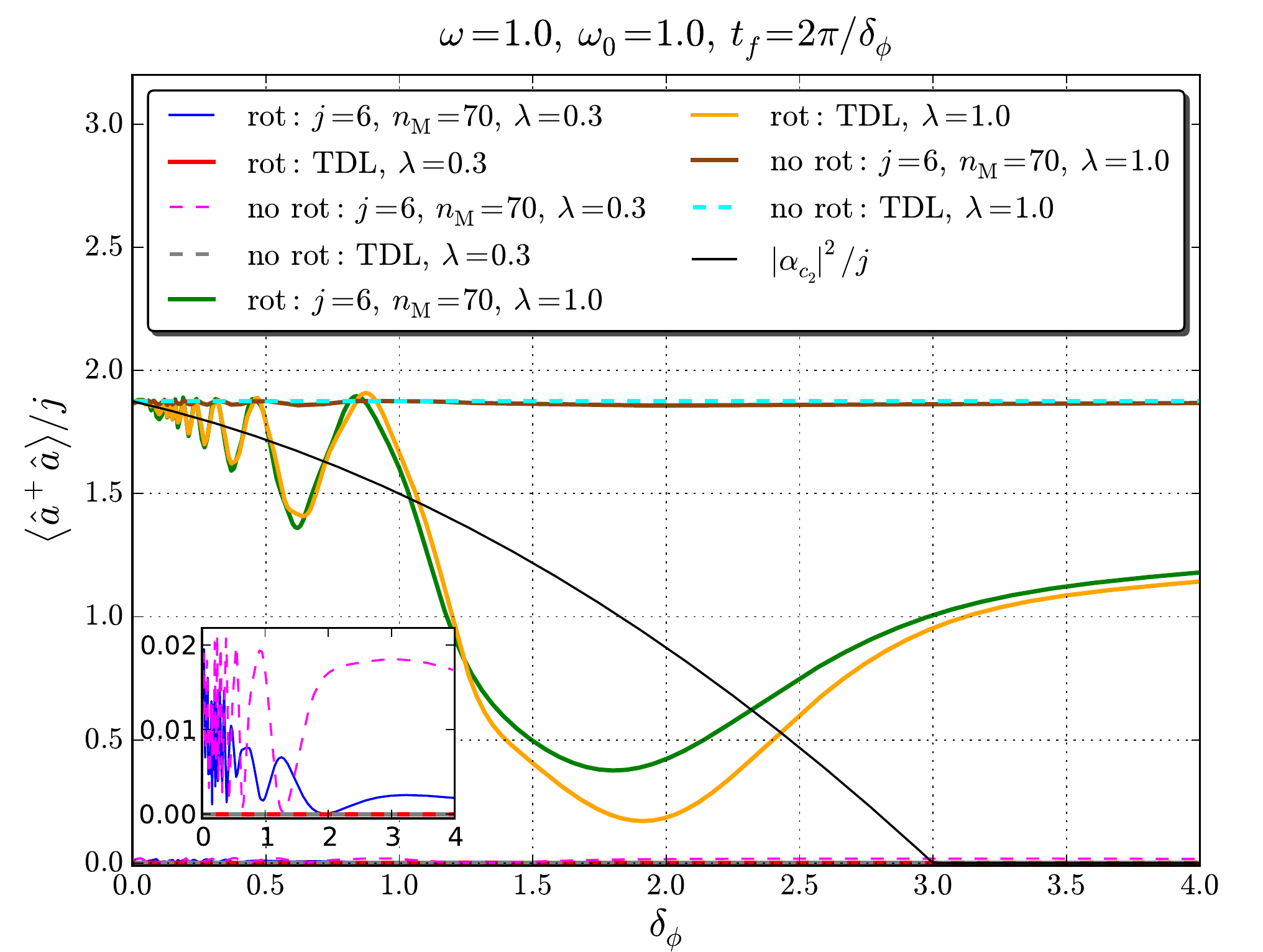}
    \caption{The driving velocity dependence of the scaled
      mean photon number after a single circle.
      }
    \label{fig:adadphics}
  \end{center}
\end{figure}

The dependence of the scaled mean photon number on the driving 
velocity after evolving the system for one circle
$t_{f}=2\pi/\delta_{\phi}$, with different atom-field coupling
parameters $\lambda$ is plotted in Fig.~\ref{fig:adadphics}.
We observe that if the critical paraboloid is not encircled $\lambda <
\lambda_{c} = \sqrt{\omega\omega_{0}}/2$, the mean photon number is zero.
But if we encircle the critical paraboloid the mean photon number is
non-zero and there should be a critical driving velocity.
Similar to the case of the stationary state of the rotated Dicke
model, where the critical atom-field coupling is given by
$\sqrt{\omega(\omega_{0}+\delta_{\phi})}/2$ which defines a critical
driving velocity $\delta_{\phi,c}^{(\mathrm{rot})}$ (see the black curve in
Fig.~\ref{fig:adadphics}: $\abs{\alpha_{c_{2}}}^{2}/j$).
Note however that in the case of one rotation we can not see an indication of presence of the 
critical driving velocity in the dynamics.

\begin{figure}[h!]
  \begin{center}
    \includegraphics[scale=0.43]{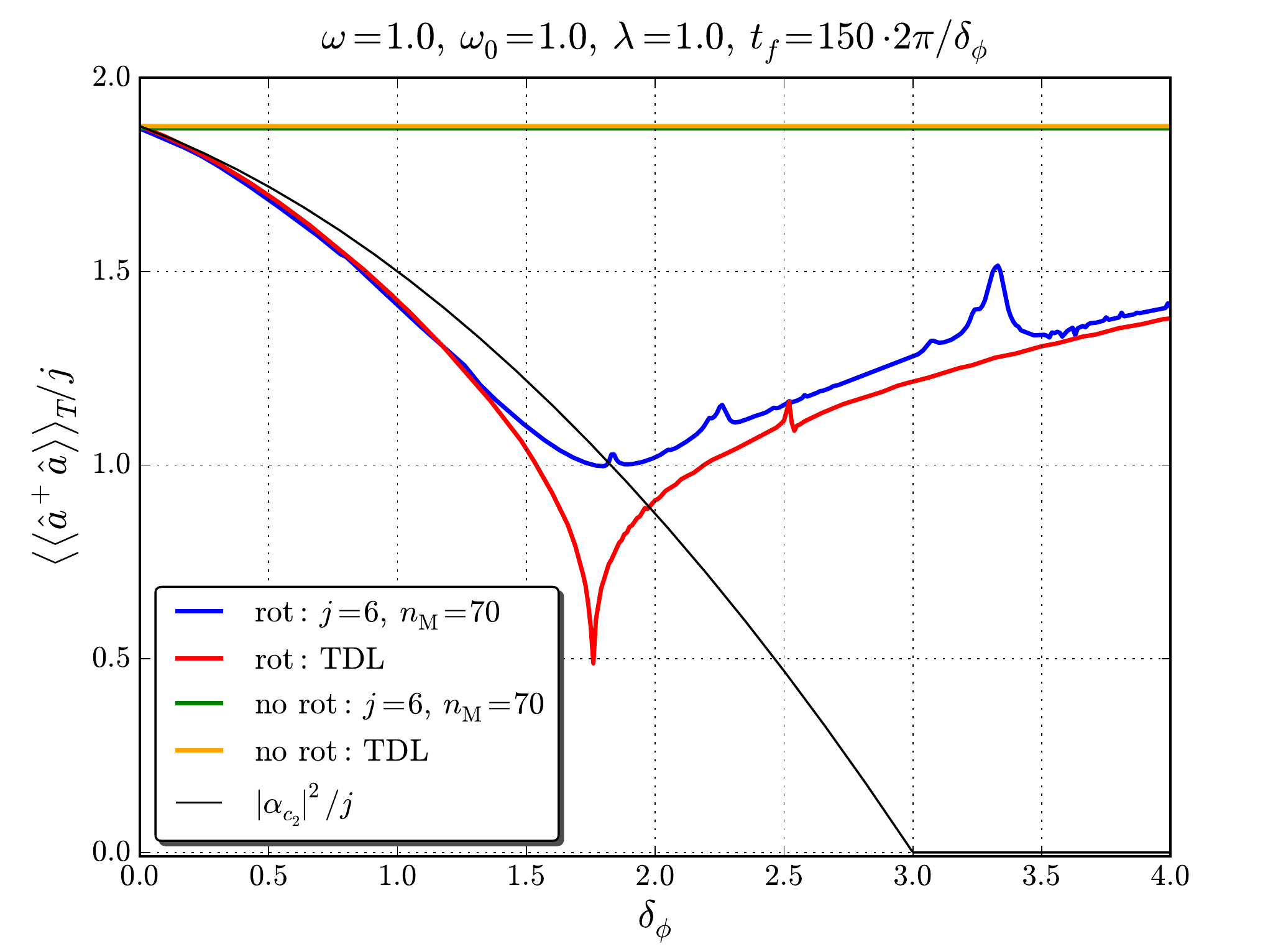}
    \caption{
      Time averaged scaled mean photon number as a function of the
      rotation velocity.
      We compare
      $\langle\langle\op{a}^{\dagger}\op{a}\rangle\rangle_{T}/j$ for
      the rotationally driven Dicke model with the one for the undriven
      Dicke model.
      In both cases we kept $t_{f}=150 \cdot 2\pi/\delta_{\phi}$.
      }
    \label{fig:adadphitacs}
  \end{center}
\end{figure}

In Fig.~\ref{fig:adadphitacs} we also plot the time averaged scaled mean
photon number which clearly illustrate the presence of a dynamical or
rotational critical driving velocity
$\delta_{\phi,c}^{(\mathrm{dyn})}$ in the TDL.

\subsubsection{Non-Equilibrium Phase Diagram}

\begin{figure}[h!]
  \begin{center}
    \begin{tabular}{c}
     \includegraphics[scale=0.7]{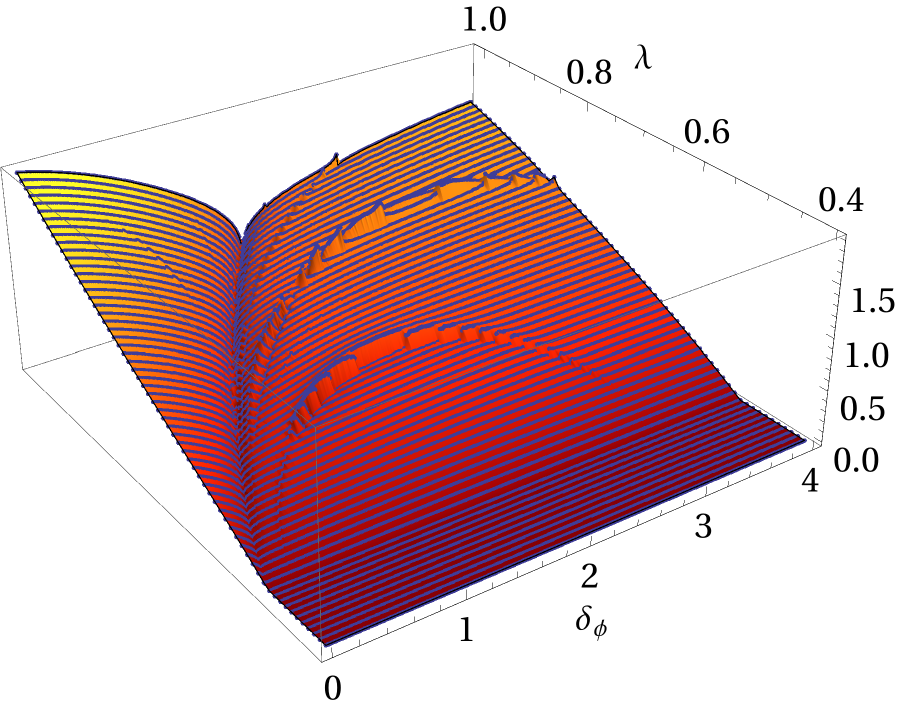} \\
     \includegraphics[scale=0.43]{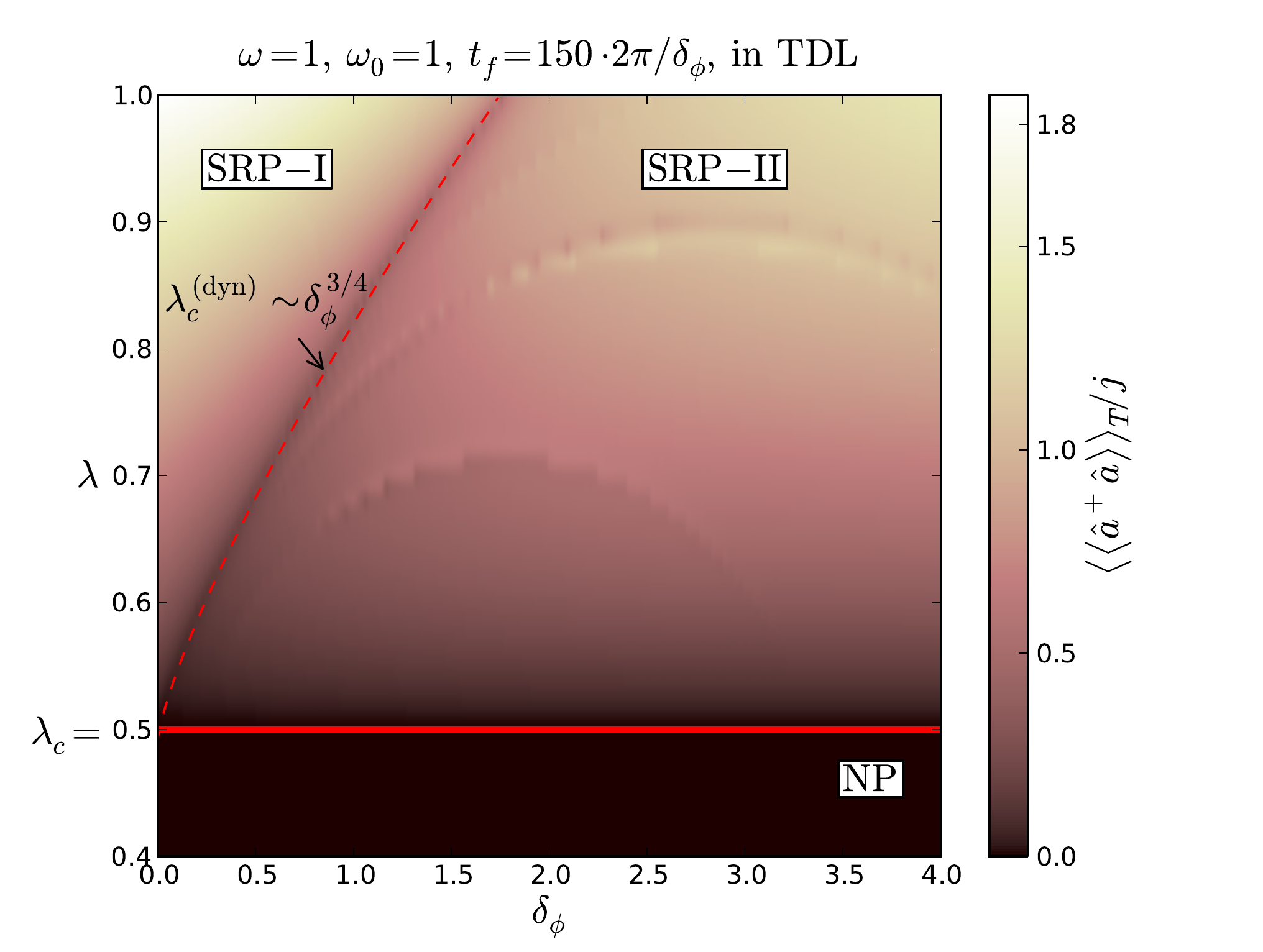}
    \end{tabular}
    \caption{
      Dependence of the time averaged scaled mean photon
      number on velocity of rotation and the atom-field coupling
      strength. It is obtained from the mean
      field equations for $\omega=\omega_{0}=1.0$ and averaged over
      a time $t_{f}=150 \cdot 2\pi/\delta_{\phi}$.
      }
    \label{fig:adaneqpdscs}
  \end{center}
\end{figure}

Summarizing our studies, in Fig.~\ref{fig:adaneqpdscs} we plot the
time averaged scaled mean photon number as a function of the driving
velocity and the atom-field coupling strength obtained from the mean
field equations~(\ref{eq:tdmfeefirst})-(\ref{eq:tdmfeelast}).
One can clearly distinguish two distinct ``meta stable'' phases in the
super-radiant phase. 
We therefore suggest this plot as a non-equilibrium phase
diagram of the rotationally driven Dicke model.
We associate this reentrant behavior with a competition between
geometric and dynamical phases.
This mechanism is described in Ref.~\cite{TPG}. 

The line in the super-radiant phase (the red dashed curve in
Fig.~\ref{fig:adaneqpdscs}) which describes a dynamical critical line,
was fitted with the function
$\lambda_{c}^{(\mathrm{dyn})}=0.5+0.327\,\delta_{\phi}^{3/4}$.
It defines a dynamical critical atom-filed coupling strength
$\lambda_{c}^{(\mathrm{dyn})}$ or reciprocally a dynamical critical
driving velocity $\delta_{\phi,c}^{(\mathrm{dyn})}$.
This should not be confused with the critical coupling
$\lambda_{c}^{(\mathrm{rot})} =
\sqrt{\omega(\omega_{0}+\delta_{\phi})}/2$, that emerges if we start
the evolution of a system from the stationary state
$\ket{\alpha_{c_{2}}}\ket{\zeta_{c_{2}}}$ (see next paragraph).

\subsubsection{Parity $\op{\Pi}$}

\begin{figure}[h!]
  \begin{center}
    \includegraphics[scale=0.43]{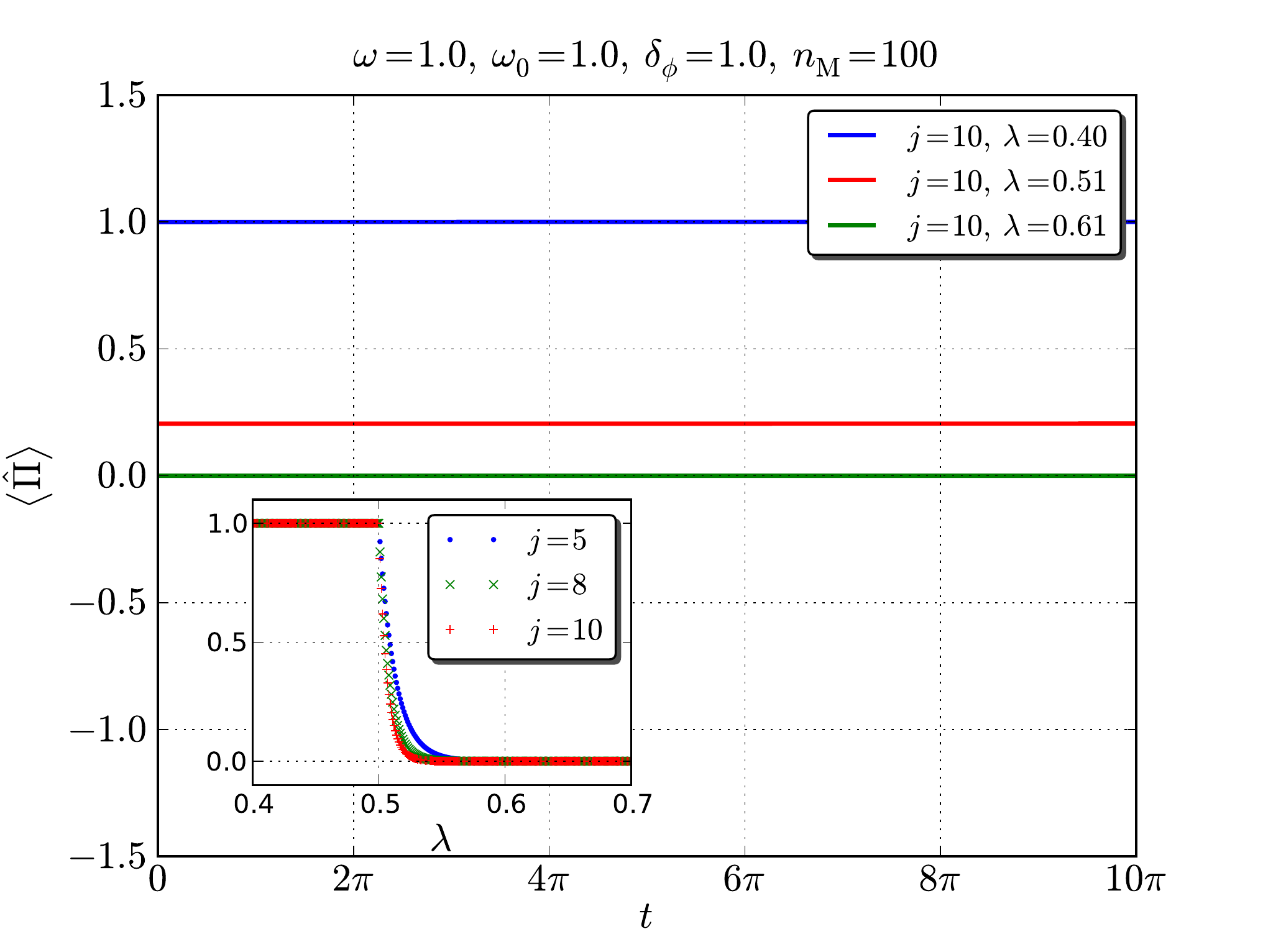}
    \caption{Time evolution of the parity operator when we start the
      dynamics in the coherent state
      $\ket{\alpha_{\mathrm{st}}}\ket{\zeta_{\mathrm{st}}}$
      for a system with a finite number of
      atoms. The inset plot shows the
      dependence on the coupling constant $\lambda$ for different
      values of $j$.}
    \label{fig:pifscs}
  \end{center}
\end{figure}

In Fig.~\ref{fig:pifscs} we show the time evolution of the parity
operator starting in the coherent state
$\ket{\alpha_{\mathrm{st}}}\ket{\zeta_{\mathrm{st}}}$
for a system with a finite number of two-level atoms.
The parity is constant in time and once we enter
a super-radiant phase $\lambda > \lambda_{c} =
\sqrt{\omega\omega_{0}}/2$ it starts to drop from one to zero.
The inset plot shows the $\lambda$ dependence of the parity and its
tendency to a step function with growing $j$.

\begin{figure}[h!]
  \begin{center}
    \includegraphics[scale=0.43]{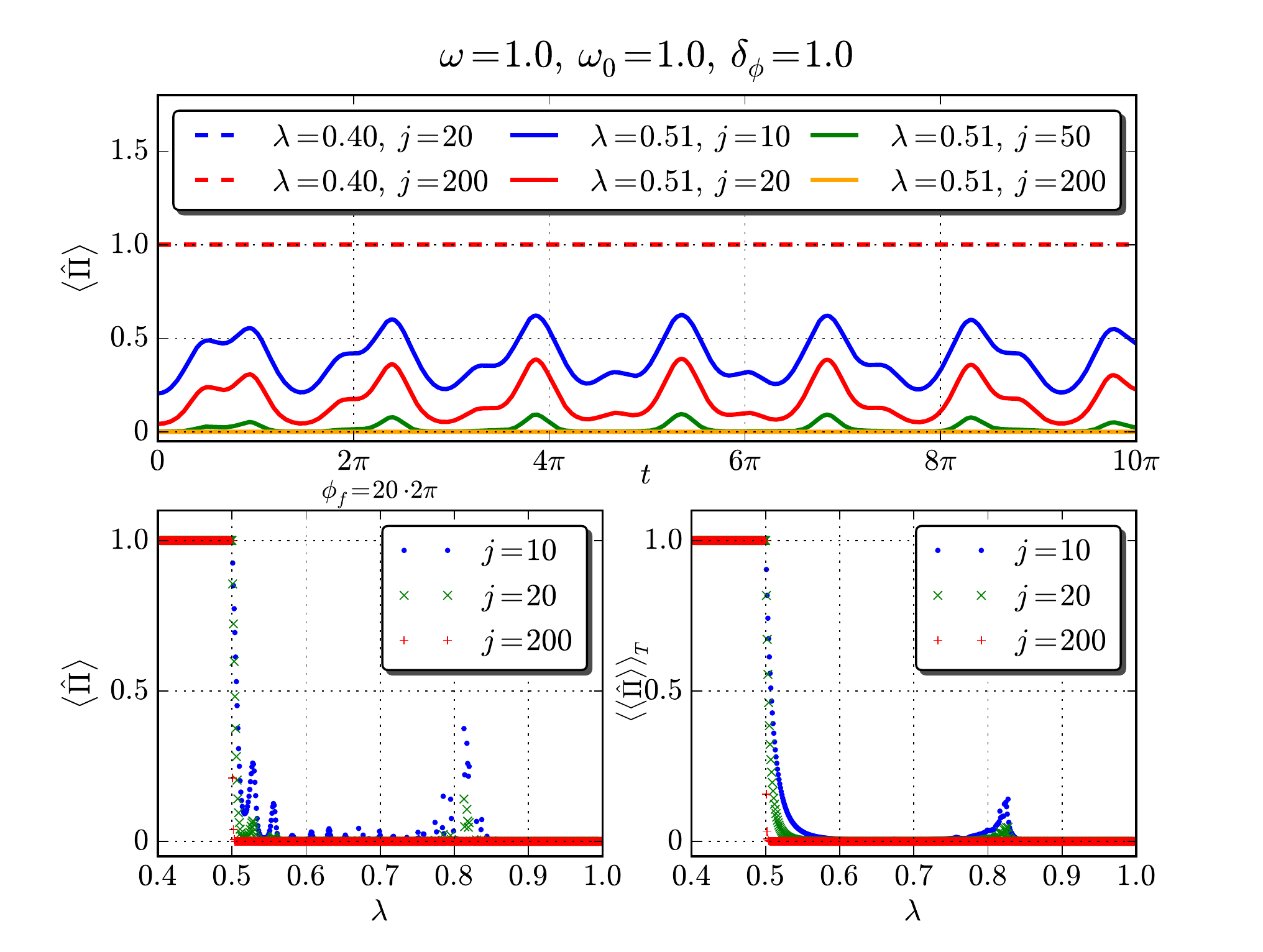}
    \caption{
      Upper panel: time evolution of the parity operator if we start the
      dynamics in the coherent state
      $\ket{\alpha_{\mathrm{st}}}\ket{\zeta_{\mathrm{st}}}$.
      Lower left panel: the dependence on the atom-field coupling constant
      $\lambda$ of the parity after $20$ circle.
      Lower right panel:
      the dependence of the time averaged parity on the coupling constant
      $\lambda$.
      The parity is obtained by solving the
      time dependent mean field equations.
      }
    \label{fig:pitdlcs}
  \end{center}
\end{figure}

The time evolution of the parity $\expect{\op{\Pi}}$ and its
dependence on the atom-field coupling $\lambda$ calculated from
the mean field equations, i.e., by Eq.~(\ref{eq:parcs}), is shown in
Fig.~\ref{fig:pitdlcs}.
The parity is not a constant function of time in the previously
observed intermediate regime of $\expect{\op{a}^{\dagger}\op{a}}/j$.
The time averaged parity as well reveals the observed dynamical critical
coupling $\lambda_{c}^{(\mathrm{dyn})}\approx0.82$.
Nevertheless, for $j\to\infty$ the parity becomes constant in time and
goes to 0 in the entire super-radiant phase.

\begin{figure}[h!]
  \begin{center}
    \includegraphics[scale=0.43]{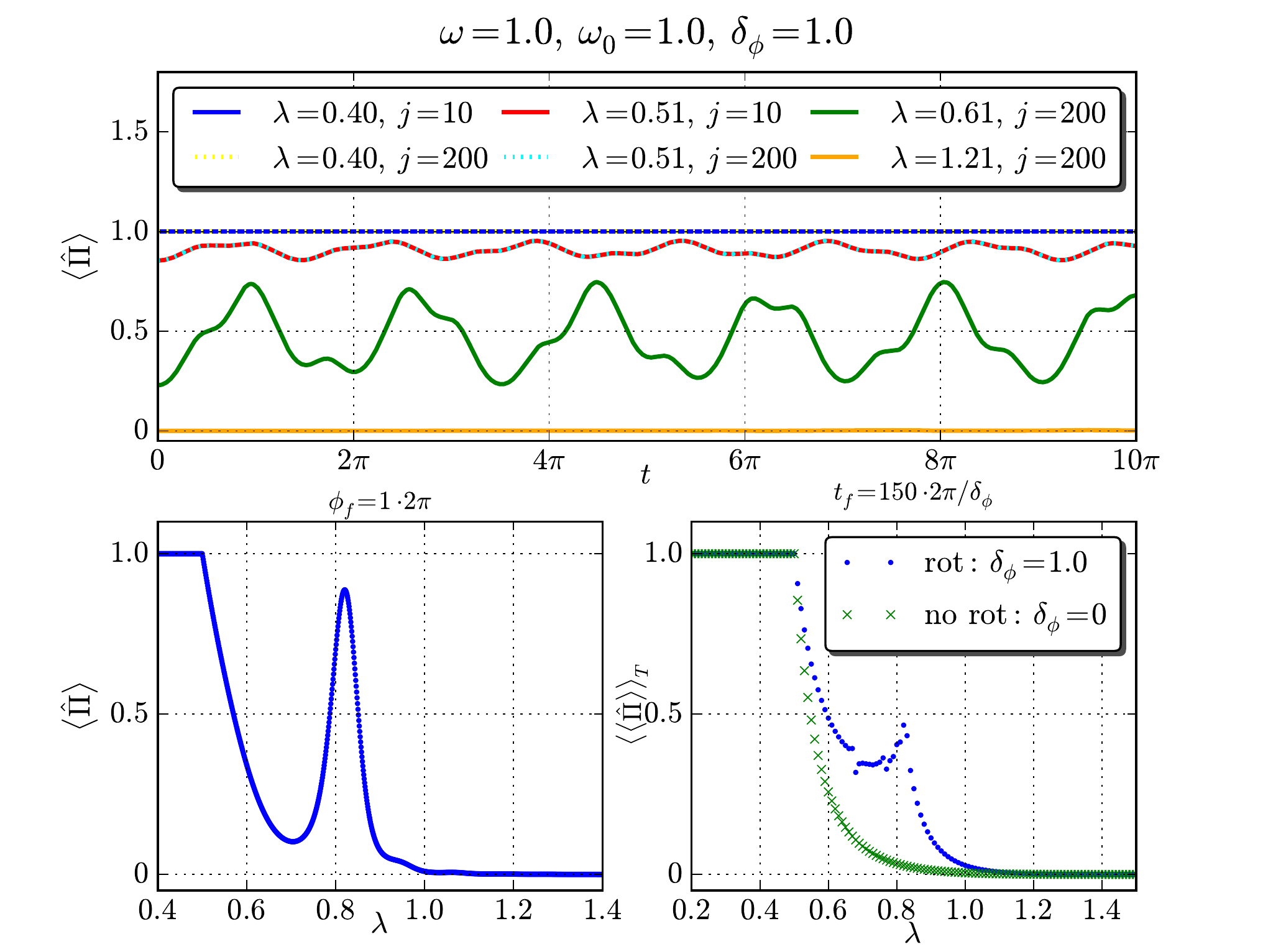}
    \caption{
      Upper panel: time evolution of the scaled parity operator if we start the
      dynamics in the coherent state
      $\ket{\alpha_{\mathrm{st}}}\ket{\zeta_{\mathrm{st}}}$.
      Lower left panel: the dependence of the scaled parity on the coupling constant
      $\lambda$ after one circle.
      Lower right panel:
      the dependence of the time averaged scaled parity on the coupling constant $\lambda$.
      The scaled parity was computed by solving the
      time dependent mean field equations.
      }
    \label{fig:pitdlcsscaled}
  \end{center}
\end{figure}

Further, in Fig.~\ref{fig:pitdlcsscaled} we also illustrate the scaled
parity, namely the parity where all the phase space coordinates
$(q_{i},p_{i})$ are rescaled by $\sqrt{j}$: $(q_{i},p_{i}) \to
(q_{i}/\sqrt{j},p_{i}/\sqrt{j})$. 
This implies
\begin{align}
&  \bra{\alpha}\bra{\zeta} \op{\Pi}_{\mathrm{sc}}
  \ket{\zeta}\ket{\alpha}  = \nonumber \\
&  \mathrm{exp}\lp \frac{1}{2j}\lp q_{2}^{2}+p_{2}^{2} \rp (e^{i\pi}-1) \rp
   \lp 1 - \frac{q_{1}^{2}+p_{1}^{2}}{2j^{2}} \rp^{2j}.
\end{align}
The scaled parity is naturally independent of the number of spins $j$
but it is no longer constant in time for couplings between $0.5$ and
$1$. 
The time averaged scaled parity shows also an intermediate regime
similar to the intermediate regime in the time averaged scaled mean
photon number.

\subsection{Stationary circle initial state: $\ket{\alpha_{c_{2}}}\ket{\zeta_{c_{2}}}$}

Here we study the mean photon number if the system is initially (at
$t_{0}=0$) prepared in the product state of the field and spin
coherent states
\be
 \ket{\psi(0)} = \ket{\alpha_{c_{2}}} \otimes \ket{\zeta_{c_{2}}},
 \label{eq:in-cond-circle}
\ee
with
\begin{align}
 &\alpha_{c_{2}} =
   \begin{cases}
    \frac{2\lambda}{\omega} \sqrt{\frac{j}{2} \lp 1 -
    \lp\frac{\Omega}{4\lambda^{2}}\rp^{2} \rp} & \text{if } \lambda \geq \frac{1}{2} \sqrt{\Omega} \\
    0 & \text{if } \lambda < \frac{1}{2} \sqrt{\Omega}
   \end{cases}, 
   \\
  &\zeta_{c_{2}} =
   \begin{cases}
    -
    \sqrt{\frac{4\lambda-\Omega}{4\lambda+\Omega}}
    & \text{if } \lambda \geq \frac{1}{2} \sqrt{\Omega} \\
    0 & \text{if } \lambda < \frac{1}{2} \sqrt{\Omega}
   \end{cases}.
\end{align}
where $\Omega=\omega(\omega_{0}+\delta_{\phi})$.
This choice of the parameters $\alpha$ and $\zeta$ of the coherent
states corresponds to the stationary ``circle'' solutions of the time
dependent mean field
equations~(\ref{eq:tdmfeefirst})-(\ref{eq:tdmfeelast}).
In the same way as in the previous section this can be obtained using
Eq.~(\ref{eq:paramalpazeta}) which gives 
\begin{align}
  q_{1}(0) &=
    \begin{cases}
     - 
     \sqrt{
      2j \lp
           1 - \frac{\Omega}{4\lambda^{2}}
         \rp
          }, & \lambda \geq \frac{1}{2} \sqrt{\Omega} \\
     0, & \lambda < \frac{1}{2} \sqrt{\Omega}
    \end{cases}
    ,
    \\
  p_{1}(0) &= 0, \\
  q_{2}(0) &=
   \begin{cases}
     \frac{2\lambda}{\omega}
     \sqrt{
           j \lp 1 - \lp
           \frac{\Omega}{4\lambda^{2}} \rp^{2} \rp
     }, & \lambda \geq \frac{1}{2} \sqrt{\Omega} \\
     0, & \lambda < \frac{1}{2} \sqrt{\Omega}
    \end{cases}
    ,
    \\
  p_{2}(0) &= 0.
\end{align}

The driving protocol is the following: we start with the system
prepared in the state $\ket{\alpha_{c_{2}}}\ket{\zeta_{c_{2}}}$.
Then we switch on the rotation at $t_{0}=0$ and let the system evolve
according to $\op{H}_{\mathrm{RD}}(t)$ until
$t_{f}=n_{\mathrm{R}}\,2\pi/\delta_{\phi}$.
Here $n_{\mathrm{R}}$ indicates the number of rotations in the space
of parameters (see Fig.~\ref{fig:dickemodeleqphdiag}).
Further, we consider the time evolution where we do not switch on the
rotation at $t_{0}=0$ but where we evolve the system according to
$\op{H}_{\mathrm{D}}$ (undriven case) in order to understand the
influence of the rotational driving.

\subsubsection{Mean photon number for the stationary coherent state}

\begin{figure}[h!]
  \begin{center}
    \includegraphics[scale=0.45]{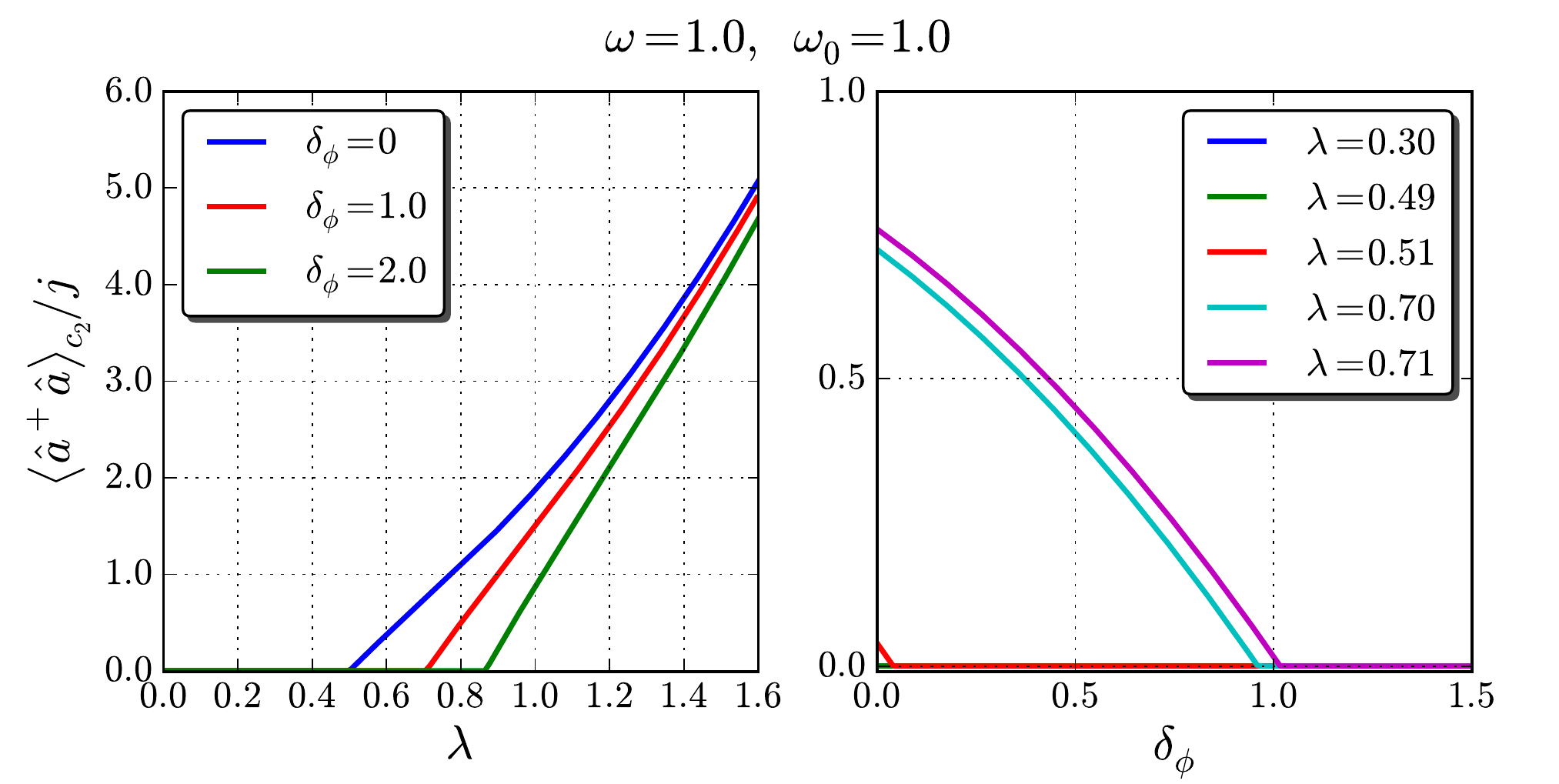}
    \caption{
      Plot of the mean photon number
      $\expect{\op{a}^{\dagger}\op{a}}_{c_{2}} / j$ as a function of
      the coupling strength $\lambda$ on the left panel and as a
      function of the rotation velocity $\delta_{\phi}$ on the right
      panel.
       }
    \label{fig:meanphotonlambdadepanddeltaphidepcp}
  \end{center}
\end{figure}

From the fixed point solution of the equations of
motions~(\ref{eq:tdmfeefirst})-(\ref{eq:tdmfeelast}) we observe that
if we start the evolution from the state (\ref{eq:in-cond-circle}),
the dynamics of the system remains bounded to the circle and
after a closed evolution $t_{f} = 2\pi/\delta_{\phi}$ the mean photon
number in the thermodynamic limit is simply given by
\begin{align}
 & \frac{1}{j} \expect{\op{a}^{\dagger}\op{a}}_{c_{2}}
 = \frac{1}{j} \bra{\zeta_{c_{2}}}\bra{\alpha_{c_{2}}}  \op{a}^{\dagger}\op{a} \ket{\alpha_{c_{2}}}\ket{\zeta_{c_{2}}}
 = \frac{1}{j} \abs{\alpha_{c_{2}}}^{2} = \frac{1}{2j} q_{2,c_{2}}^{2} 
 \nonumber \\
 & =
 \left\{
  \begin{array}{l l}
    \frac{1}{2}
    \lp \frac{2\lambda}{\omega} \rp^{2}
    \lb 1 - \lp \frac{\Omega}{4\lambda^{2}} \rp^{2} \rb 
    & \quad \text{if $\lambda \geq \frac{1}{2} \sqrt{\Omega}$} \\
    0 & \quad \text{if $\lambda <  \frac{1}{2} \sqrt{\Omega}$} \\
  \end{array}
 \right.
\end{align}
We also observe now that the critical coupling, which marks the
quantum phase transition, is shifted by the amount of the
rotation velocity as $\lambda_{c}^{(\mathrm{rot})}=\frac{1}{2}
\sqrt{\omega(\omega_{0}+\delta_{\phi})}$.
In Fig.~\ref{fig:meanphotonlambdadepanddeltaphidepcp} (left) we plot
$\expect{\op{a}^{\dagger}\op{a}}_{c_{2}} / j$
in order to illustrate the shift of the critical coupling caused by
the rotation.
Further, in Fig.~\ref{fig:meanphotonlambdadepanddeltaphidepcp} (right)
we draw the velocity dependence of $\expect{\op{a}^{\dagger}\op{a}}_{c_{2}} / j$ for circles
$\mathcal{C}_{2}$ inside the critical paraboloid and compare it with the
situation corresponding to circles $\mathcal{C}_{1}$ that encircle
the critical paraboloid.
One can see that the critical driving velocity is defined by
$\delta_{\phi,c}^{(\mathrm{rot})} =
\frac{4\lambda^{2}}{\omega}-\omega_{0}$.
Above this critical velocity the mean photon number is zero.
This allows one to probe the equilibrium quantum
critical point $\lambda_{c}=\sqrt{\omega\omega_{0}}/2$ in an
indirect way by encircling critical paraboloid in the parameter space and measuring
the dependence of mean photon number as a function of a driving velocity.

\subsubsection{Deviation from stationary coherent state}

\begin{figure}[h!]
  \begin{center}
    \includegraphics[scale=0.43]{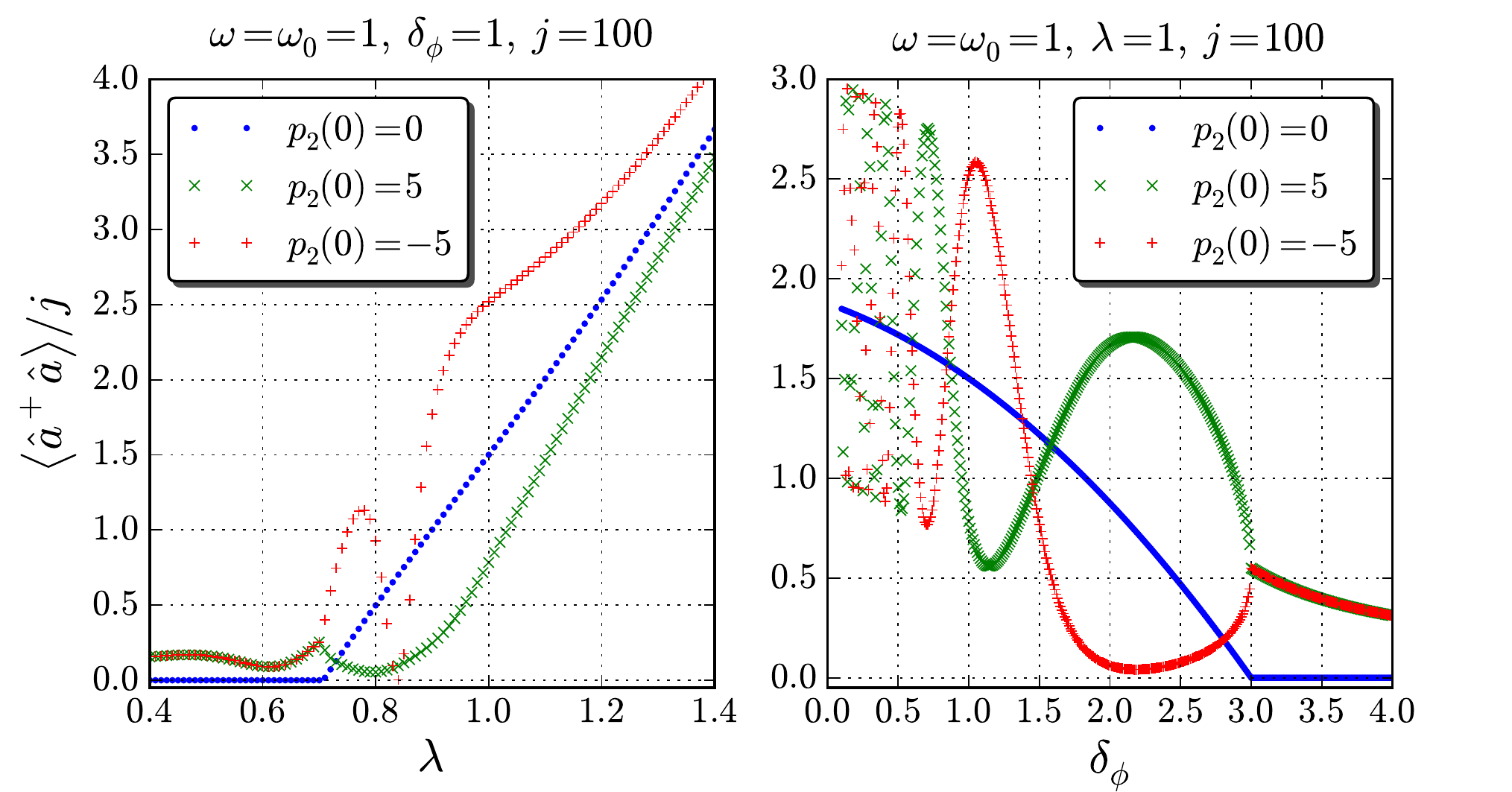}
    \caption{
      The mean photon number
      $\expect{\op{a}^{\dagger}\op{a}}_{c_{2}} / j$ as a
      function of the coupling strength $\lambda$ on the left and as a
      function of the rotation velocity $\delta_{\phi}$ on the right
      starting on the fixed point $(q_{1,c_{2}}, p_{1,c_{2}}, q_{2,c_{2}},
      p_{2,c_{2}})$ (blue dots) and with small deviation from this fixed
      point (red $+$ and green $\times$).
       }
    \label{fig:meanphotonlambdadepanddeltaphidepdeviation}
  \end{center}
\end{figure}

The natural question which immediately arises is what happens if the
initial state is not exactly given by the fixed point conditions,
namely if the initial state is not a coherent state
\begin{align}
  \ket{\alpha(t)} &=
  e^{-\frac{1}{2}\alpha\alpha^{\ast}} \sum_{n=0}^{\infty}
  \frac{\alpha^{n}}{\sqrt{n!}} \ket{n} \\
  \ket{\zeta(t)} &=
  \frac{1}{(1+\zeta\zeta^{\ast})^{j}}
  \sum_{m=-j}^{j} \zeta^{m+j} \sqrt{\binom{2j}{m+j}} \ket{j,m}
\end{align}
with $\alpha(0)=\alpha_{c_{2}}$ and $\zeta(0)=\zeta_{c_{2}}$.
In Fig.~(\ref{fig:meanphotonlambdadepanddeltaphidepdeviation}) we plot
the scaled mean photon number obtained from solving the time
dependent mean field equations but this time with the initial condition
that slightly differ from the fixed point.
%

\subsubsection{Time dependence of the mean photon number}

\begin{figure}[h!]
  \begin{center}
    \includegraphics[scale=0.45]{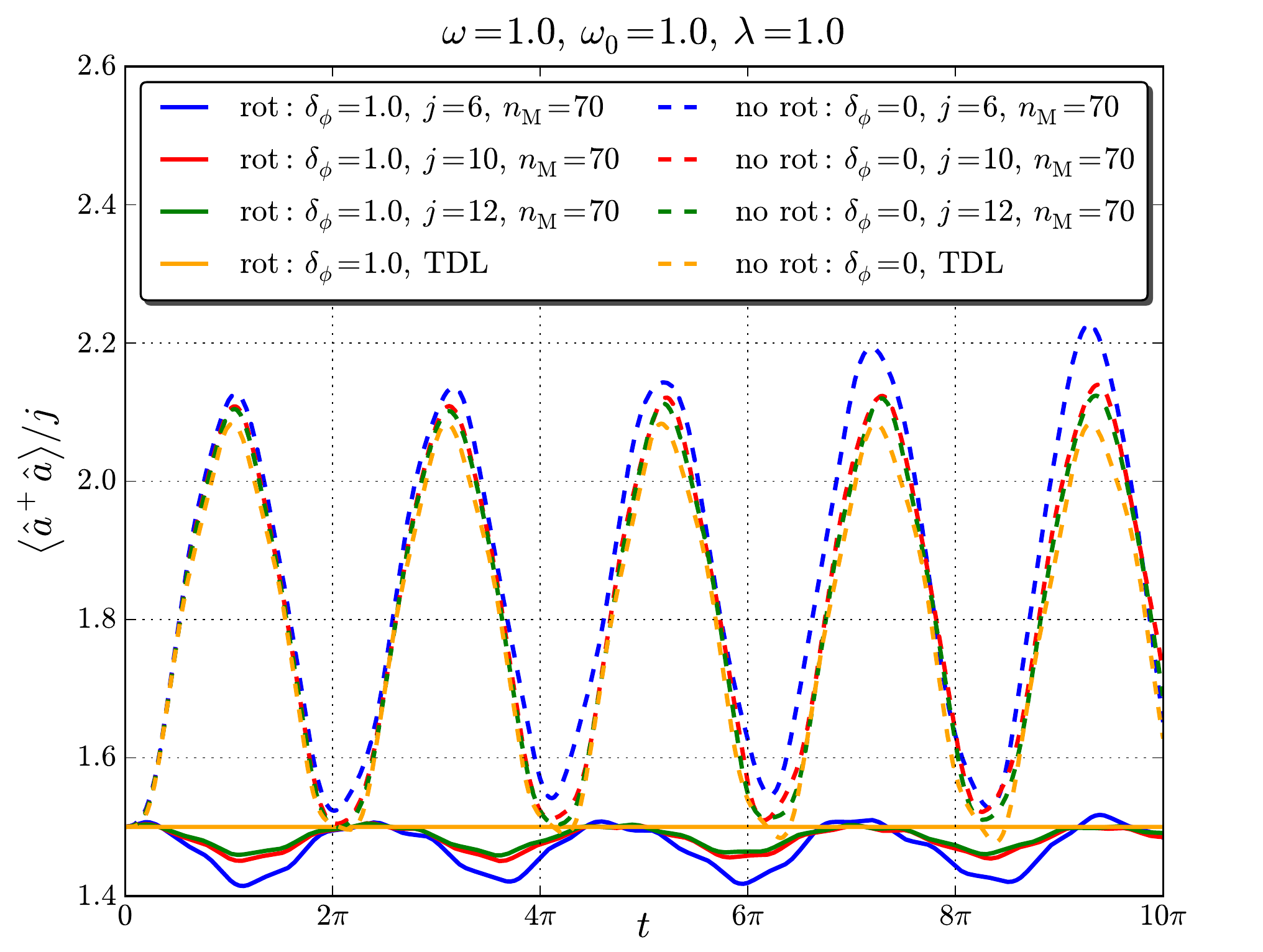}
    \caption{
      Time dependence of the scaled mean photon number,
      $\expect{\op{a}^{\dagger}\op{a}}/j$, for different values of
      $j$, a coupling $\lambda=1.0$ and for $\omega=\omega_{0}=1.0$.
      The solid lines represent the time evolution with
      rotational driving and the dashed lines give the time evolution
      of the Dicke model without rotational driving.
      The evolution time $t_{f}$ in both cases is the same.
      }
    \label{fig:adascsftimedep}
  \end{center}
\end{figure}

In Fig.~\ref{fig:adascsftimedep} we illustrate the time evolution of
the scaled mean photon number.
In the TDL ($j\to\infty$) we used the time dependent mean field
equations~(\ref{eq:tdmfeefirst})-(\ref{eq:tdmfeelast}) in order to
compute $\expect{\op{a}^{\dagger}\op{a}}(t)/j$ (orange curve) while for
a finite number of two-level systems we used the Chebyshev technique
to calculate the time evolution.
We compare the time evolution of the scaled mean photon
number of the rotationally driven Dicke model (solid lines in
Fig.~\ref{fig:adascsftimedep}) with the time evolution
of the scaled mean photon number of the Dicke model with no rotation
of the pseudo-spin vector around the $z$ axis (dashed lines in
Fig.~\ref{fig:adascsftimedep}).

As expected, if we start in the coherent state
$\ket{\alpha_{c_{2}}}\ket{\zeta_{c_{2}}}$ the time evolution of
$\expect{\op{a}^{\dagger}\op{a}}/j$ is constant in the TDL (solid
orange line in Fig.~\ref{fig:adascsftimedep}).
However, for finite $j$ one can see that the scaled mean photon number
is not time independent, but increasing $j$ it tends to a constant
value (the amplitude of the oscillation decreases). 
In the case where we let the system evolve without rotational driving,
the scaled mean photon number shows pronounced oscillatory behavior
compared to the rotational driving which is not completely $2\pi$
periodic.

\subsubsection{Dependence of the mean photon number on the atom-field
  coupling strength $\lambda$}

In this subsection we illustrate the dependence of the scaled mean
photon numbers on the atom-filed coupling strength,
$\expect{\op{a}^{\dagger}\op{a}}(\lambda)/j$.
The system is initially prepared in the product coherent state
$\ket{\alpha_{c_{2}}}\ket{\zeta_{c_{2}}}$.
First we consider only one revolution in parameter space,
$t_{f}=2\pi/\delta_{\phi}$.
\begin{figure}[h!]
  \begin{center}
    \includegraphics[scale=0.45]{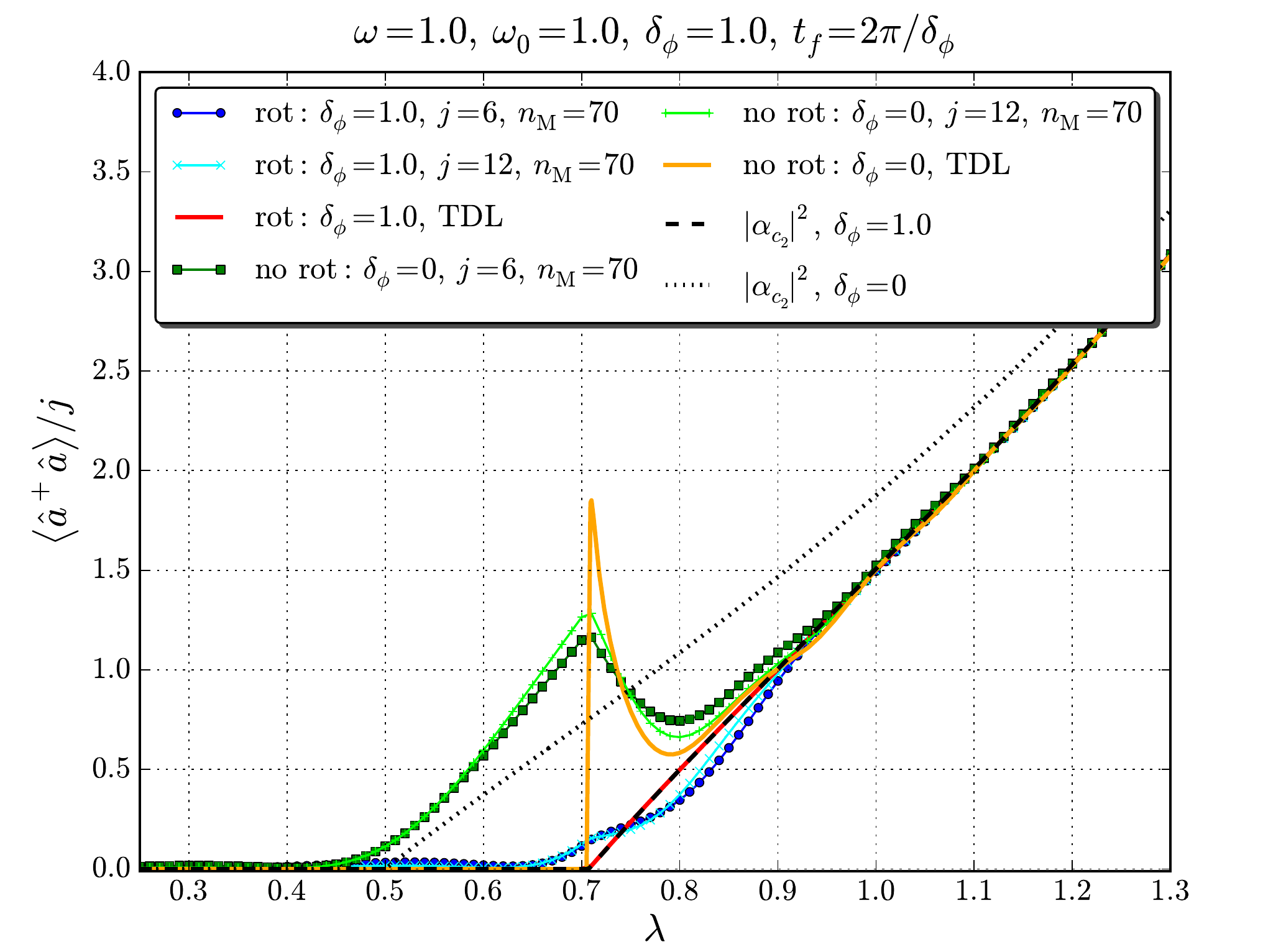}
    \caption{
      The scaled mean photon number as a function of the coupling
      strength $\lambda$.
      We compare $\expect{\op{a}^{\dagger}\op{a}}/j$
      for the rotationally driven Dicke model with the one for undriven Dicke
      model ($t_{f}=2\pi/\delta_{\phi}$ was kept the same in both cases). 
      }
    \label{fig:adascsflambdadep}
  \end{center}
\end{figure}
For rotational driving the system remains in the same
state and there is no significant deviation from the static behavior
of the scaled mean photon number.
However, if the system is evolved according to the usual Dicke model
$\op{H}_{\mathrm{D}}$ a peak is observed at
$\lambda_{c}^{(\mathrm{rot})}$. 
For $\lambda<\lambda_{c}^{(\mathrm{rot})}$ there is a large discrepancy
between the finite size calculation and the mean field solution.

\begin{figure}[h!]
  \begin{center}
    \includegraphics[scale=0.45]{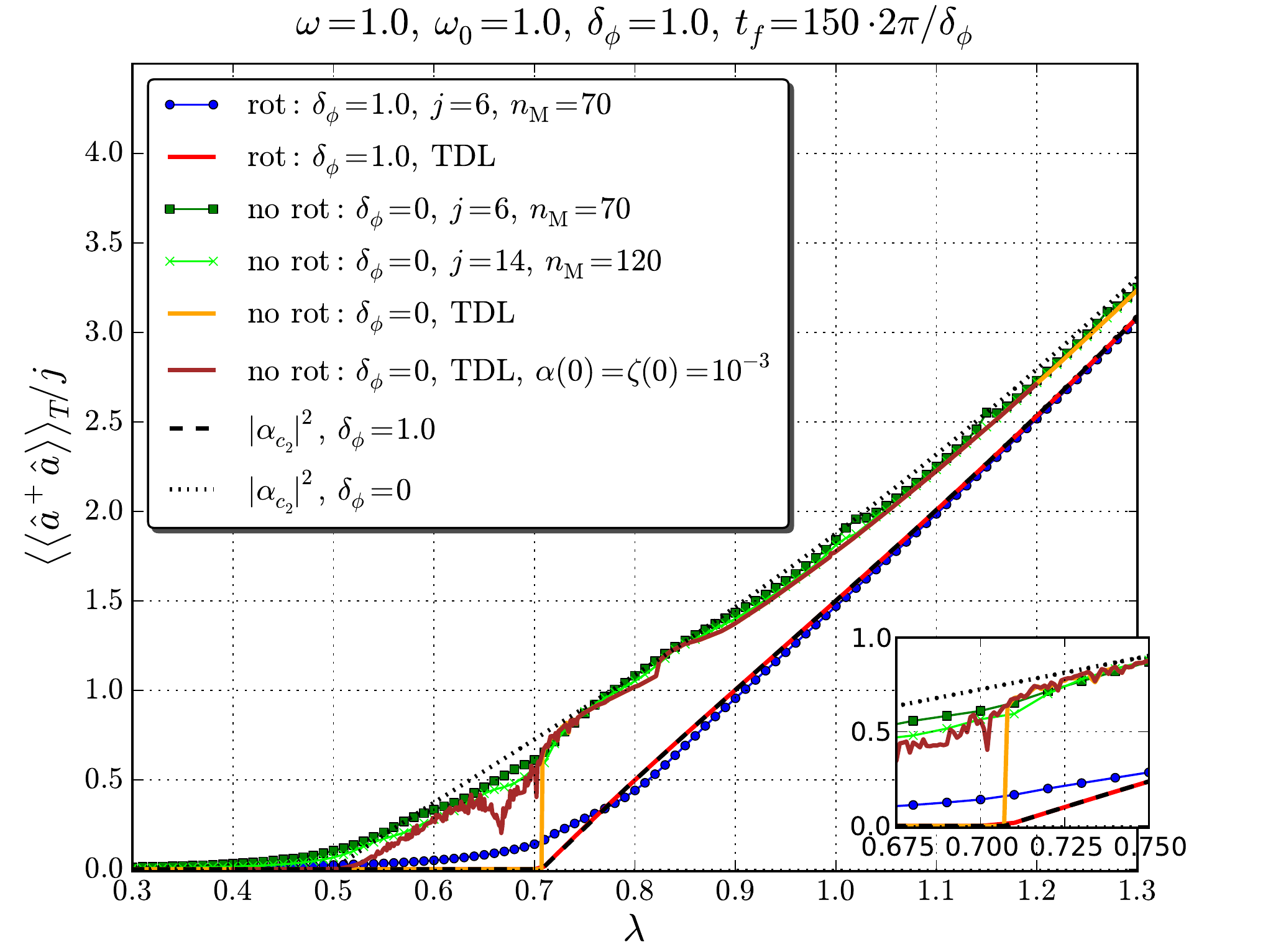}
    \caption{
      The time averaged scaled mean photon number as a function of the coupling
      strength $\lambda$.
      We compare $\langle\expect{\op{a}^{\dagger}\op{a}}\rangle_{T}/j$
      for the rotationally driven Dicke model and the unrotated Dicke
      model ($t_{f}$ was kept the same in both cases). 
      }
    \label{fig:adascsflambdadepta}
  \end{center}
\end{figure}

In Fig.~\ref{fig:adascsflambdadepta} we show the time averaged scaled
mean photon number, $\langle\expect{\op{a}^{\dagger}\op{a}}\rangle_{T}/j$ as defined in Eq.~\ref{eq:timeavesmph} for the
fixed rotation velocity $\delta_{\phi}=1.0$.
For a finite number of two-level systems $N=2j$ the time propagation
was calculated using the Chebyshev scheme~\cite{Tal1984} and in the
TDL we solved the time-dependent mean field equations.

For the initial state $\ket{\alpha_{c_{2}}}\ket{\zeta_{c_{2}}}$ which
corresponds to the stationary state of $\op{H}_{\mathrm{RD}}$ we
effectively observe no difference between
$\abs{\alpha_{c_{2}}}^{2}/j$ and the time evolved computations.
However, when the system is evolved according to $\op{H}_{\mathrm{D}}$ we
again observe that below $\lambda_{c}^{(\mathrm{rot})}$ the finite size
results do not coincide in both cases.
This can be corrected if we introduce small
fluctuations to the semi-classical description by setting the initial
conditions of the mean field
equations~(\ref{eq:tdmfeefirst})-(\ref{eq:tdmfeelast}) not strictly
to zero but
$\alpha(0)=\zeta(0)=\alpha^{\ast}(0)=\zeta^{\ast}(0)=10^{-3}$ (brown
line).

\subsubsection{Dependence of the mean photon number on the rotation velocity $\delta_{\phi}$}

\begin{figure}[h!]
  \begin{center}
    \includegraphics[scale=0.45]{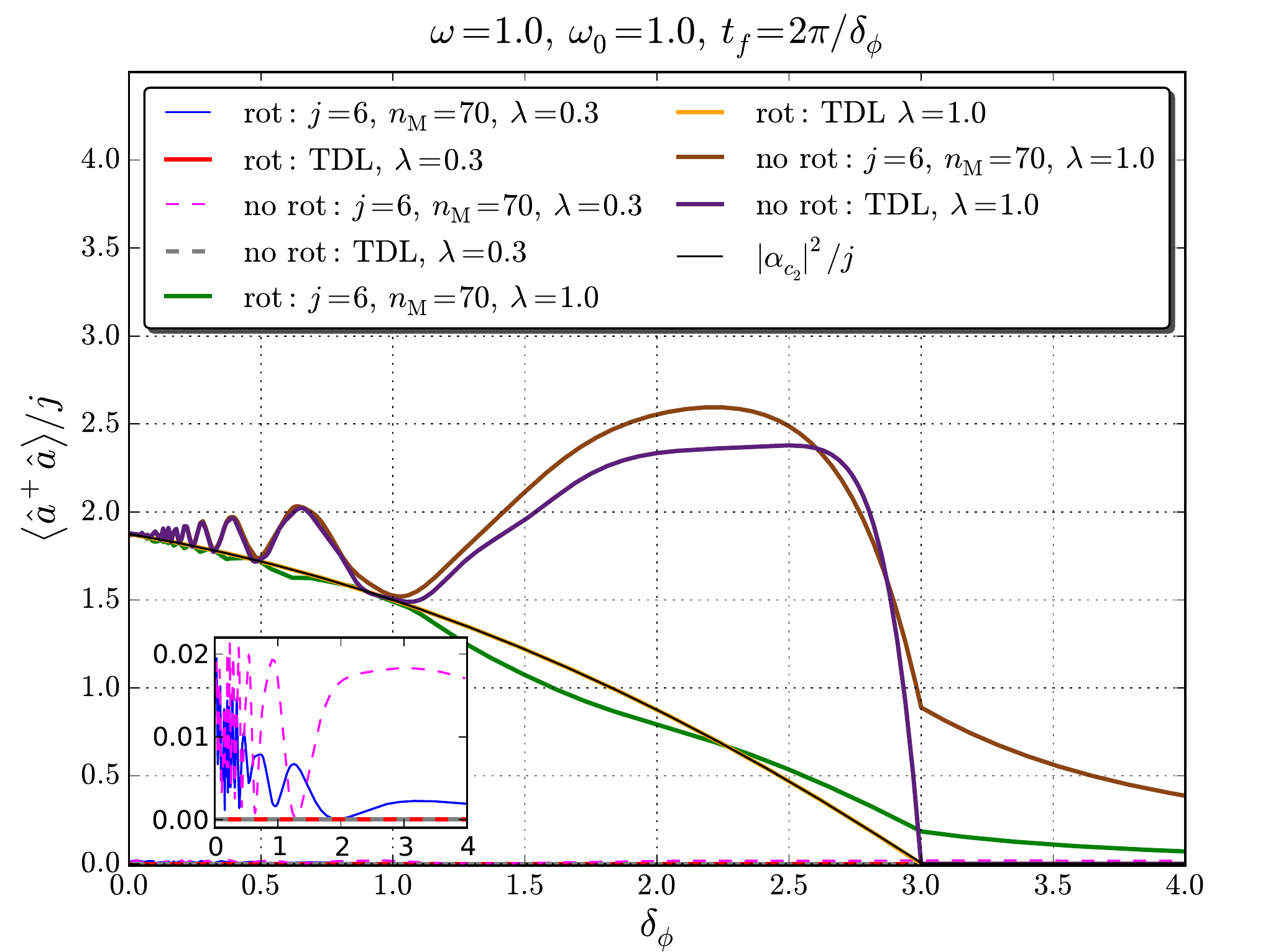}
    \caption{
      The scaled mean photon number as a function of the rotation
      velocity $\delta_{\phi}$. We compare $\expect{\op{a}^{\dagger}\op{a}}/j$
      for the rotationally driven Dicke model and for the undriven Dicke
      model with $t_{f}=2\pi/\delta_{\phi}$ in both cases.
      }
    \label{fig:adascsfdphi}
  \end{center}
\end{figure}

In this subsection we illustrate the dependence of the scaled mean
photon number on the rotation velocity $\delta_{\phi}$ after one
revolution in parameter space, $t_{f}=2\pi/\delta_{\phi}$, which is shown in
Fig.~\ref{fig:adascsfdphi}.
We observe that if the critical paraboloid is encircled, the scaled
mean photon number is different from zero and the critical
driving velocity is given by $\delta_{\phi,c}^{(\mathrm{rot})} =
\frac{4\lambda^{2}}{\omega}-\omega_{0}$.
This basically follows from the rotation
velocity shift in the rotated critical coupling
$\lambda_{c}^{(\mathrm{rot})}=\sqrt{\omega(\omega_{0}+\delta_{\phi})}/2$.
This allows one to probe the equilibrium quantum critical point
$\lambda_{c}=\sqrt{\omega\omega_{0}}/2$ from a distance in parameter space. 

\begin{figure}[h!]
  \begin{center}
    \includegraphics[scale=0.45]{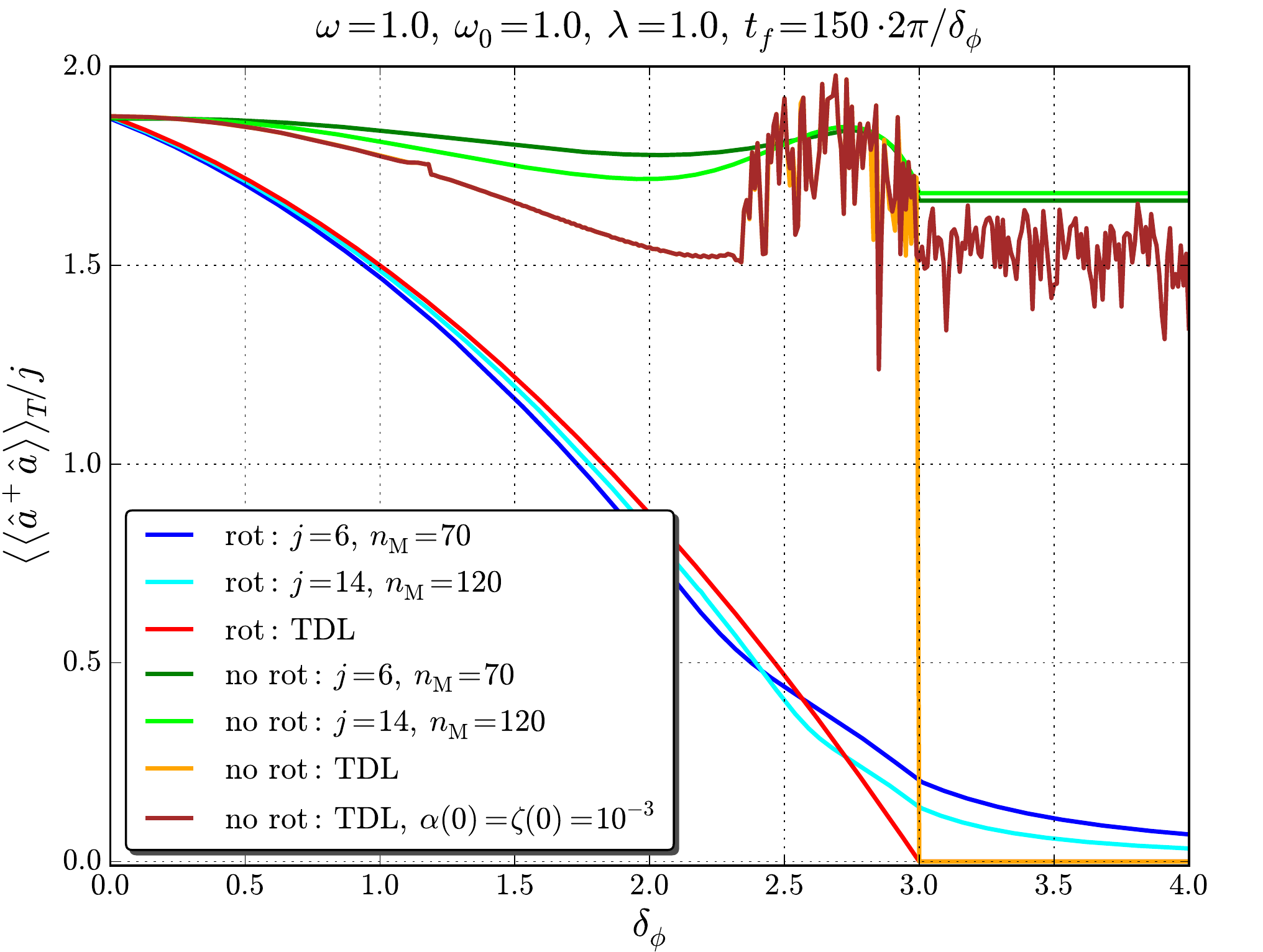}
    \caption{
      The time averaged scaled mean photon number as a function of the rotation
      velocity $\delta_{\phi}$. We compare $\langle\expect{\op{a}^{\dagger}\op{a}}\rangle_{T}/j$
      for the rotationally driven Dicke model with the results for the undriven Dicke
      model averaged over $t_{f}=150\cdot 2\pi/\delta_{\phi}$ in both cases. 
      }
    \label{fig:adascsfdphita}
  \end{center}
\end{figure}

The velocity dependence of the time averaged scaled mean photon number
(Fig.~\ref{fig:adascsfdphita}) clearly reveals the critical driving
velocity $\delta_{\phi,c}^{(\mathrm{rot})}$. However, if we
suddenly, at $t_{0}=0$, switch on the undriven evolution
(with $\op{H}_{\mathrm{D}}$)
the mean field solution is zero for
$\delta_{\phi}>\delta_{\phi}^{(\mathrm{rot})}$ while the numerical
calculation for a finite number of atoms is non zero in this region.
Again, this can be corrected if we allow small fluctuations
in the semi-classical description by setting the initial conditions
of the mean field equations not strictly to zero:
$\alpha(0)=\zeta(0)=\alpha^{\ast}(0)=\zeta^{\ast}(0)=10^{-3}$ in the
corresponding range of the parameters.

\subsubsection{Non-Equilibrium Phase Diagram}

\begin{figure}[h!]
  \begin{center}
    \begin{tabular}{c}
     \includegraphics[scale=0.7]{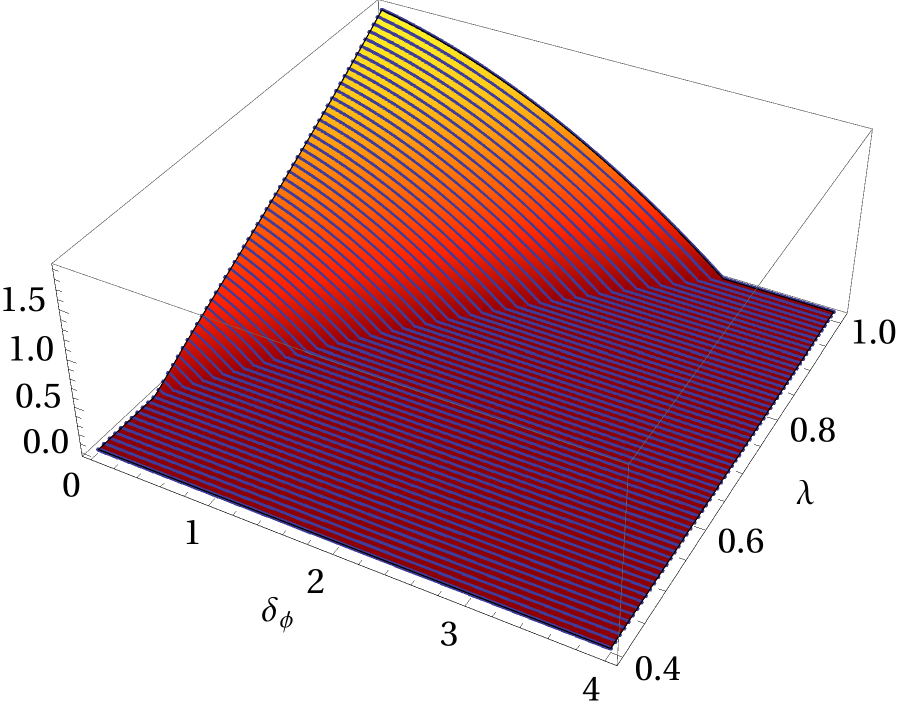} \\
     \includegraphics[scale=0.43]{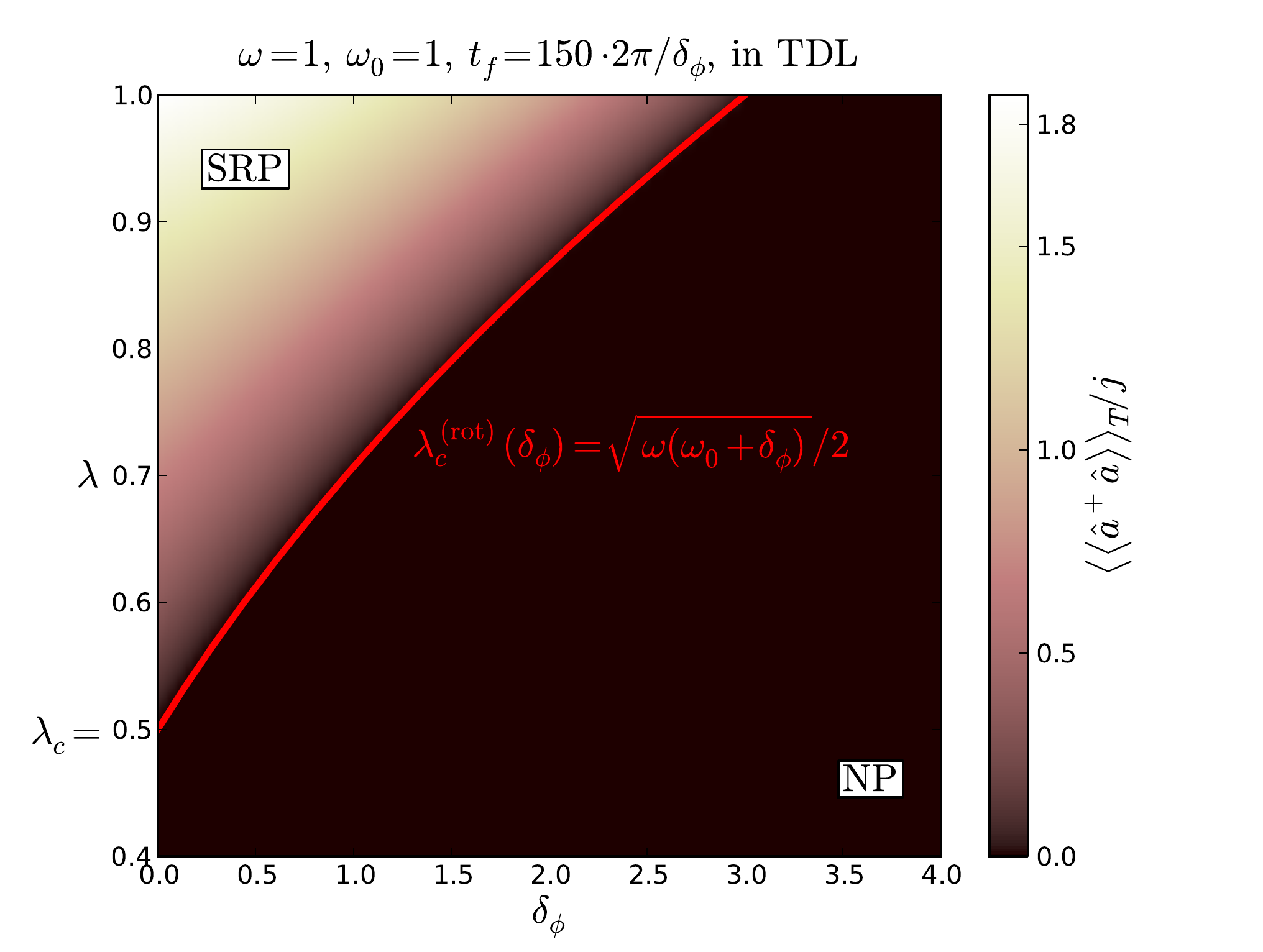}
    \end{tabular}
    \caption{The
      time averaged scaled mean photon number as a function of the rotation velocity and the atom-field coupling
      strength. Results are obtained from the mean
      field equations for $\omega=\omega_{0}=1.0$ and averaged over
      a time $t_{f}=150 \cdot 2\pi/\delta_{\phi}$.}
    \label{fig:adaneqpdscsf}
  \end{center}
\end{figure}

We summarize the dependence of the time averaged scaled mean photon
number on the atom field-coupling strength $\lambda$ and the rotation
velocity $\delta_{\phi}$ in Fig.~\ref{fig:adaneqpdscsf}.
Here we see that the rotational driving
shifts the super-radiant phase transition by the amount given by the rotation
velocity. 
The critical line (solid red line in Fig.~\ref{fig:adaneqpdscsf}) is defined by
$\lambda_{c}^{(\mathrm{rot})}=\sqrt{\omega(\omega_{0}+\delta_{\phi})}$
or by $\delta_{\phi,c}^{(\mathrm{rot})} = \frac{4\lambda^{2}}{\omega}-\omega_{0}$.

\subsubsection{Parity $\op{\Pi}$}

\begin{figure}[h!]
  \begin{center}
    \includegraphics[scale=0.43]{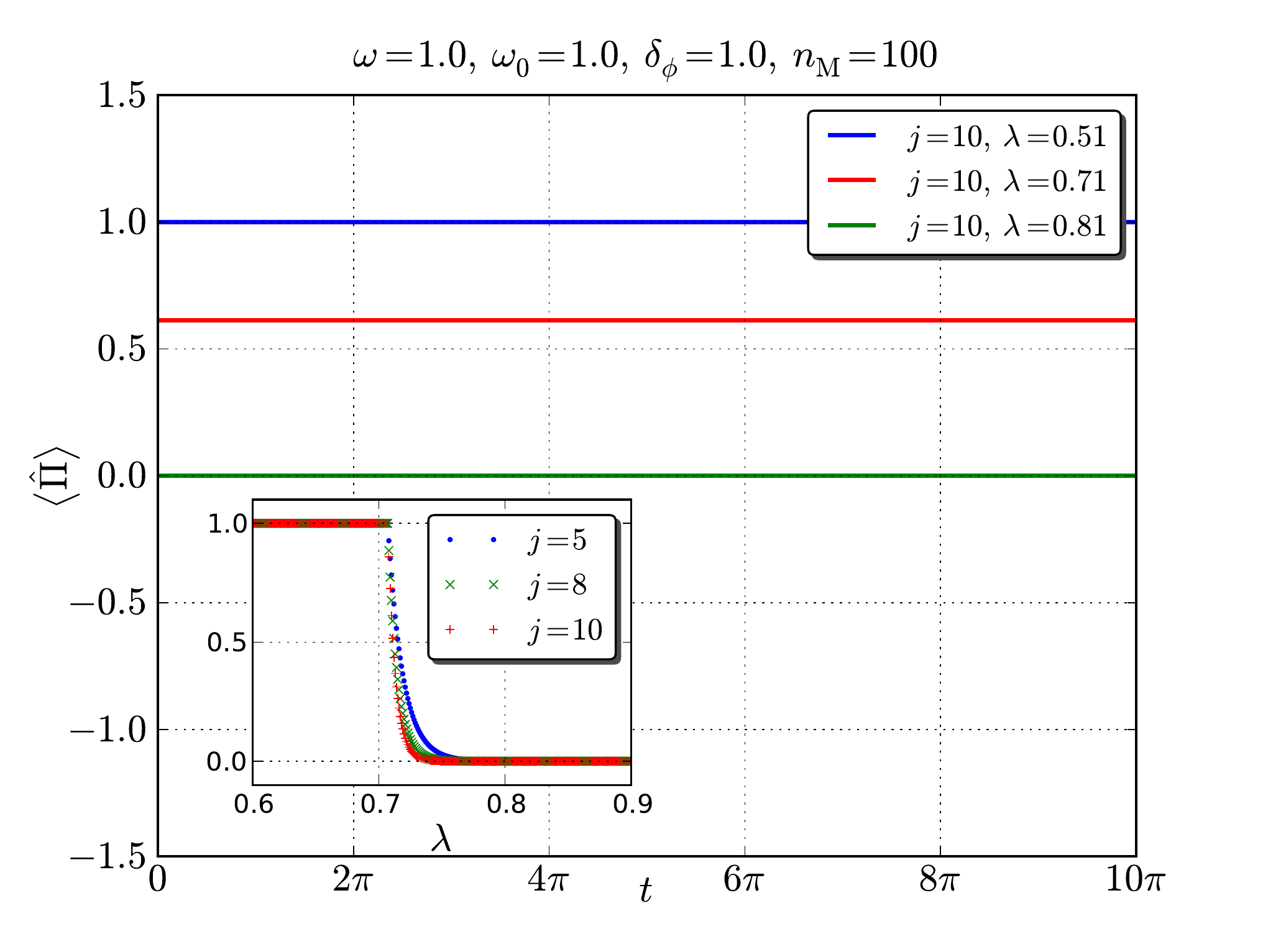}
    \caption{
      Time evolution of the parity operator starting from the coherent state
      $\ket{\alpha_{c_{2}}}\ket{\zeta_{c_{2}}}$ for a system of finite
      number of two-level atoms.
      The inset plot shows the dependence on the coupling constant
      $\lambda$ for different values of $j$.}
    \label{fig:pifscsf}
  \end{center}
\end{figure}

The time evolution of the parity operator
$\bra{\psi(t)}\op{\Pi}\ket{\psi(t)}$ is calculated for the stationary
circle initial state $\ket{\alpha_{c_{2}}}\ket{\zeta_{c_{2}}}$.
Results are shown in Fig.~\ref{fig:pifscsf} for a system with a finite
number of two-level atoms.
Note that the parity is constant in time and jumps from one to zero at the
critical atom-field coupling strength $\lambda_{c}^{(\mathrm{rot})}$.

\begin{figure}[h!]
  \begin{center}
    \includegraphics[scale=0.43]{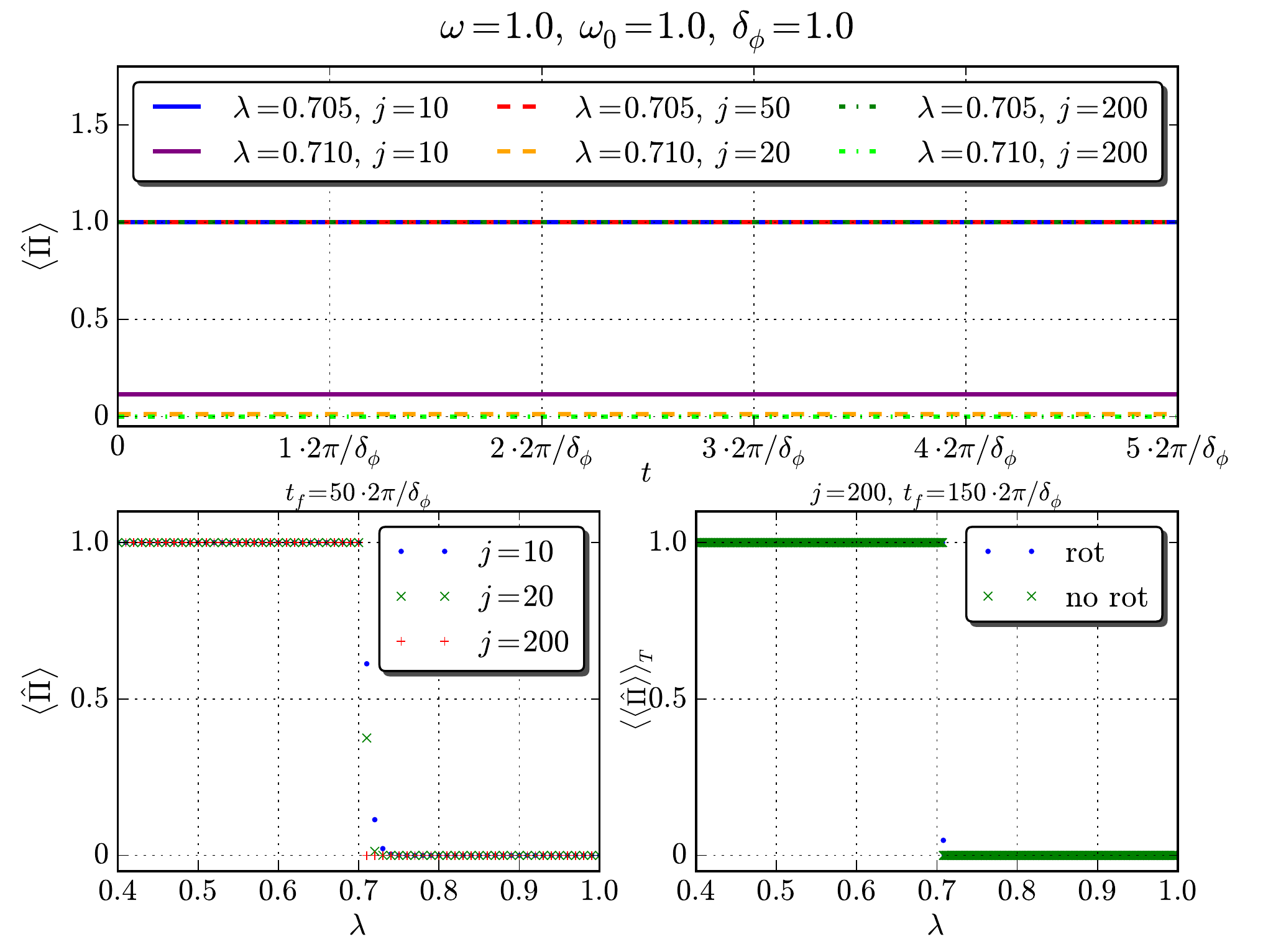}
    \caption{
      Upper panel: time evolution of the parity $\expect{\op{\Pi}}$
      starting from the coherent state
      $\ket{\alpha_{c_{2}}}\ket{\zeta_{c_{2}}}$.
      Lower left panel: dependence of $\expect{\op{\Pi}}$ on the coupling constant $\lambda$ for different
      values of $j$ after $50$ circles.
      Lower right panel: time averaged parity as a function of the
      coupling $\lambda$. The time evolution is calculated from the
      time dependent mean field equations.}
    \label{fig:pitdlcsf}
  \end{center}
\end{figure}

In the TDL $(j\to\infty)$ we use the solutions of the mean field
equations to calculate the time evolution of the parity, Fig.~\ref{fig:pitdlcsf}. 
We also illustrate the time evolution of the scaled parity
(where as previously we rescaled all the phase space coordinates by $\sqrt{j}$).
\begin{figure}[h!]
  \begin{center}
    \includegraphics[scale=0.43]{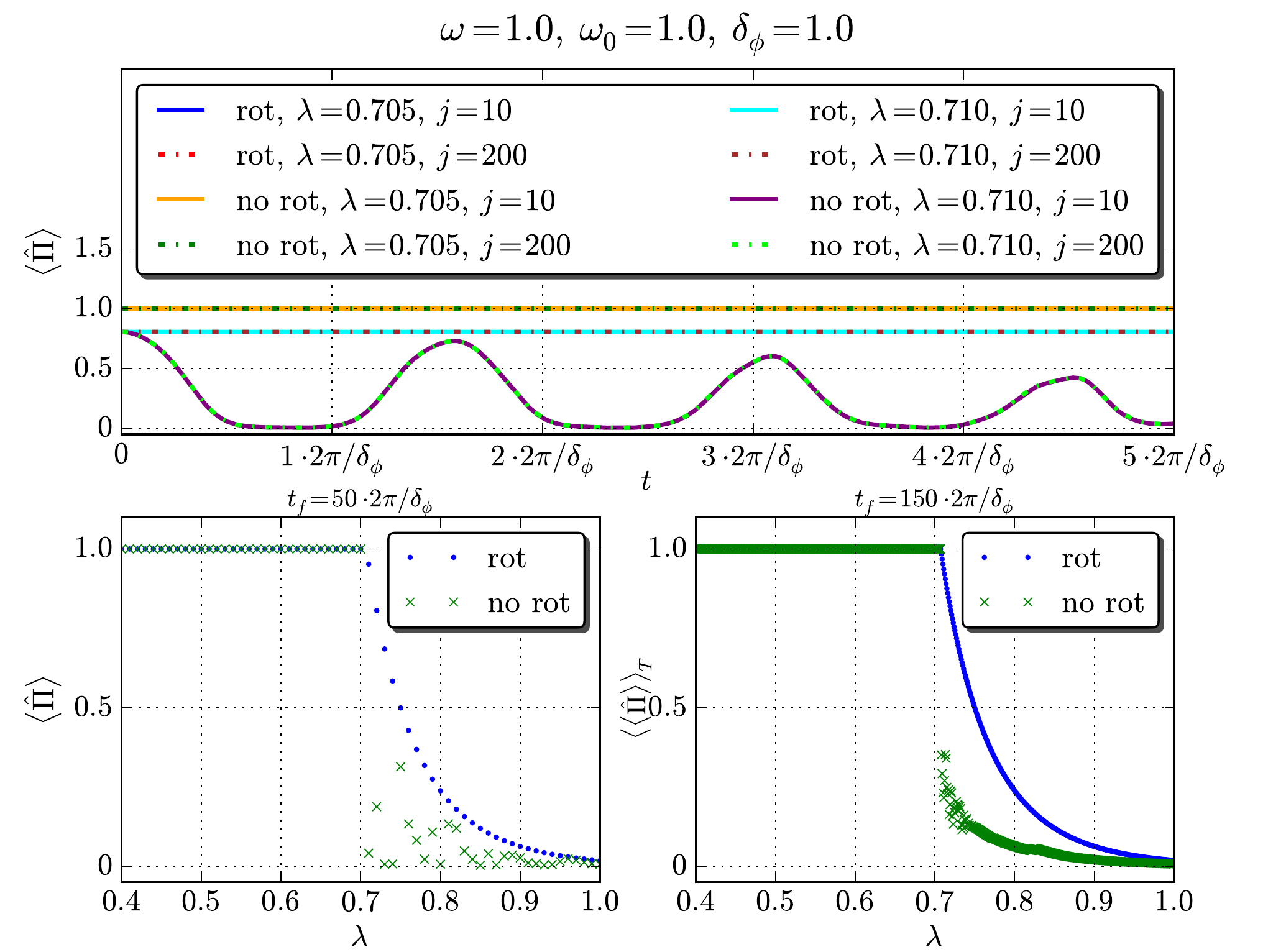}
    \caption{
       Upper panel: time evolution of the scaled parity operator starting from 
       the coherent state
       $\ket{\alpha_{c_{2}}}\ket{\zeta_{c_{2}}}$.
       Lower left panel: the
       dependence on the coupling constant $\lambda$ of the scaled
       parity.
       Lower right panel: the time averaged scaled parity as a
       function of $\lambda$.
      }
    \label{fig:pitdlsccsf}
  \end{center}
\end{figure}

\subsection{Fock initial state: $\ket{n=0}\ket{j,m=-j}$}

Here we consider the evolution of the mean photon number if we
initially prepare the systems in the Fock state with no photons $n=0$
while the spin state has the lowest weight, $m=-j$, 
\be
 \ket{\psi(t_{0}=0)} = \ket{0} \otimes \ket{j,-j}.
\ee
We switch on the rotation at $t_{0}=0$ and let the system evolve
until $t_{f}=n_{\mathrm{R}} \, 2 \pi/\delta_{\phi}$,
where we measure the mean photon number
$\expect{\op{a}^{\dagger}\op{a}}(t_{f})/j$.
We also compare it with the evolution of the un-rotated system.
In terms of the coherent states this initial state corresponds to 
$\alpha=0$ and $\zeta=0$.
The mean photon number is calculated in the TDL using the mean field
equations~(\ref{eq:tdmfeefirst})-(\ref{eq:tdmfeelast}) with the
initial conditions
$\alpha(0)=\alpha^{\ast}(0)=\zeta(0)=\zeta^{\ast}(0)=0$.

\subsubsection{Time evolution of the mean photon number}

\begin{figure}[h!]
  \begin{center}
    \includegraphics[scale=0.43]{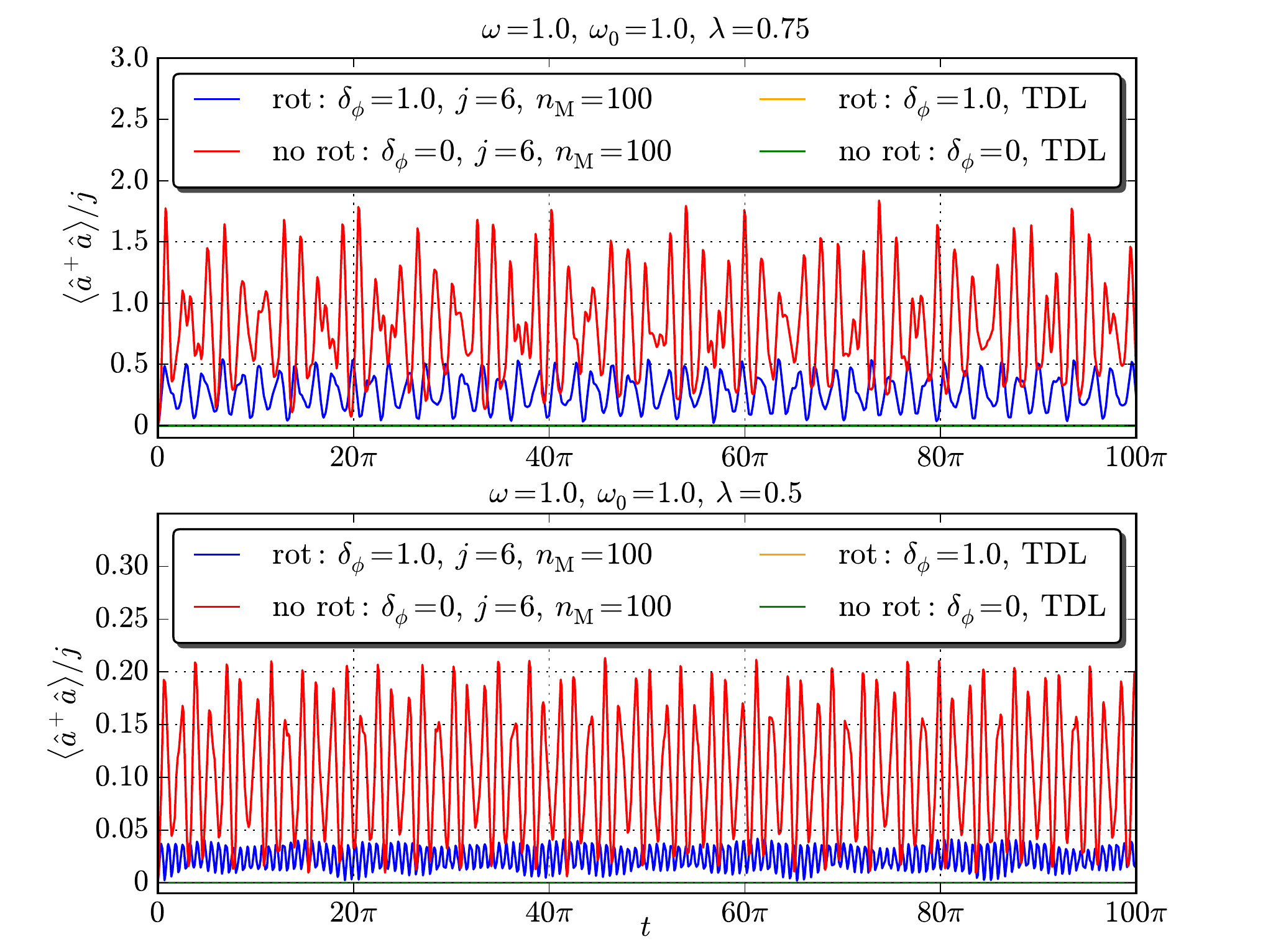}
    \caption{
      Time dependence of the scaled mean photon number on a resonance
      $\omega = \omega_{0} = 1.0$ for different values of the coupling
      strength $\lambda$. The system is initially prepared in the
      state $\ket{0}\ket{j,-j}$.
      }
    \label{fig:adatimesljno}
  \end{center}
\end{figure}

In Fig.~\ref{fig:adatimesljno} we show the time evolution of the
scaled mean photon number, $\expect{\op{a}^{\dagger}\op{a}}/j$.
It compares the scaled mean photon calculated for a system of
finite size ($j=6$) by the Chebyshev scheme with and without rotations (red
and blue curve respectively) with results for the TDL ($j\to\infty$) with and
without rotations (orange and green curve respectively).

In the static case the mean photon number is explicitly zero for this
initial state, however for the driven situation we observe a critical
coupling above which the scaled mean photon number becomes
macroscopic. 
We observe that for the initial Fock state, $\ket{0}\ket{j,-j}$, the
scaled mean photon number in the rotated case (we switch on the
rotation at $t_{0}=0$) is still microscopic for
$\lambda=0.5$ whereas if no rotation is switched on the scaled mean
photon number becomes macroscopic at
$\lambda_{c}=\sqrt{\omega\omega_{0}}/2=0.5$.
Therefore, the critical coupling $\lambda_{c}$ is shifted by
the amount given by the applied rotation velocity $\delta_{\phi}$. 
Thus for the driven case
$\lambda_{c}^{(\mathrm{rot})}=\sqrt{\omega(\omega_{0}+\delta_{\phi})}/2$.

Note that in the TDL the mean field equations do not
reproduce the finite size calculations of the scaled mean photon
number.

\subsubsection{Dependence of the mean photon number on the atom-field coupling strength $\lambda$}

Accordingly, we study the dependence of the scaled mean photon number
on the atom-field coupling strength $\lambda$.
First only for one closed circle in parameter space $\phi_{f}=2\pi$,
i.e., $t_{f}=2\pi/\delta_{\phi}$.
\begin{figure}[h!]
  \begin{center}
    \includegraphics[scale=0.43]{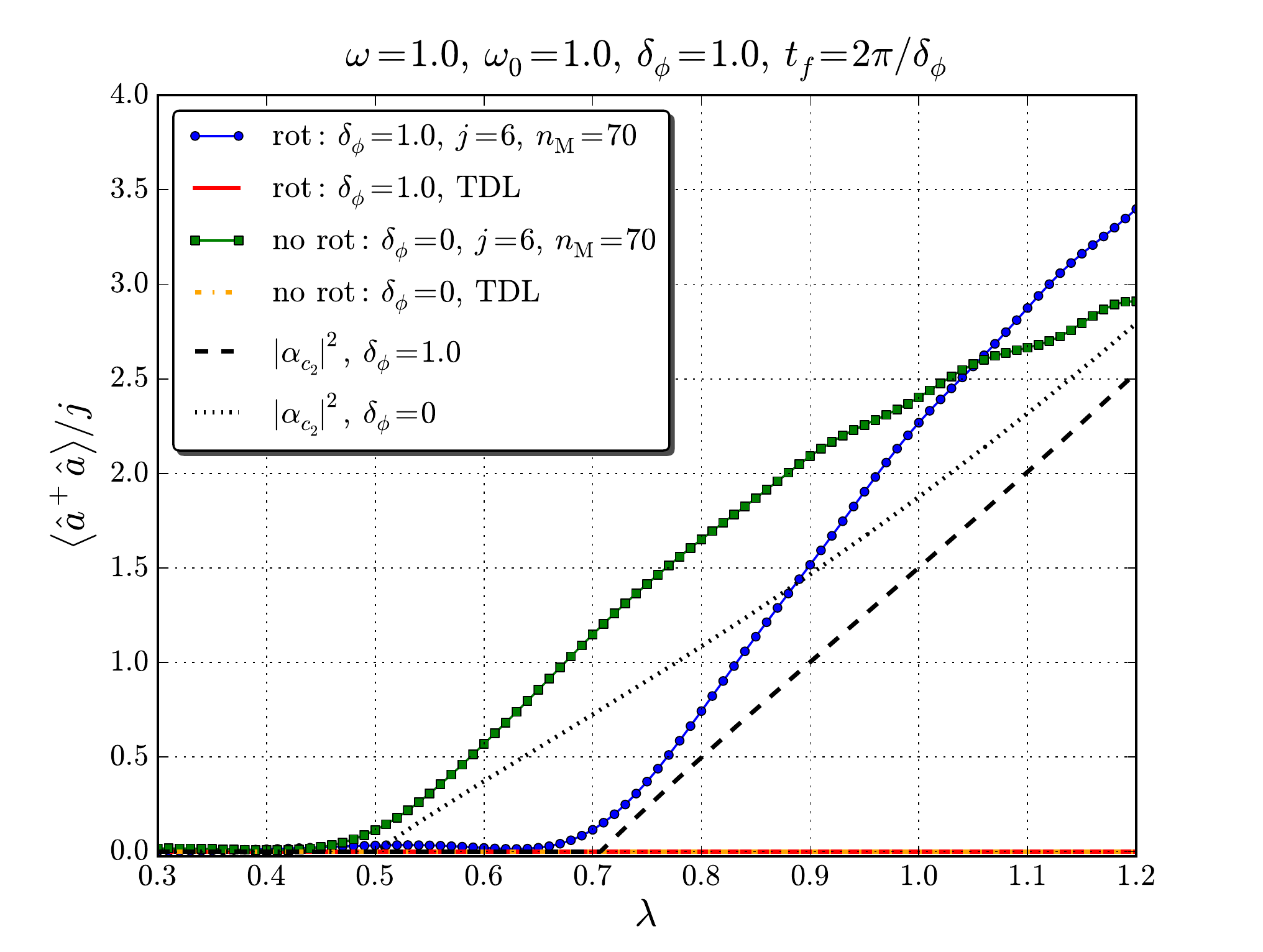}
    \caption{
      Scaled mean photon number dependence on the atom-field coupling strength on resonance
      $\omega = \omega_{0} = 1.0$ for a rotation velocity
      $\delta_{\phi}=1.0$. The system at
      $t_{0}=0$ is prepared in the state $\ket{n=0}\ket{j,m=-j}$.
      }
    \label{fig:adalamsljno}
  \end{center}
\end{figure}
The scaled mean photon number becomes macroscopic for a
coupling strength greater than some critical coupling.
In the unrotated case this is given by
$\lambda_{c}=\sqrt{\omega\omega_{0}}/2$ while for a rotational driving
it is given by
$\lambda_{c}^{(\mathrm{rot})}=\sqrt{\omega(\omega_{0}+\delta_{\phi})}/2$.
In the TDL the mean field solutions give zero.

\begin{figure}[h!]
  \begin{center}
    \includegraphics[scale=0.43]{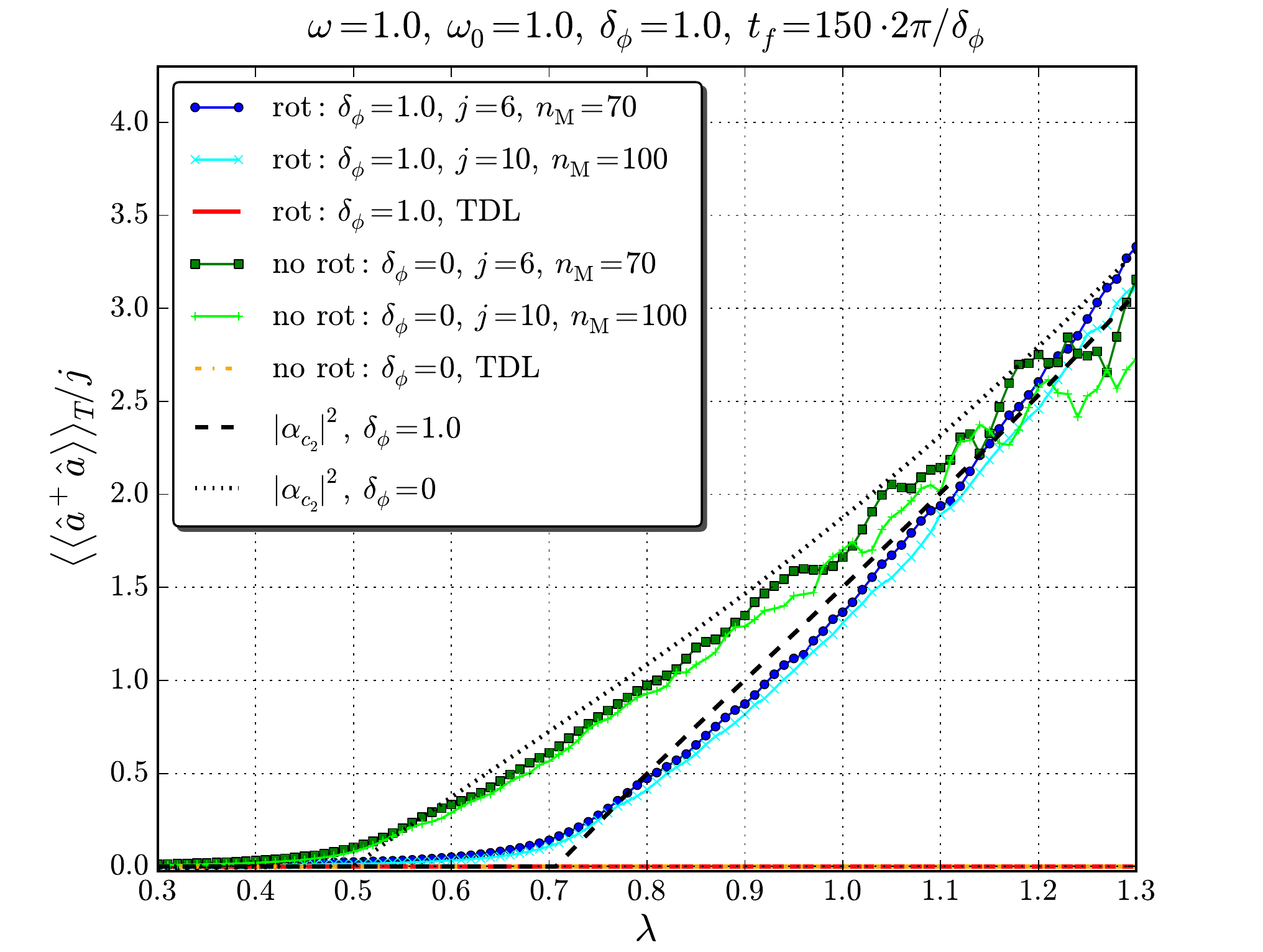}
    \caption{
      Time averaged scaled mean photon number
      on resonance
      $\omega = \omega_{0} = 1.0$ for a rotation velocity $\delta_{\phi}=1.0$.
      The system at
      $t_{0}=0$ is prepared in the state
      $\ket{n=0}\ket{j,-j}$ and evolved until $t_{f}=150 \cdot
      2\pi/\delta_{\phi}$. The mean field solutions
      in the TDL remain strictly zero.
      }
    \label{fig:adalamtasljno}
  \end{center}
\end{figure}

By looking at the time evolution of the scaled mean photon number one
can observe rather irregular structure of the oscillations.
Motivated by this in Fig.~(\ref{fig:adalamtasljno}) we also plot
the time averaged scaled mean
photon number as a function of $\lambda$ for multiple rotations
$t_{f}=150 \cdot 2\pi/\delta_{\phi}$.
This plot shows that if we prepare the initial state to be a Fock
state $\ket{n=0}\ket{j,m=-j}$ and evolve the system in time we observe
two distinct phases.
Below a critical coupling strength the time averaged scaled mean
photon number is zero (normal phase) while above a critical coupling
the time averaged scaled mean photon number becomes non-zero
(super-radiant phase).
Rotational driving shifts this critical coupling strength as
$\lambda_{c}^{(\mathrm{rot})}=\sqrt{\omega(\omega_{0}+\delta_{\phi})}/2$.  
We therefore conclude here that the evolution starting from this state
is similar to the one corresponding to the initial state
$\ket{\alpha_{c_{2}}}\ket{\zeta_{c_{2}}}$.

\subsubsection{Dependence of the mean photon number on the rotation velocity $\delta_{\phi}$}

\begin{figure}[h!]
  \begin{center}
    \includegraphics[scale=0.43]{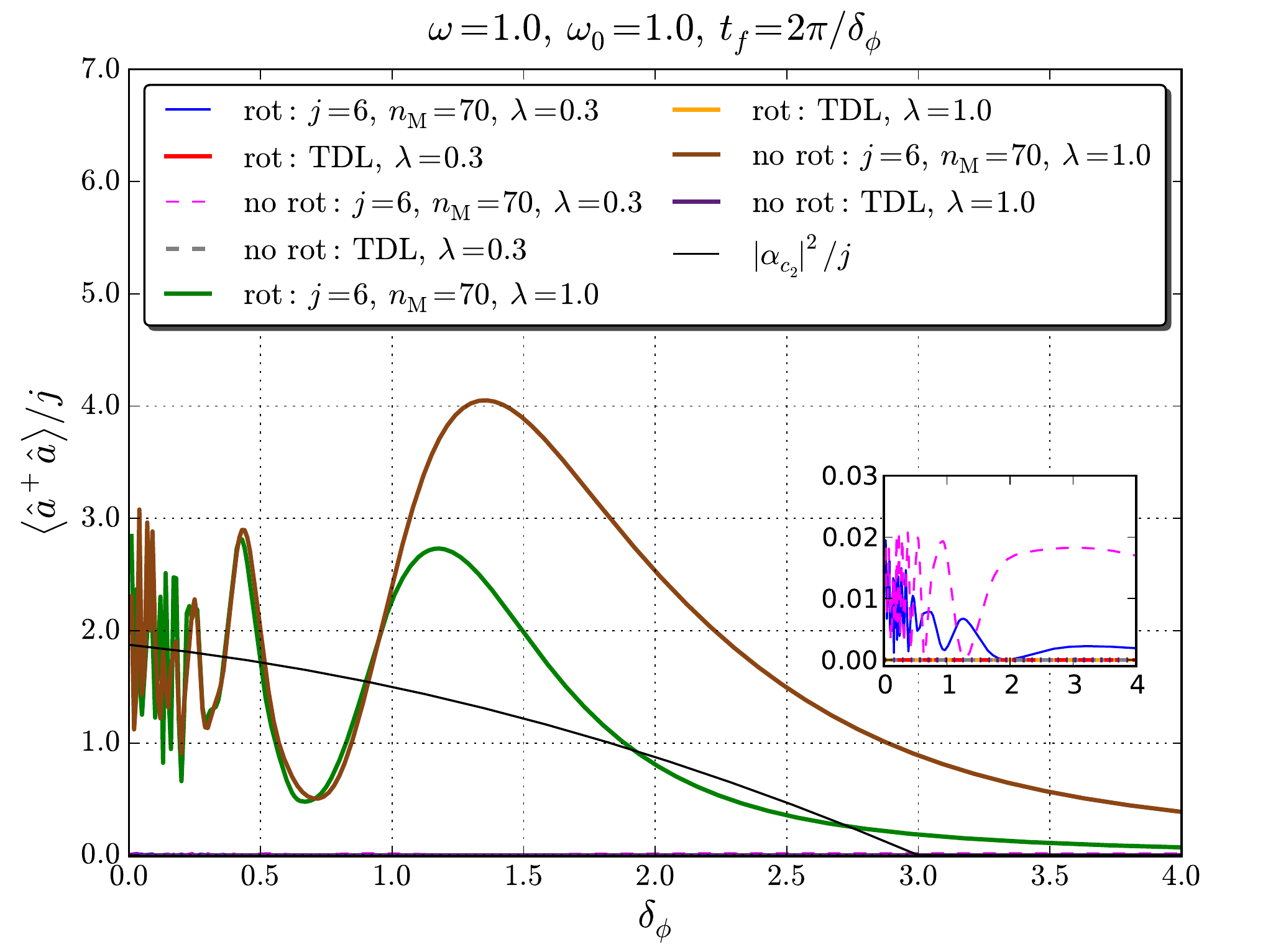}
    \caption{
      Dependence of $\expect{\op{a}^{\dagger}\op{a}}/j$ on rotation velocity $\delta_{\phi}$ 
      after one circular evolution in the parameter space for a fixed
      atom-field coupling $\lambda=1.0$.
      The mean field solution does not reproduce the finite size
      calculations it is strictly zero.
      }
    \label{fig:adadphisljno}
  \end{center}
\end{figure}

\begin{figure}[h!]
  \begin{center}
    \includegraphics[scale=0.43]{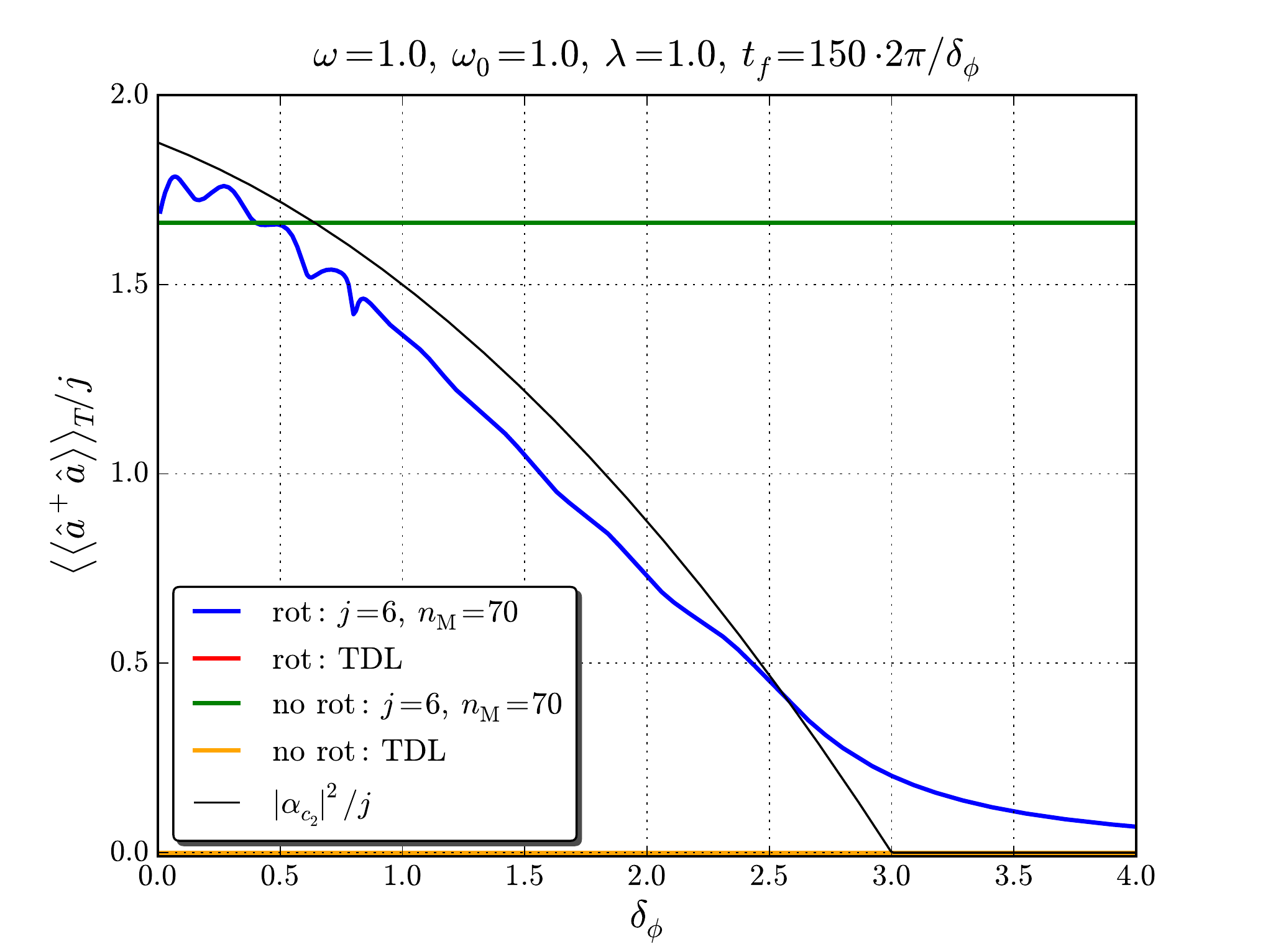}
    \caption{
      Dependence of the time averaged scaled mean
      photon number on the driving velocity at resonance $\omega =
      \omega_{0} = 1.0$ and for $\lambda=1.0$.
      At $t_{i}=0$ the system is prepared in the state
      $\ket{n=0}\ket{j,m=-j}$ and evolved up to $t_{f}=150 \cdot 2\pi/\delta_{\phi}$.
      The mean field solutions always stays zero.
      }
    \label{fig:adadphitasljno}
  \end{center}
\end{figure}

Here we look into the dependence of the scaled mean photon number on
the rotation velocity $\delta_{\phi}$,
$\expect{\op{a}^{\dagger}\op{a}}(\delta_{\phi})/j$ when
the system is initially prepared in the Fock state
$\ket{\psi(0)}=\ket{n=0}\ket{j,m=-j}$.
Again first, we consider only one rotation in the parameter space,
$t_{f}=2\pi/\delta_{\phi}$ and compare it to the undriven
evolution (see Fig.~\ref{fig:adadphisljno}).
We observe that if we encircle the critical paraboloid ($\mathcal{C}_{2}$)
the scaled mean photon number shows velocity dependence.
Fig.~\ref{fig:adadphitasljno} shows the time averaged scaled mean photon number
$\langle\expect{\op{a}^{\dagger}\op{a}}\rangle_{T}/j$ as a function of
the rotation velocity $\delta_{\phi}$.
It was averaged over a time interval of $t_{f}=150 \cdot
2\pi/\delta_{\phi}$. 

It is clear that there is a critical driving velocity for rotational evolution given by
$\delta_{\phi,c}^{(\mathrm{rot})}=\frac{4\lambda^{2}}{\omega}-\omega_{0}$
allowing to probe two different non-equilibrium phases of the Dicke
model.
On the other hand, in the case of the time evolution of the undriven Dicke
Hamiltonian the time averaged scaled mean photon number becomes
constant (if we encircle the critical paraboloid).
This shows that the presence of the non-equilibrium transition in the
Dicke model can depend on the initial state.

\subsubsection{Parity $\op{\Pi}$}

\begin{figure}[h!]
  \begin{center}
    \includegraphics[scale=0.43]{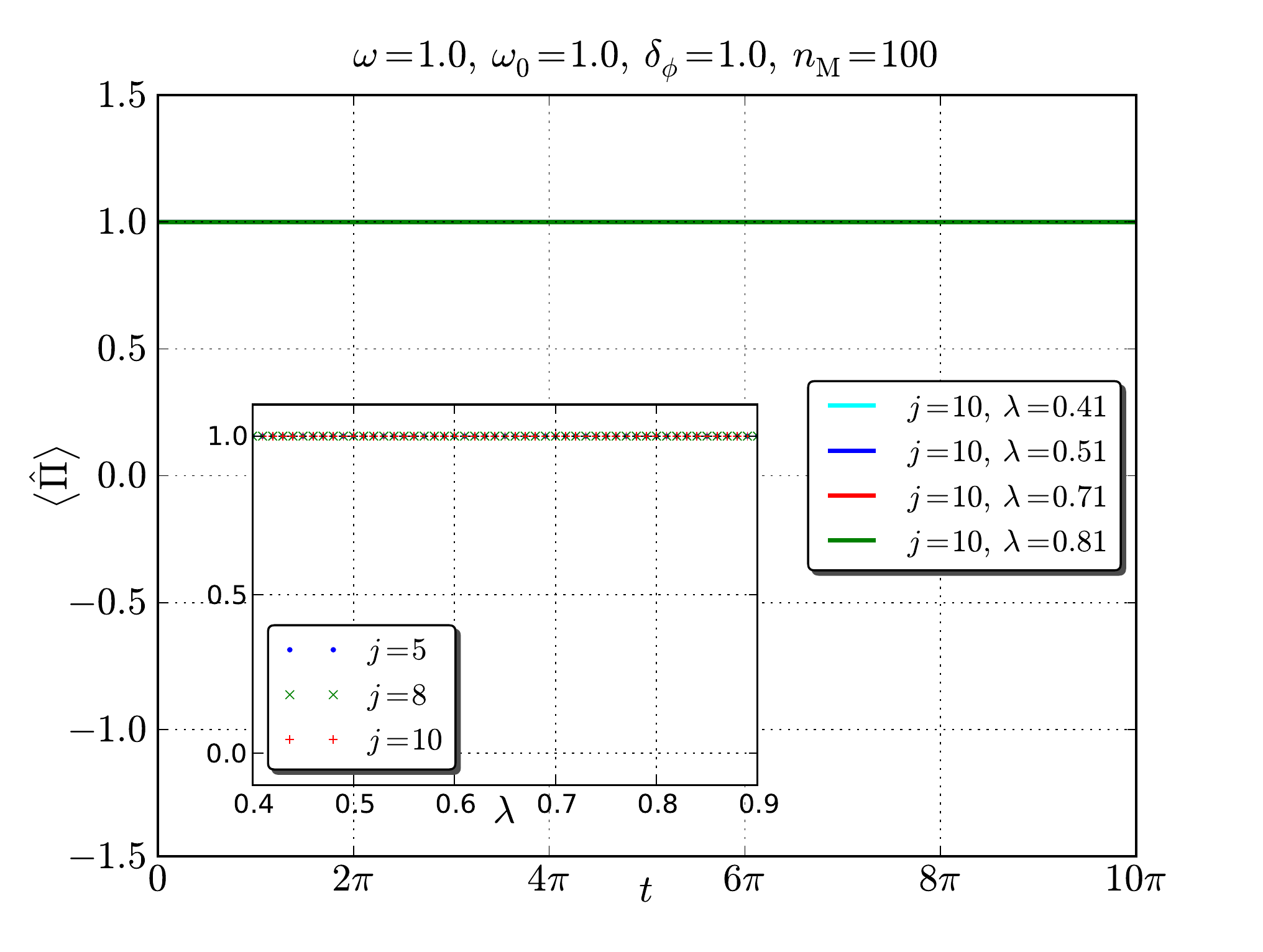}
    \caption{
      Time evolution of the parity operator when the system is driven
      from the initial Fock state $\ket{n=0}\ket{j,m=-j}$
      for a pseudo spin of $j=10$.
      The inset shows the dependence on the atom-file coupling
      strength $\lambda$ for different pseudo-spin length $j$.
      }
    \label{fig:pitimelamfssljno}
  \end{center}
\end{figure}

Here we study the time evolution of the parity operator if the
system is initially prepared in the Fock state.
In Fig.~\ref{fig:pitimelamfssljno} we show the time
evolution of the parity operator $\expect{\op{\Pi}}$ for different
coupling strength at fixed finite spin $j=10$.
In this case the parity is always constant in time with no sign of
change to zero at a critical coupling.
The system always remains in the state with the same parity $+1$. 

\begin{figure}[h!]
  \begin{center}
    \includegraphics[scale=0.43]{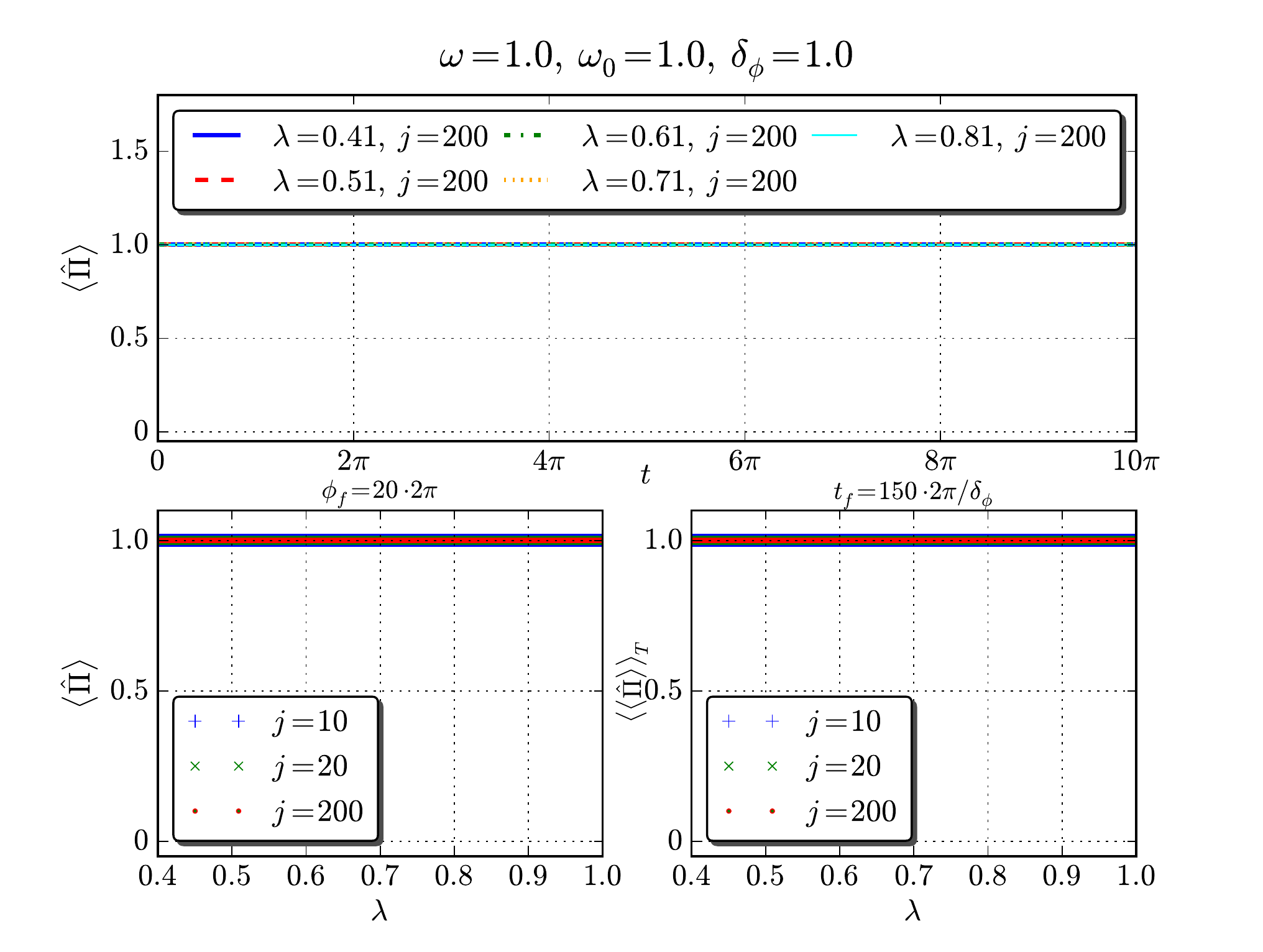}
    \caption{
      Upper panel: time evolution of the parity operator for the Fock
      initial state $\ket{n=0}\ket{j,m=-j}$.
      Lower left panel: dependence of the parity on the coupling
      constant $\lambda$ after $20$ circles.
      Lower right panel: dependence of the time averaged parity
      $\langle\langle\op{\Pi}\rangle\rangle_{T}$ on the
      atom-field coupling $\lambda$.
      }
    \label{fig:pitimelamfssljnotdl}
  \end{center}
\end{figure}

In the TDL ($j\to\infty$) one can use the solutions of the mean field
equations.
In this case the parity is also unchanged (see
Fig.~\ref{fig:pitimelamfssljnotdl}) which can be traced back to the
fact that the initial conditions 
$\alpha(0)=\alpha^{\ast}(0)=\zeta(0)=\zeta^{\ast}(0)=0$ allow no
quantum fluctuations and therefore do not evolve in time.
Also in the case of the scaled phase space coordinates
$(q_{1},p_{1},q_{2},p_{2}) \to 1/\sqrt{j}(q_{1},p_{1},q_{2},p_{2})$
the parity is always $+1$.

%
\subsection{Nearly Fock initial state: $\ket{\alpha(0)=10^{-\epsilon}}\ket{\zeta(0)=10^{-\epsilon}}$}

In the previous subsection we looked into the rotational dynamics of the
Dicke model with the initial state given by the Fock state with no
photons $n=0$ and lowest weight spin state $m=-j$,
$\ket{\psi(0)}=\ket{n=0}\ket{j,m=-j}$.
This correspond to zero parameters of the coherent states
$\alpha=0$ and $\zeta=0$.
This state does not evolve in time in the semi-classical limit.
As shown in the previous paragraph the mean field equations did not
reproduce the results calculated numerically for a system with a
finite number of two-level atoms.

Motivated by this we study the effect of small perturbation of the
initial conditions.
We consider a situation when the parameters $\alpha$
and $\zeta$ are not strictly zero but small and parametrized as
\be
  \alpha(0) = 10^{-\epsilon}, \qquad \zeta(0) = 10^{-\epsilon}
\ee
where $\epsilon$ is large.
Consider the following initial coherent state
\be
  \ket{\psi(t_{0}=0)}
  =
  \ket{\alpha(0) = 10^{-\epsilon}} \otimes \ket{\zeta(0) = 10^{-\epsilon}},
\ee
which differs slightly from the Fock state with no photons and lowest
weight state: for example, for $j=10$ and $n_{\mathrm{M}}=100$ we find
for the overlap
$\bracket{\alpha=10^{-3},\zeta=10^{-3}}{n=0,m=-j}=0.99999$.

We start the non-equilibrium evolution by switching on the rotation at
$t_{0}=0$ and evolving the system until $t_{f}=n_{\mathrm{R}} \, 2 \pi/\delta_{\phi}$.
Again we will compare the rotationally driven evolution and the free
evolution.

\subsubsection{Time evolution of the mean photon number}

\begin{figure}[h!]
  \begin{center}
    \includegraphics[scale=0.43]{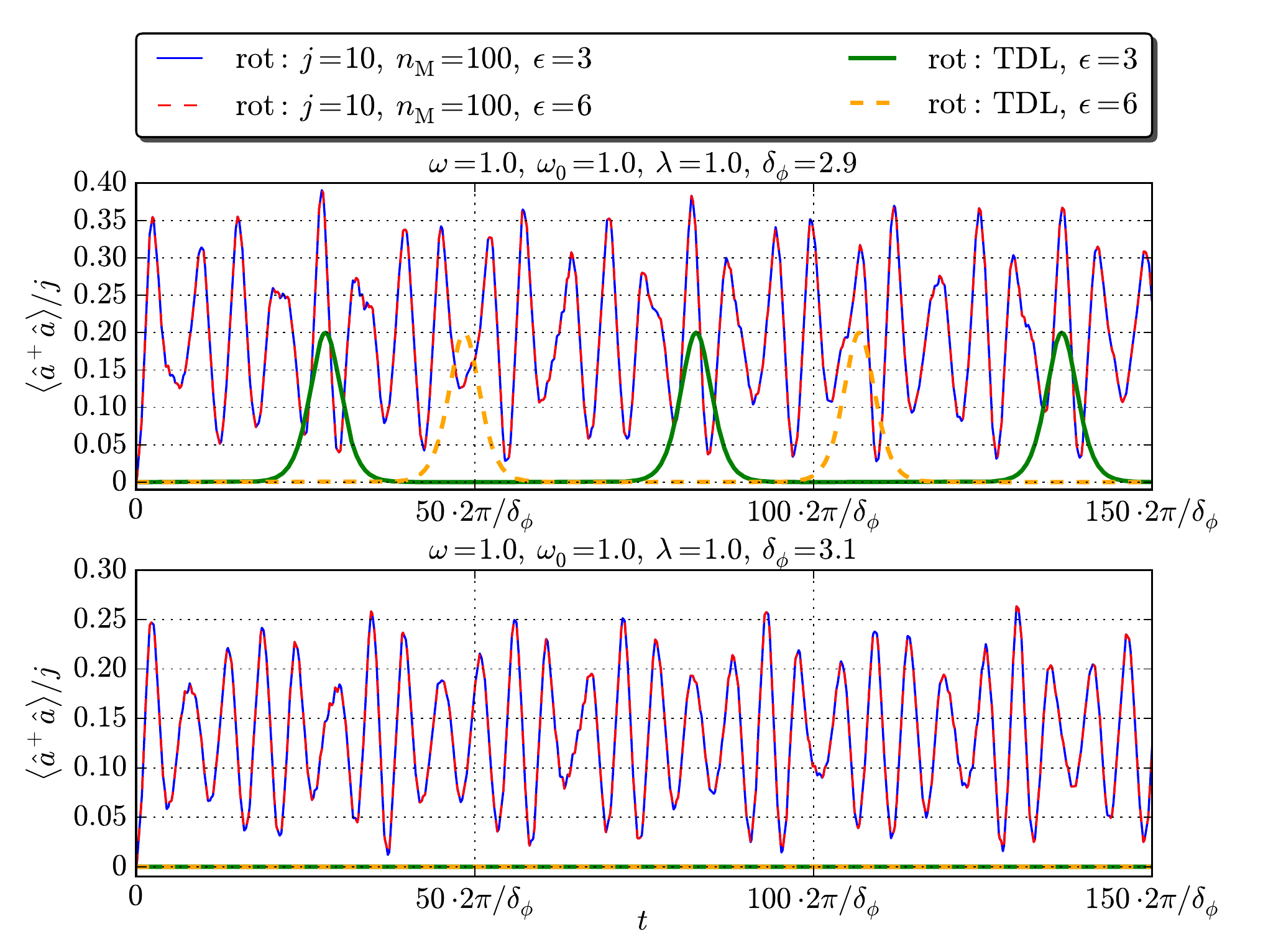}
    \caption{
      Time dependence of the scaled mean photon number on resonance
      $\omega = \omega_{0} = 1.0$ and 
      $\lambda=1.0$ for two different rotations velocities: one is slightly
      below the critical driving velocity $\delta_{\phi,c}=3.0$ while the second one 
      is just above it.
      The system is initially prepared in the coherent state
      $\ket{\alpha(0)=10^{-\epsilon}}\ket{\zeta(0)=10^{-\epsilon}}$, with
      $\epsilon=3$ and $\epsilon=6$ for rotational driving.
      }
    \label{fig:adatimesljnoe}
  \end{center}
\end{figure}

The time evolution of the scaled mean photon number,
$\expect{\op{a}^{\dagger}\op{a}}/j$, is depicted in
Fig.~(\ref{fig:adatimesljnoe}).
It compares the scaled mean photon number obtained for a finite system
using the numerical Chebyshev scheme
(blue and red curve) and the results of the TDL computed from the mean
field equations~(\ref{eq:tdmfeefirst})-(\ref{eq:tdmfeelast}) (green and
orange curve).
We observe that the initial state
$\ket{\alpha(0)=10^{-\epsilon}}\ket{\zeta(0)=10^{-\epsilon}}$ for different
$\epsilon$ does generate the same dynamics as the initial Fock state
$\ket{n=0}\ket{j,m=-j}$ for a finite number of two-level atoms.
However, in the TDL the mean field equations show
macroscopic excitations if we encircle the criticality (here
$\lambda=1.0 > \lambda_{c} = 0.5$)
and drive the system slower than the critical driving velocity
$\delta_{\phi,c}=\frac{4\lambda^{2}}{\omega}-\omega_{0}$.
We note that the ``soliton train'' appearing in this figure (green and
yellow lines) is somewhat reminiscent to the findings of
Ref.~\cite{Yuzbashyan},~\cite{Faribault}.
Note however that the model we study here is non-integrable. 

\begin{figure}[h!]
  \begin{center}
    \includegraphics[scale=0.43]{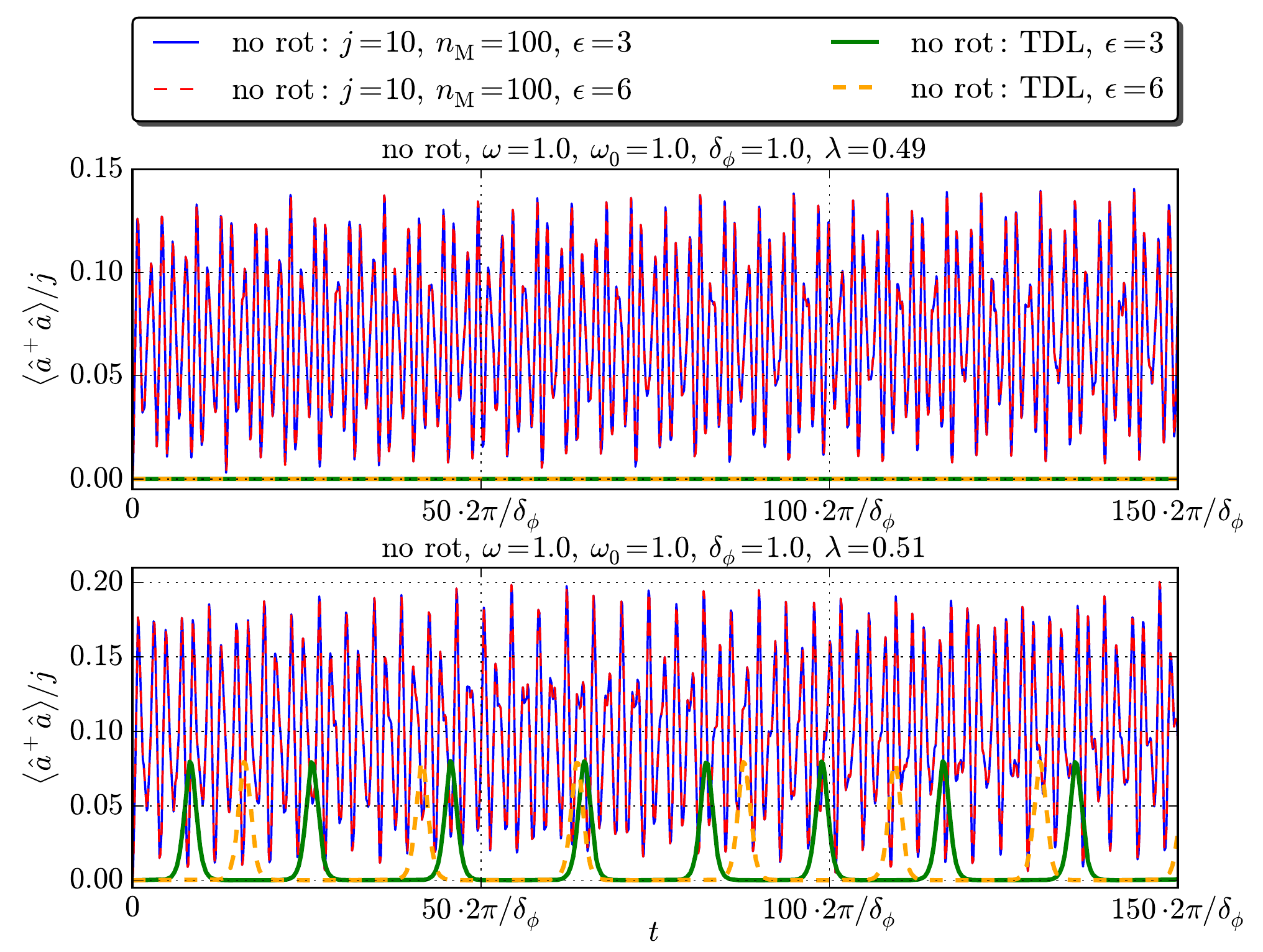}
    \caption{
      Time dependence of the scaled mean photon number on resonance
      $\omega = \omega_{0} = 1.0$ for two different coupling strength
      $\lambda$: one slightly
      below the critical driving velocity $\lambda_{c}=\frac{\sqrt{\omega\omega_{0}}}{2}=0.5$ and
      one just above it.
      The system is initially prepared in the coherent state
      $\ket{\alpha(0)=10^{-\epsilon}}\ket{\zeta(0)=10^{-\epsilon}}$, with
      $\epsilon=3$ and $\epsilon=6$ and evolved with undriven Hamiltonian.
      }
    \label{fig:adatimesljnoenorot}
  \end{center}
\end{figure}

In Fig.~(\ref{fig:adatimesljnoenorot}) we show the time evolution of
the scaled mean photon number with no rotational driving, i.e., the
time evolution is governed by the usual Dicke Hamiltonian
$\op{H}_{\mathrm{D}}$.
Again, the system is initially prepared in the coherent state
$\ket{\alpha(0)=10^{-\epsilon}}\ket{\zeta(0)=10^{-\epsilon}}$.

Summarizing these results we observe that the initial state
$\ket{\alpha(0)=10^{-\epsilon}}\ket{\zeta(0)=10^{-\epsilon}}$ for different
$\epsilon$ and finite $j$ reproduces the same dynamics as for the initial
Fock state $\ket{n=0}\ket{j,m=-j}$.
In the TDL the mean field equations show a macroscopic
excitations of the mean photon number in the super-radiant phase
($\lambda>\lambda_{c}$) if we use the initial conditions
$\alpha(0)=\alpha^{\ast}(0)=\zeta(0)=\zeta^{\ast}(0)=10^{-\epsilon}$.
This means that small fluctuations present in the initial state make
the quasi-classical description possible.

\subsubsection{Dependence of the mean photon number on the atom-field coupling strength $\lambda$}

\begin{figure}[h!]
  \begin{center}
    \includegraphics[scale=0.43]{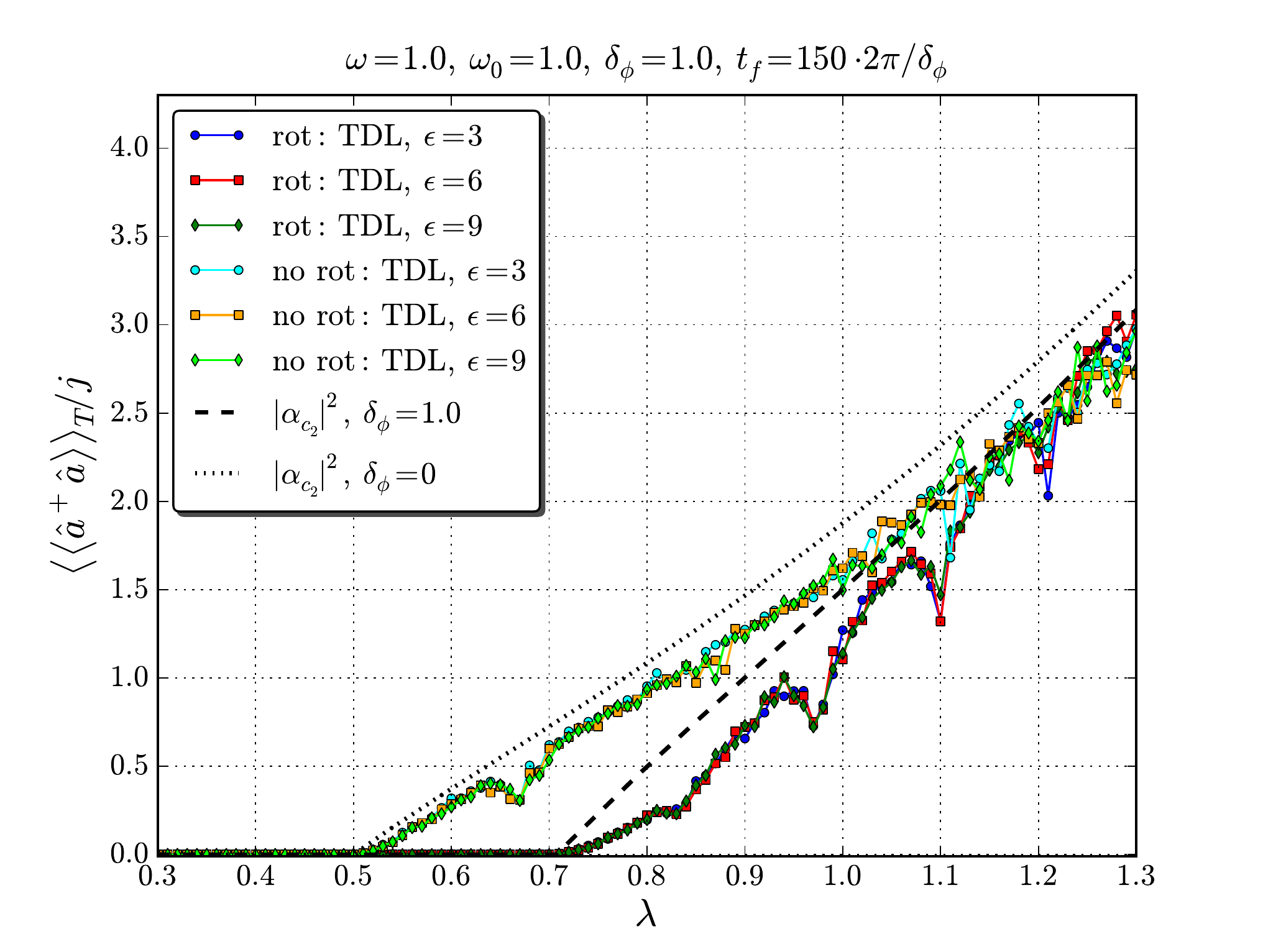}
    \caption{
      Time-averaged
      mean photon number $\langle \expect{\op{a}^{\dagger}\op{a}}
      \rangle_{T}/j$ as a function of the coupling strength.
      In the TDL ($j \to \infty$) obtained by solving
      the mean field
      equations~(\ref{eq:tdmfeefirst})-(\ref{eq:tdmfeelast}) with the
      initial conditions
      $\alpha(0)=\alpha^{\ast}(0)=\zeta(0)=\zeta^{\ast}(0)=10^{-\epsilon}$.       }
    \label{fig:adalamtasljnoe}
  \end{center}
\end{figure}

In Fig.~\ref{fig:adalamtasljnoe} we plot $\langle
\expect{\op{a}^{\dagger}\op{a}} \rangle_{T}/j$ as a function $\lambda$ 
for the driven model (blue, red, green) and for the undriven one
(cyan, orange, lime).
We again observe a critical coupling strength $\lambda_{c}$ at
which the number of excited photons becomes macroscopic.
In the case of rotational driving the critical coupling is shifted by the amount of
the applied rotation velocity $\delta_{\phi}$.
The $\lambda$-dependence of $\langle
\expect{\op{a}^{\dagger}\op{a}} \rangle_{T}/j$ for the initial 
Fock state of the Dicke model with a finite number of atoms looks
similar to the one obtained in the TDL with
$\alpha(0)=\alpha^{\ast}(0)=\zeta(0)=\zeta^{\ast}(0)=10^{-\epsilon}$.
However, the main differences with a driven model are (i) appearance of dips and 
(ii) difference in the scaling form of the curve when $\lambda \to
\lambda_{c}^{(\mathrm{rot})}$.

\subsubsection{Dependence of the mean photon number on the rotation velocity $\delta_{\phi}$}

\begin{figure}[h!]
  \begin{center}
    \includegraphics[scale=0.43]{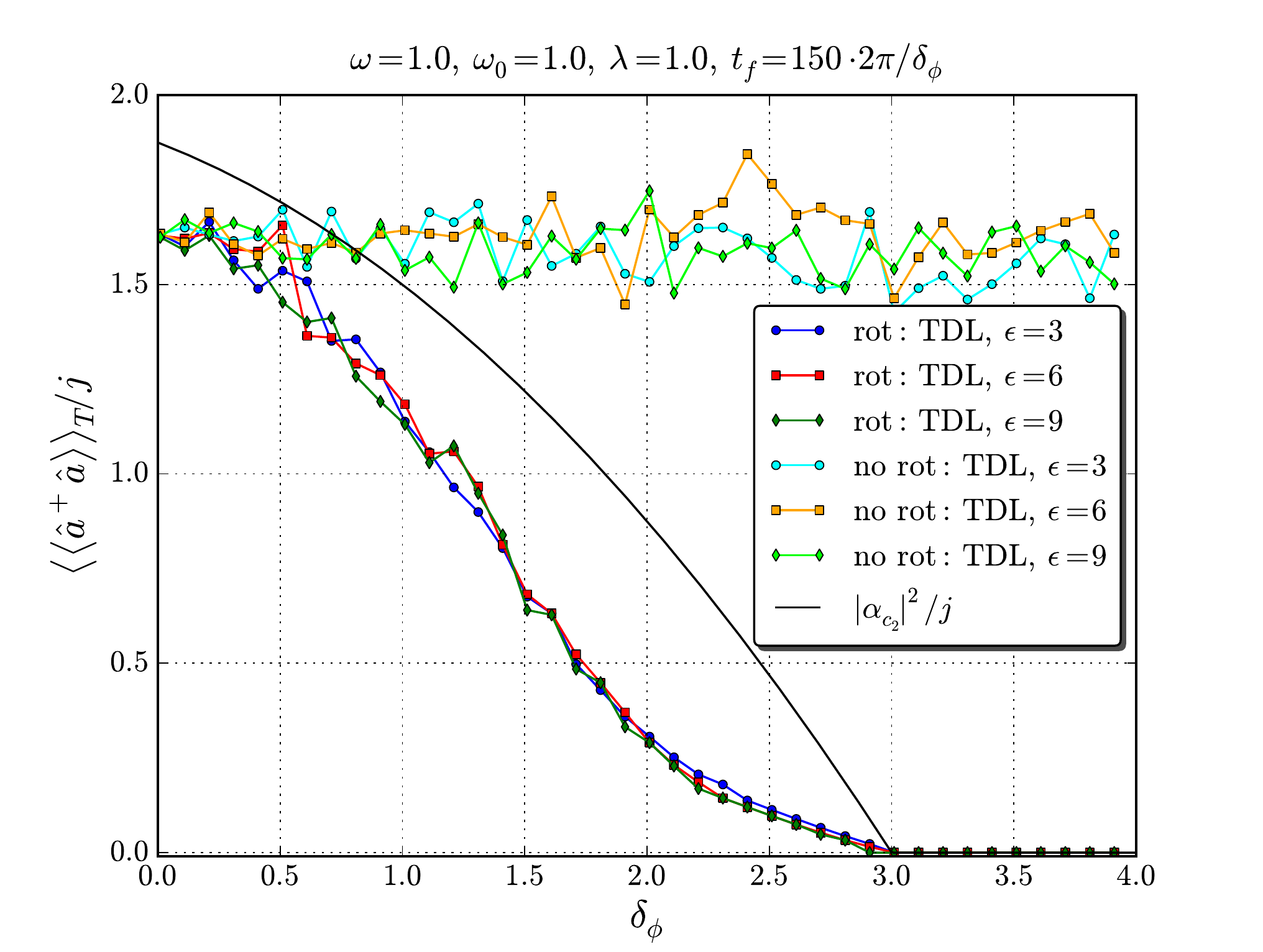}
    \caption{
      Time-averaged mean photon number $\langle
      \expect{\op{a}^{\dagger}\op{a}} \rangle_{T}/j$ as a function of
      the driving velocity.
      The TDL ($j \to \infty$) is obtained by solving
      the mean field
      equations~(\ref{eq:tdmfeefirst})-(\ref{eq:tdmfeelast}) with the
      initial conditions
      $\alpha(0)=\alpha^{\ast}(0)=\zeta(0)=\zeta^{\ast}(0)=10^{-\epsilon}$.
      }
    \label{fig:adadphitasljnoe}
  \end{center}
\end{figure}

The dependence of $\langle \expect{\op{a}^{\dagger}\op{a}}
\rangle_{T}/j$ on $\delta_{\phi}$ for a driven (blue, red, green) and
undriven (cyan, orange, lime) systems is shown in
Fig.~(\ref{fig:adadphitasljnoe}).
When the critical paraboloid (path $\mathcal{C}_{2}$) is encircled,
there is a critical driving velocity given by
$\delta_{\phi,c}^{(\mathrm{rot})}=\frac{4\lambda^{2}}{\omega}-\omega_{0}$. 
This is in contrast to the TDL with the initial conditions
$\alpha(0)=\alpha^{\ast}(0)=\zeta(0)=\zeta^{\ast}(0)=0$ when the
mean photon number was always zero.
In the case of undriven model the time averaged scaled
mean photon number fluctuates around a constant
value.

\subsubsection{Parity $\op{\Pi}$}

\begin{figure}[h!]
  \begin{center}
    \includegraphics[scale=0.43]{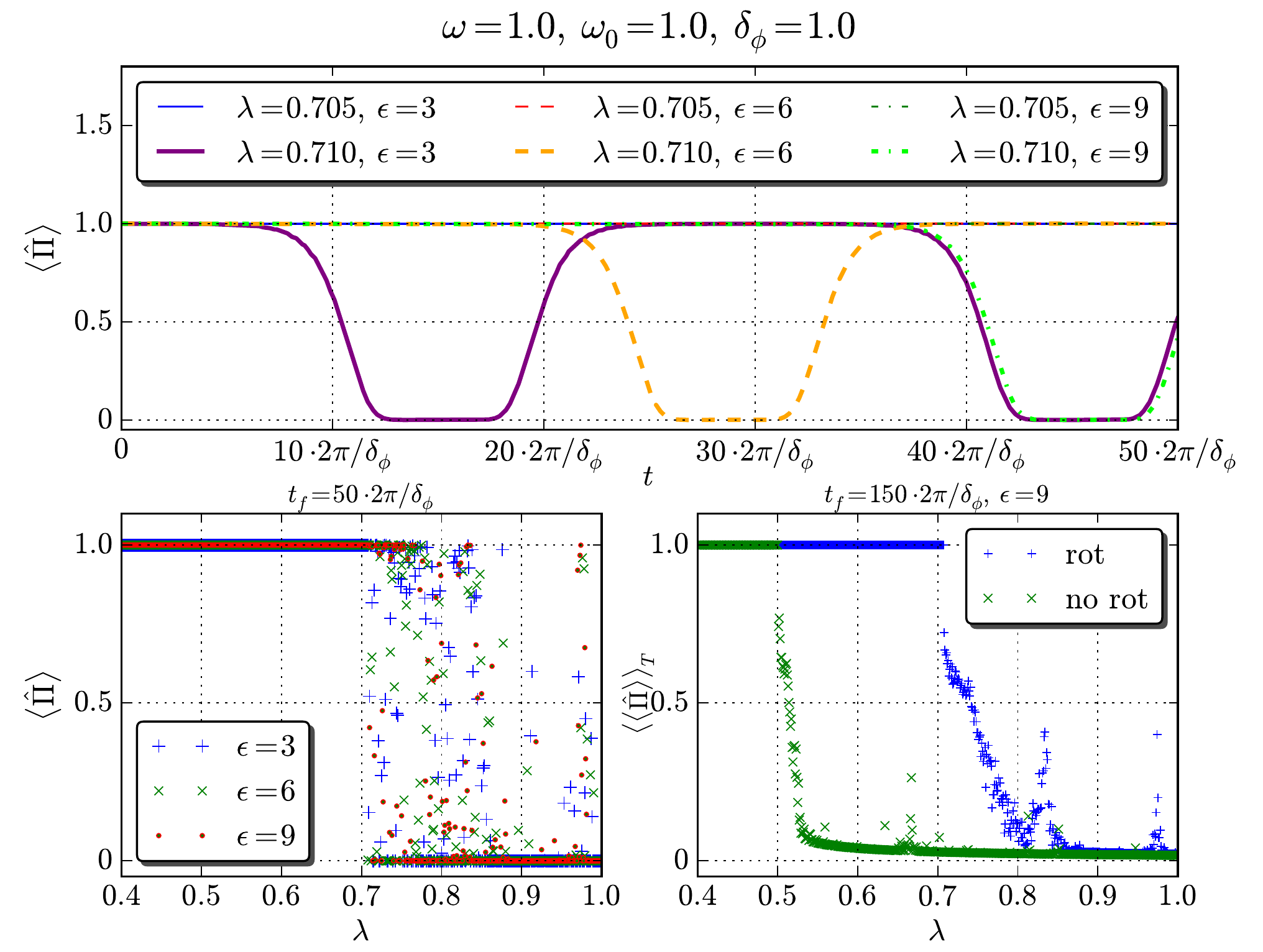}
    \caption{
      Upper panel:
      time evolution of the parity operator for the driven system starting from the nearly-Fock state.
      Lower left panel: 
      the parity after $50$ circles as a function of the coupling.
      Lower right plot: the same as left panel but averaged over $150$ circles.
      }
    \label{fig:pitimelamtdlsljnoepsilon}
  \end{center}
\end{figure}
\begin{figure}[h!]
  \begin{center}
    \includegraphics[scale=0.43]{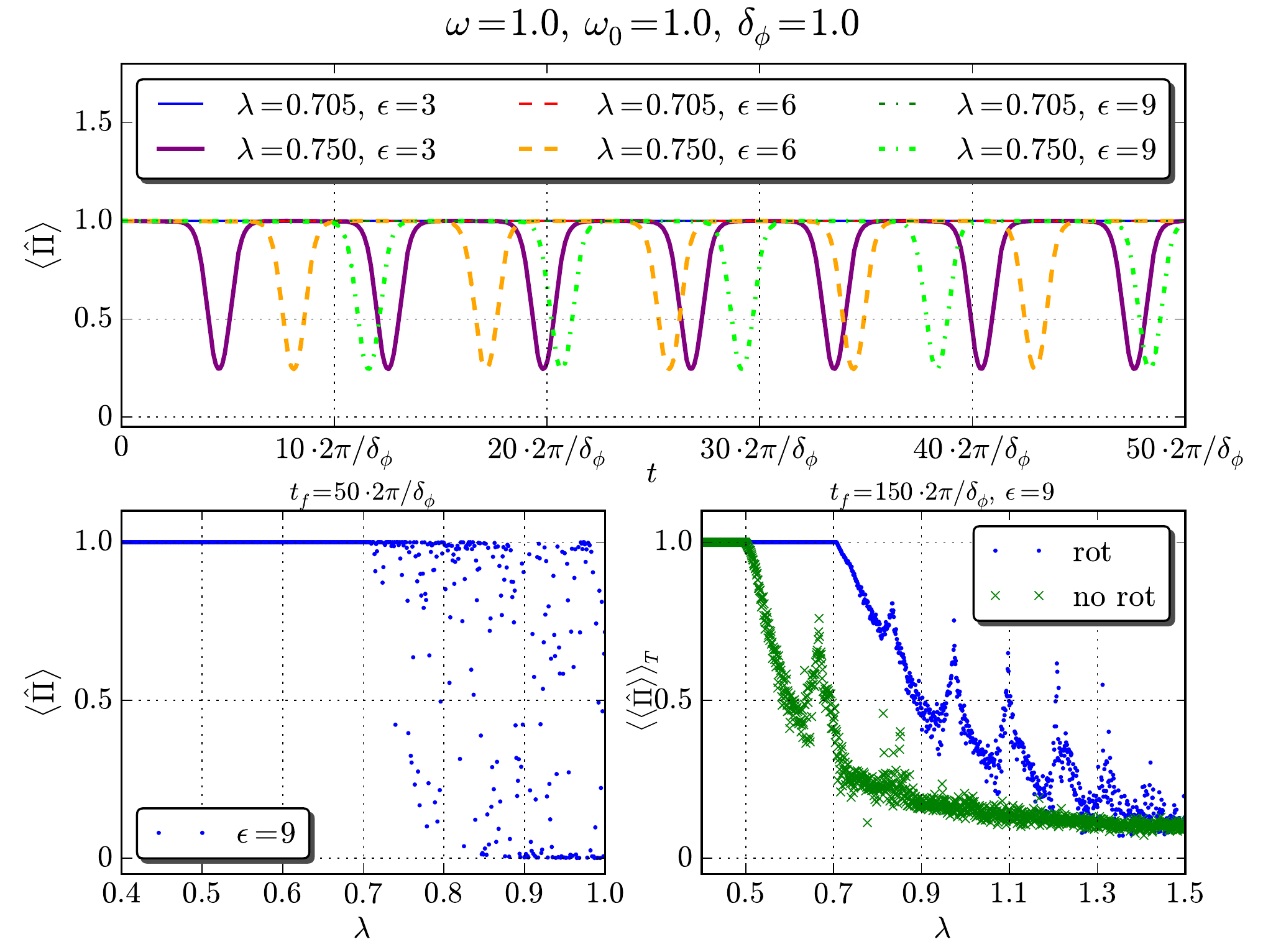}
    \caption{
      Upper panel:
      time evolution of the scaled parity operator when we start the
      rotational dynamics with the system prepared in the nearly-Fock state.
      Lower left panel: 
      the scaled parity as a function of $\lambda$ after $50$ circles.
      Lower right panel: the same as the left one but averaged over $150$ circles.
      }
    \label{fig:pitimelamtdlsljnoscepsilon}
  \end{center}
\end{figure}

Using the initial state
$\ket{\alpha(0)=10^{-\epsilon}}\ket{\zeta(0)=10^{-\epsilon}}$ with
$\epsilon>2$ for a system of a finite number of atoms we find the same
behavior, $\expect{\op{\Pi}}=+1$, for any time and coupling strength as
for the initial Fock state $\ket{n=0}\ket{j,m=-j}$.
On the other hand, in the TDL ($j\to\infty$) using the solutions of
the mean field equations~(\ref{eq:tdmfeefirst})-(\ref{eq:tdmfeelast})
we compute the time evolution of the parity and observe
(Fig.~\ref{fig:pitimelamtdlsljnoepsilon}) that for a coupling
$\lambda>\lambda_{c}^{(\mathrm{rot})}$ the time evolution shows a
smooth jumps from $+1$ to $0$.
We also show the parity calculated using the rescaled phase space
coordinates (see Fig.~\ref{fig:pitimelamtdlsljnoscepsilon}).

\subsection{Ground state of the un-rotated Dicke model as initial state: $\ket{\mathrm{GS}}$}

Here we consider the evolution of a system initially prepared in 
the ground state 
\be
 \ket{\psi(t_{0}=0)}=\ket{\mathrm{GS}}.
\ee
First, we numerically compute the ground state $\ket{\mathrm{GS}}$
of the Dicke Hamiltonian $\op{H}_{\mathrm{D}}$ by
truncating the bosonic Hilbert 
space up to $n_{\mathrm{M}}$ while maintaining the full Hilbert
space of the pseudo-spin $j$. 
Then, at $t_{0}=0$, we start to evolve the system according to the
rotated Dicke model $\op{H}_{\mathrm{RD}}(t)$.
Next we measure the scaled mean photon number
$\expect{\op{a}^{\dagger}\op{a}}$ as
\begin{align}
  \expect{\op{a}^{\dagger}\op{a}} =&
  \bra{\psi(t_{f})} \op{a}^{\dagger}\op{a} \ket{\psi(t_{f})}
  \nonumber \\
  =&
  \bra{\mathrm{GS}} e^{i\op{H}_{\mathrm{ROT}}t_{f}} \,
  \op{a}^{\dagger}\op{a} \, e^{-i\op{H}_{\mathrm{ROT}}t_{f}} \ket{\mathrm{GS}},
\end{align}
where the final time is, as before, given by
$t_{f}=n_{\mathrm{R}} \, 2 \pi/\delta_{\phi}$.  
We also consider the undriven evolution governed by the
evolution of the usual Dicke model $\op{H}_{\mathrm{D}}$.

\subsubsection{Time evolution of the mean photon number}

\begin{figure}[h!]
  \begin{center}
    \includegraphics[scale=0.43]{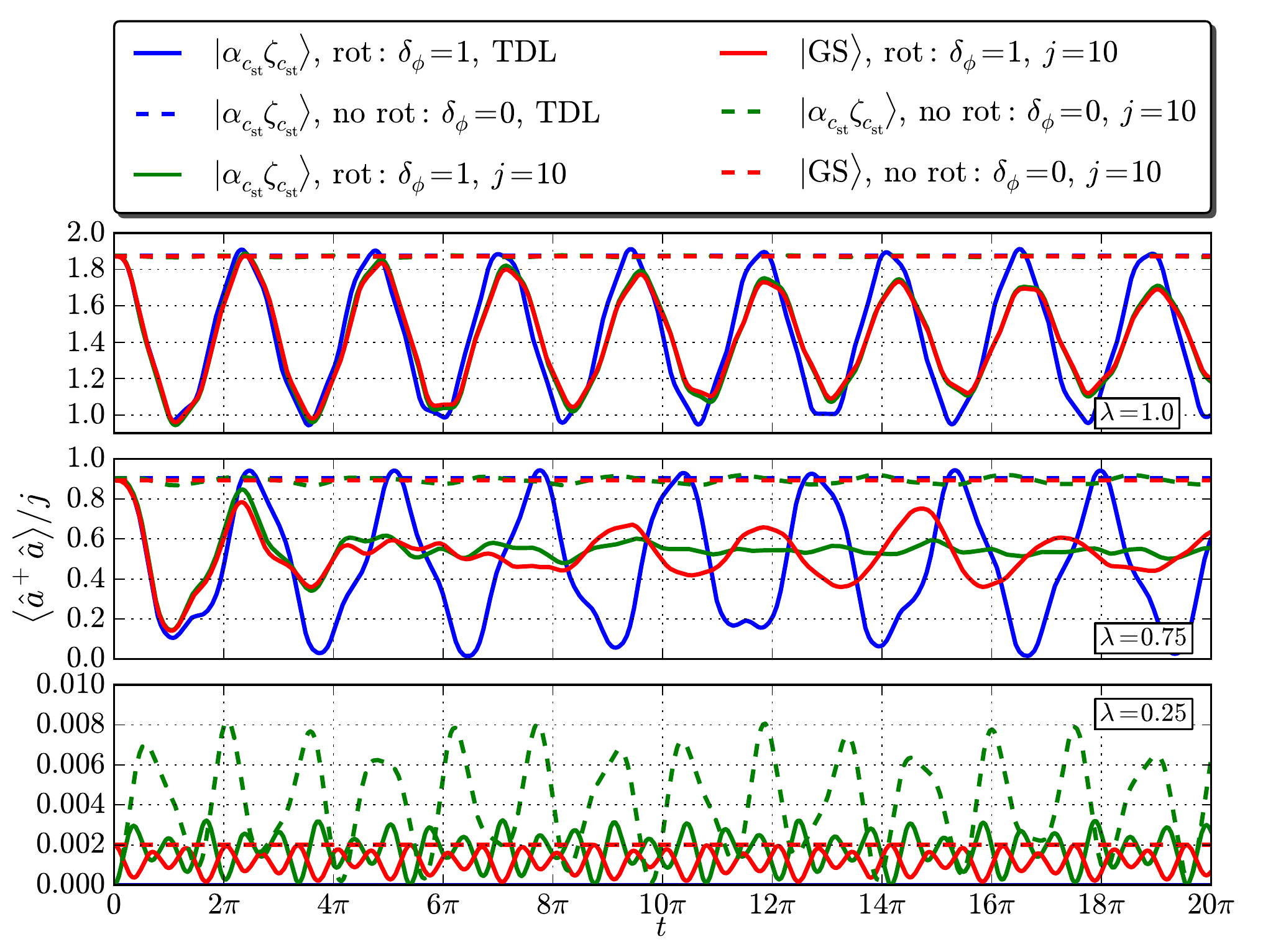}
    \caption{
      Time dependence of the scaled mean photon number on resonance
      $\omega = \omega_{0} = 1.0$ for different values of the coupling
      strength $\lambda$.
      Here we compare the results for different initial states: $\ket{\mathrm{GS}}$ and
      $\ket{\alpha_{c_{\mathrm{st}}}}\ket{\zeta_{c_{\mathrm{st}}}}$.
      The ground state $\ket{\mathrm{GS}}$ is obtained numerically and
      we choose $n_{\mathrm{M}}=100$.
      }
    \label{fig:adatimesgs}
  \end{center}
\end{figure}

The time dependence of the scaled mean photon number is shown in
Fig.~\ref{fig:adatimesgs}.
Here we compare the results for the initial ground state
$\ket{\mathrm{GS}}$ (red curves) with the results for the initial
coherent state with the parameters $\alpha_{c_{\mathrm{st}}}$ and
$\zeta_{c_{\mathrm{st}}}$ (blue and green curves).
The dashed lines show the situation for the undriven evolution.

For the rotationally driven case (solid lines) we observe an oscillatory
behavior of the scaled mean photon number.
For a finite number of atoms the time evolution
of the scaled mean photon number for the
ground state $\ket{\mathrm{GS}}$ is close to the one for the coherent
state $\ket{\alpha_{c_{\mathrm{st}}}}\ket{\zeta_{c_{\mathrm{st}}}}$.
For large atom-field couplings they become identical.
Thus, the coherent state
$\ket{\alpha_{c_{\mathrm{st}}}}\ket{\zeta_{c_{\mathrm{st}}}}$ is a
good approximation for the ground state of the system even in the
driven case.
This phenomenon is noticed already in~\cite{Gilmore1993}.
In the undriven case the time evolution of the mean photon number for
the ground state is constant and corresponds to the value found in the
TDL
\be
  \bra{\mathrm{GS}} \op{a}^{\dagger} \op{a} \ket{\mathrm{GS}} =
  \left\{
  \begin{array}{l l}
    \frac{1}{2}
    \lp \frac{2\lambda}{\omega} \rp^{2}
    \lb 1 - \lp \frac{\omega\omega_{0}}{4\lambda^{2}} \rp^{2} \rb 
    & \quad \text{if $\lambda \geq \frac{1}{2} \sqrt{\omega\omega_{0}}$} \\
    0 & \quad \text{if $\lambda <  \frac{1}{2} \sqrt{\omega\omega_{0}}$} \\
  \end{array}.
 \right.
\ee

\subsubsection{Dependence of the mean photon number on the atom-field coupling strength $\lambda$}

\begin{figure}[h!]
  \begin{center}
    \includegraphics[scale=0.43]{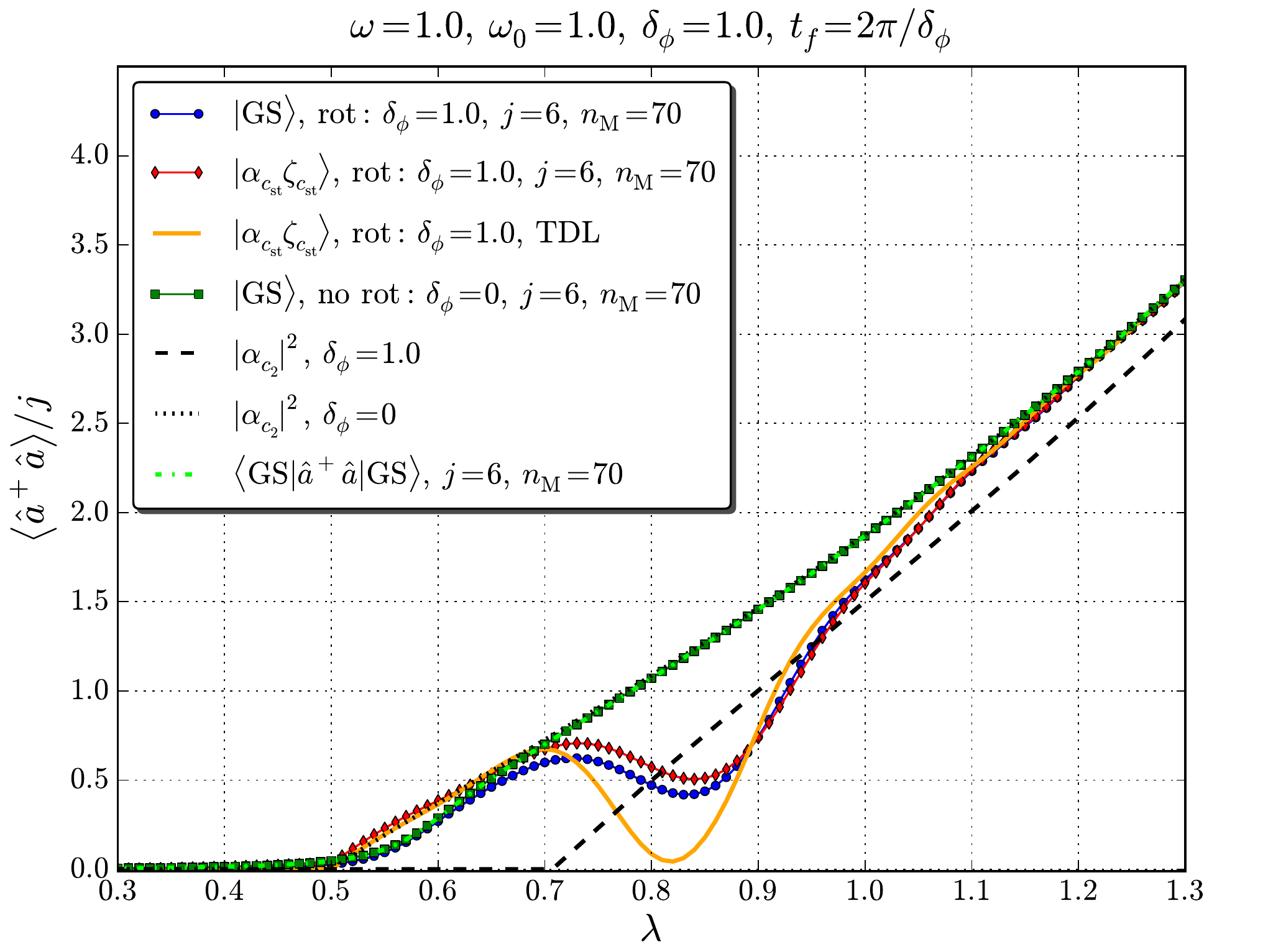}
    \caption{
      The scaled mean photon number as a function of the atom-field
      coupling strength: here we compare the mean field solution
      ($j\to\infty$) for the initial coherent state
      $\ket{\alpha_{c_{\mathrm{st}}}}\ket{\zeta_{c_{\mathrm{st}}}}$ (yellow solid
      line) with the time evolution calculated numerically 
      for a system with a finite number of atoms ($j=6$) for the
      initial state chosen as ground state $\ket{\mathrm{GS}}$ (blue
      line) and the coherent state (red line).
      We also compare it with the undriven case (green line) and the
      static (dashed lime).
      }
    \label{fig:adalamsgs}
  \end{center}
\end{figure}

Here we study the dependence of the scaled mean photon number on the
atom-field coupling strength $\lambda$.
First, we consider one revolution in parameter space,
$t_{f}=2\pi/\delta_{\phi}$.
This case is shown for a fixed rotation velocity $\delta_{\phi}=1.0$ at 
resonance $\omega=\omega_{0}=1.0$, in Fig.~\ref{fig:adalamsgs}.
We also show the dependence of the scaled mean
photon number on $\lambda$ for the undriven Dicke model (green curve), where the
final time was kept the same $t_{f}=2\pi/\delta_{\phi}$.

As we have already seen the time dependence of the scaled mean photon
number calculated for the initial ground state
$\ket{\mathrm{GS}}$ and for the initial coherent 
state $\ket{\alpha_{c_{\mathrm{st}}}}\ket{\zeta_{c_{\mathrm{st}}}}$
give qualitatively the same results.
For the ground initial state we observe a minimum in the super-radiant
phase (in Fig.~\ref{fig:adalamsgs} at $\lambda\approx0.82$).
This indicates a competition between the two stationary states or the
ground states of the rotationally driven and undriven Dicke
Hamiltonians.
One further interesting point is the difference in scaling behavior
close to the critical point $\lambda_{c}=\sqrt{\omega\omega_{0}}/2$:
This behavior is different for the initial ground and coherent states.
For the ground state it follows the same scaling behavior as in the
static and unrotated case.

\begin{figure}[h!]
  \begin{center}
    \includegraphics[scale=0.43]{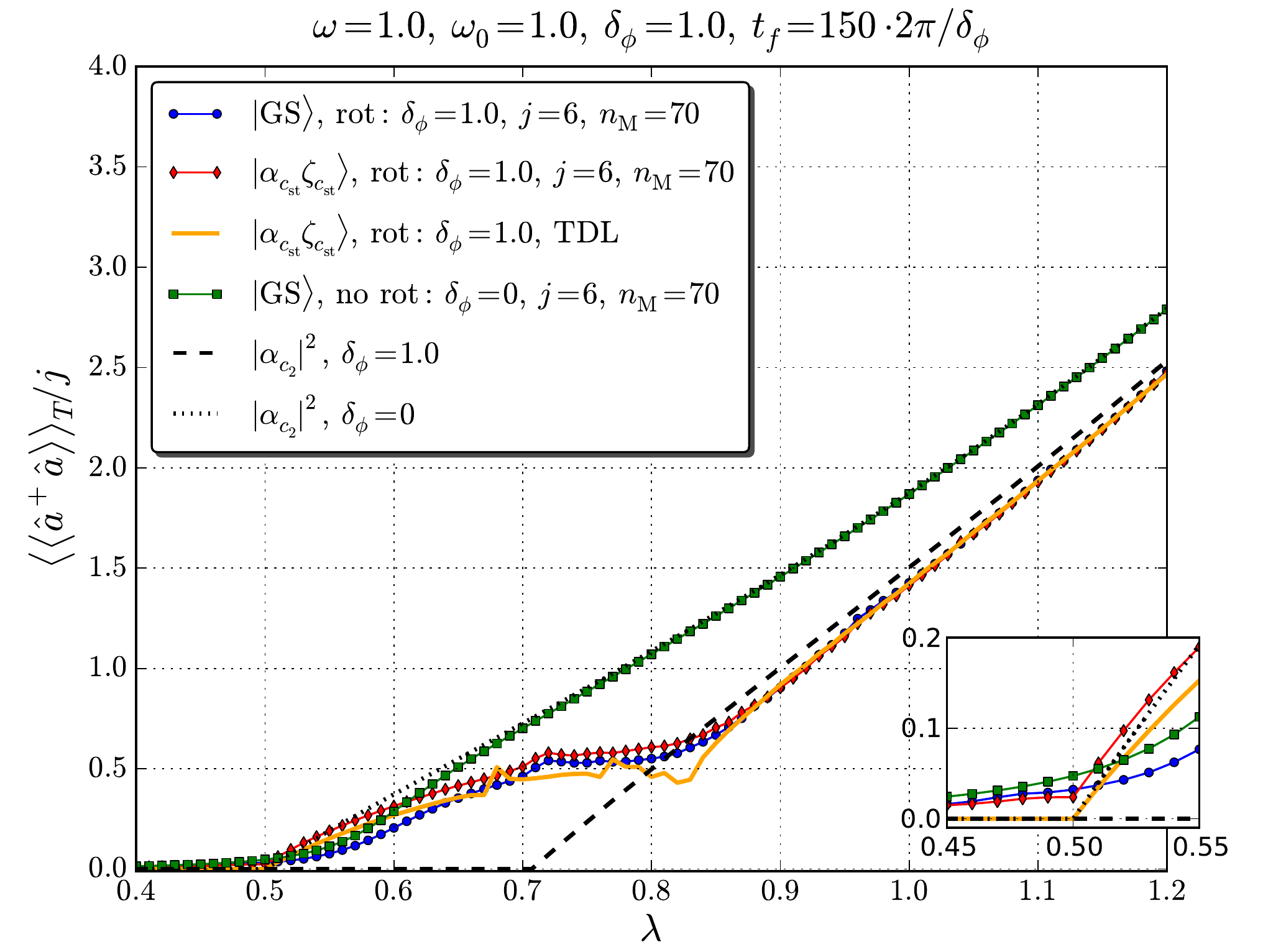}
    \caption{
      The time averaged scaled mean photon number
      $\langle\expect{\op{a}^{\dagger}\op{a}}\rangle_{T}/j$ at
      resonance $\omega = \omega_{0} = 1.0$ as a function of the
      coupling $\lambda$.
      The system is initially prepared in the 
      ground state $\ket{\mathrm{GS}}$ which is calculated numerically
      and then evolved until $t_{f}=150 \cdot
      2\pi/\delta_{\phi}$ (blue curve).
      The result is compared with the one obtained for the initial coherent
      state (red and yellow curve) and with the one corresponding to undriven system
      (green line).
      }
    \label{fig:adalamtasgs}
  \end{center}
\end{figure}

So far we considered the mean photon number after one circle
$\phi_{f}=2\pi$, but the situation with many rotations 
$\phi_{f}\to\infty$ represent an interesting limit.
In this case the interplay of dynamical and
geometrical phase may be a source some new effects~\cite{TPG}.
Considering the case of many rotations in Fig.~\ref{fig:adatimesgs} it
is natural to take the time average of the scaled mean photon number
as already defined in Eq.~(\ref{eq:timeavesmph}).
The time averaged scaled mean photon number $\langle\langle
\op{a}^{\dagger}\op{a} \rangle \rangle_{T} / j$ is depicted in
Fig.~\ref{fig:adalamtasgs}.
Here we averaged over a time interval of $t_{f}=150 \cdot
2\pi/\delta_{\phi}$
for the resonance conditions $\omega = \omega_{0} =
1.0$ and kept the rotation velocity to be $\delta_{\phi}=1.0$.

We qualitatively observe the same behavior for the time
averaged scaled mean photon number for the two different initial states,
namely for the ground state $\ket{\mathrm{GS}}$ and for the coherent state
$\ket{\alpha_{c_{\mathrm{st}}}}\ket{\zeta_{c_{\mathrm{st}}}}$.
Moreover, the time averaged scaled mean photon number shows a
different scaling for the two initial states when approaching the equilibrium critical
coupling $\lambda_{c}=\sqrt{\omega\omega_{0}}/2$.

\subsubsection{Dependence of the mean photon number on the rotation velocity $\delta_{\phi}$}

\begin{figure}[h!]
  \begin{center}
    \includegraphics[scale=0.43]{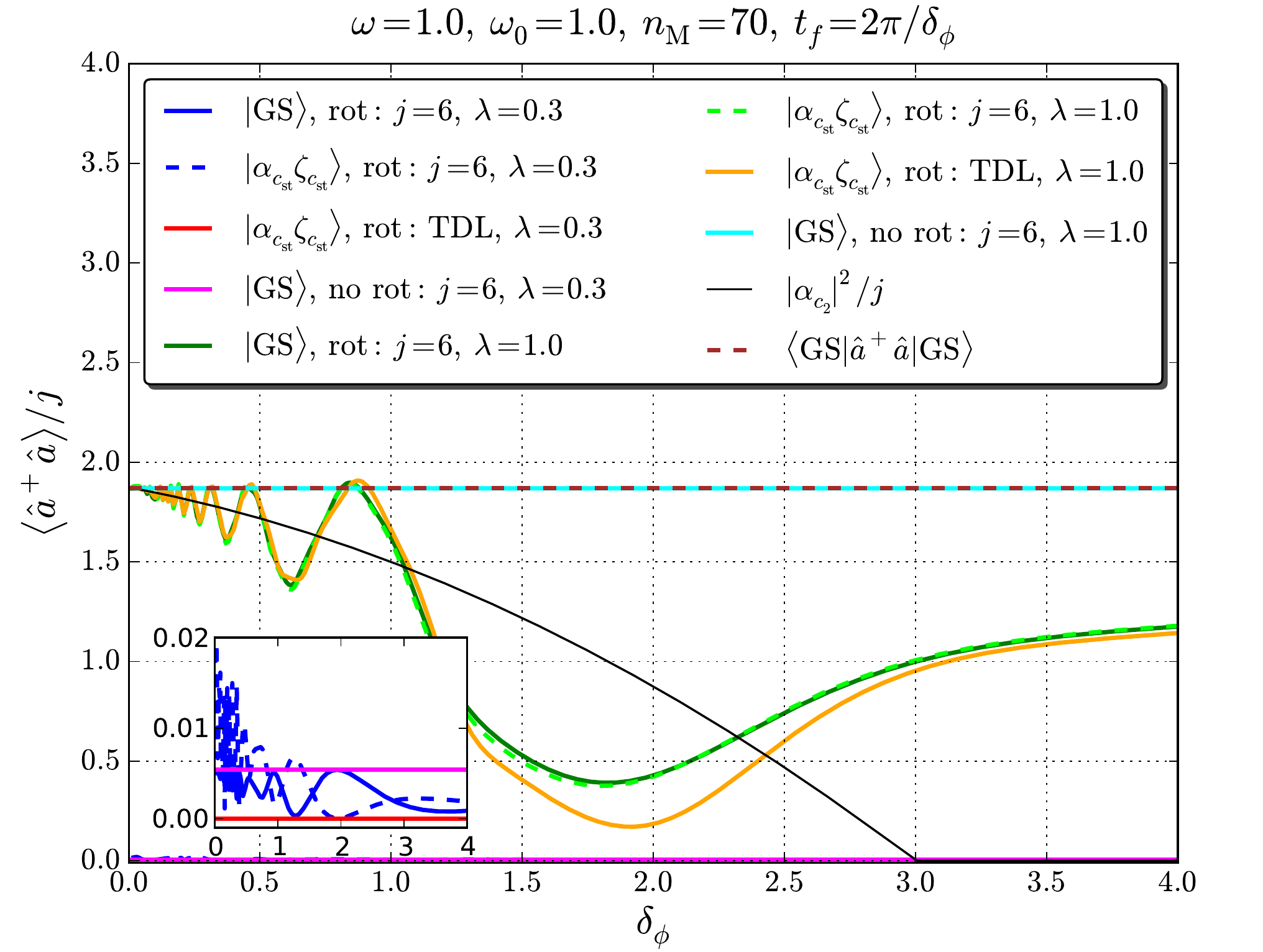}
    \caption{
      The scaled mean photon number on a
      resonance $\omega = \omega_{0} = 1.0$ for $\lambda=0.3$ and
      $\lambda=1.0$ as a function of the driving velocity.
      We compare $\expect{\op{a}^{\dagger}\op{a}}(t_{f},\delta_{\phi})/j$
      for two different initial states: $\ket{\mathrm{GS}}$ and
      $\ket{\alpha_{c_{\mathrm{st}}}}\ket{\zeta_{c_{\mathrm{st}}}}$
      for the driven and undriven situation (with
      $t_{f}=2\pi/\delta_{\phi}$ in both cases).
      }
    \label{fig:adadphisgs}
  \end{center}
\end{figure}

In this subsection we illustrate the dependence of the scaled mean
photon number on the rotation velocity $\delta_{\phi}$.
First, in Fig.~\ref{fig:adadphisgs} we consider one revolution,
$t_{f}=2\pi/\delta_{\phi}$ for different atom-field couplings
and for $\omega = \omega_{0} = 1.0$.
We plot the velocity dependence of the scaled mean photon number for
rotational driving once for the initial ground state
$\ket{\mathrm{GS}}$ (solid blue and solid green curve) and for the
initial coherent state
$\ket{\alpha_{c_{\mathrm{st}}}}\ket{\zeta_{c_{\mathrm{st}}}}$ (dashed
blue and green curves and solid red and orange curves).
We also show the undriven situation (pink and cyan) and the static
(brown dashed curve).
The behavior of 
$\expect{\op{a}^{\dagger}\op{a}}/j$ for the two different initial state
$\ket{\mathrm{GS}}$ and
$\ket{\alpha_{c_{\mathrm{st}}}}\ket{\zeta_{c_{\mathrm{st}}}}$ is
qualitatively similar. 
Namely that if the equilibrium critical paraboloid is not
encircled $\lambda<\lambda_{c}=\sqrt{\omega\omega_{0}}/2$ then the
mean photon number is zero.
But if we encircle it ($\lambda>\lambda_{c}=\sqrt{\omega\omega_{0}}/2$) the mean photon
number becomes macroscopic.
There is no critical driving velocity in the driven
case.
We remind that for the stationary state of the rotated Dicke model (see black
curve in Fig.~\ref{fig:adadphisgs}) $\delta_{\phi,c}^{(\mathrm{rot})}=\frac{4\lambda^{2}}{\omega}-\omega_{0}$.

\begin{figure}[h!]
  \begin{center}
    \includegraphics[scale=0.43]{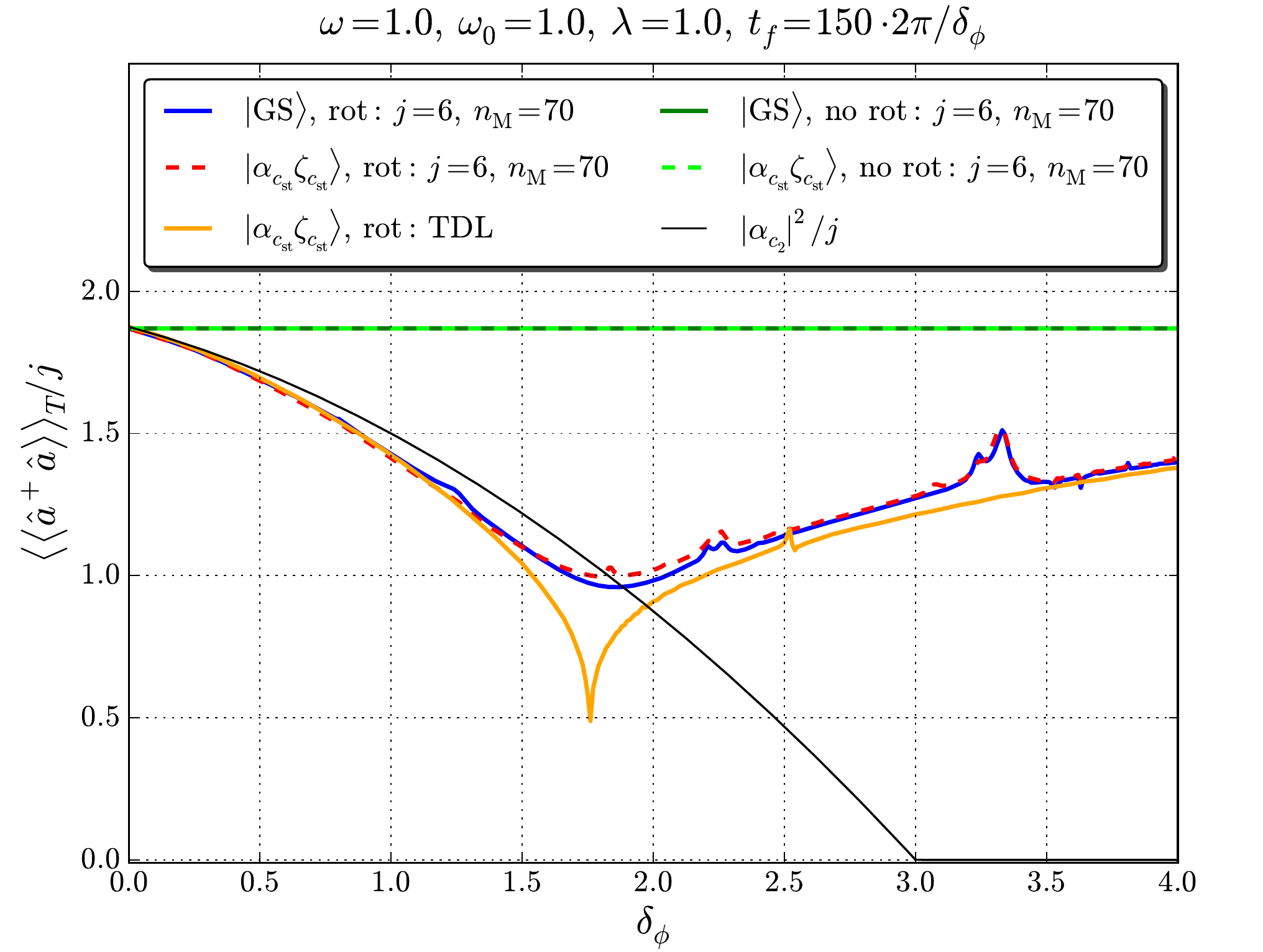}
    \caption{
      The time averaged scaled mean
      photon number on resonance $\omega = \omega_{0} = 1.0$ for
      $\lambda=1.0$ as a function of the driving velocity. 
      We plot $\expect{\op{a}^{\dagger}\op{a}}(t_{f},\delta_{\phi})/j$
      for two different initial states: $\ket{\mathrm{GS}}$ and
      $\ket{\alpha_{c_{\mathrm{st}}}}\ket{\zeta_{c_{\mathrm{st}}}}$
      for the driven and undriven situations (with
      $t_{f}=150 \cdot 2\pi/\delta_{\phi}$ in both cases).
      }
    \label{fig:adadphitasgs}
  \end{center}
\end{figure}

Next in Fig.~\ref{fig:adadphitasgs} we consider
$\langle\expect{\op{a}^{\dagger}\op{a}}\rangle_{T}/j$ as a function of
$\delta_{\phi}$ averaged over a time span of $t_{f} = 150 \cdot
2\pi/\delta_{\phi}$ at resonance $\omega=\omega_{0}=1.0$ for
the protocol which encircles the equilibrium critical paraboloid
$\lambda=1.0$.
We show $\langle\expect{\op{a}^{\dagger}\op{a}}\rangle_{T}/j$ 
for the two initial states $\ket{\mathrm{GS}}$ and
$\ket{\alpha_{c_{\mathrm{st}}}}\ket{\zeta_{c_{\mathrm{st}}}}$.
Also we compare the driven and undriven situations. 

To summarize, we see that the dynamics of the mean photon number for
the initial ground state $\ket{\mathrm{GS}}$ and for the initial coherent state
$\ket{\alpha_{c_{\mathrm{st}}}}\ket{\zeta_{c_{\mathrm{st}}}}$ are very similar.
Evolution starting from these initial states for a finite number of atoms has a tendency to
develop two new sub-phases in the super radiant phase.
Similarly, evolution from both states develops a new dynamical critical coupling
$\lambda_{c}^{(\mathrm{dyn})}$ or critical driving
velocity $\delta_{\phi,c}^{(\mathrm{dyn})}$ in the thermodynamic
limit.
This phenomenon is traced back to the interplay of geometric and
dynamical phases in dynamics. 
Finally, we remark that the non-equilibrium phase
diagram for the initial coherent state
$\ket{\alpha_{c_{\mathrm{st}}}}\ket{\zeta_{c_{\mathrm{st}}}}$ is the
same as for the initial ground state $\ket{\mathrm{GS}}$.

\subsubsection{Parity $\op{\Pi}$}

\begin{figure}[h!]
  \begin{center}
    \includegraphics[scale=0.43]{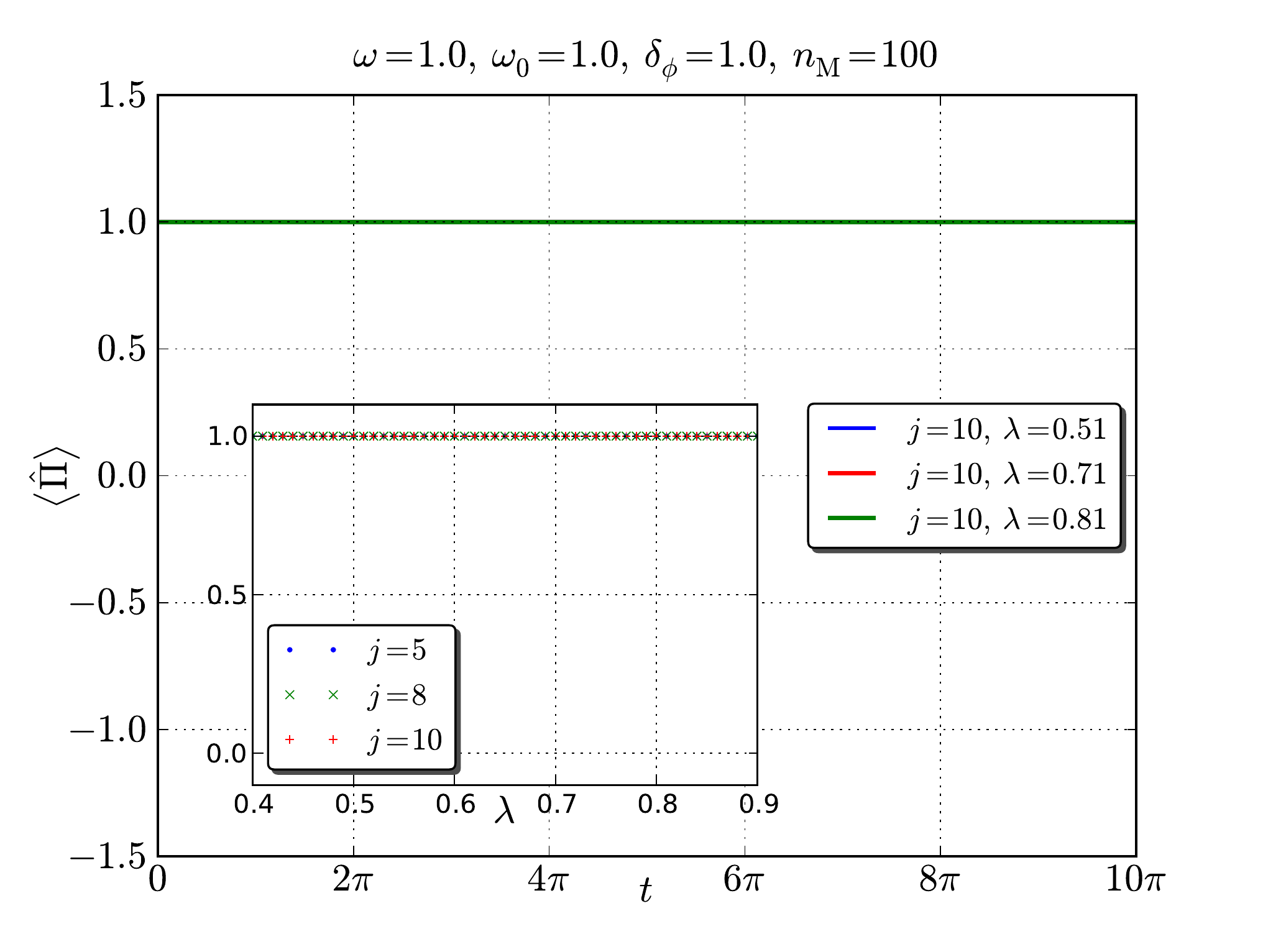}
    \caption{
      Time evolution of the parity operator when the
      dynamics is started from the ground state
      of the unrotated Dicke model
      $\ket{\mathrm{GS}}$.
      The inset plot shows the dependence on the atom-file coupling
      strength $\lambda$ for different pseudo-spin length $j$.
      }
    \label{fig:pitimelamfssgs}
  \end{center}
\end{figure}

Here we illustrate the time evolution of the parity operator
$\bra{\psi(t)}\op{\Pi}\ket{\psi(t)}$.
The result, for a situation when the initial state is the ground state
of the unrotated Dicke model, $\ket{\mathrm{GS}}$, while the evolution is given by 
$\op{H}_{\mathrm{RD}}(t)$, is shown in 
Fig.~\ref{fig:pitimelamfssgs}.
The calculation was done numerically according to the Chebyshev scheme
for a finite number of atoms $j$. 

One can clearly see that the parity is always constant in time.
More interesting is that there is no transition to zero, at a critical
coupling.
Apparently the system always remains in the state with the same parity
equal to $+1$.

\section{Conclusion}

In conclusion, we studied the dynamics of the rotationally driven
Dicke model beyond the rotating wave approximation.
We calculated the time evolution of the mean photon number and the
expectation values of the parity operator of the system for
different initial states: (i) stationary Dicke coherent state
$\ket{\alpha_{c_{\mathrm{st}}}}\ket{\zeta_{c_{\mathrm{st}}}}$, (ii) the
stationary coherent state
$\ket{\alpha_{c_{2}}}\ket{\zeta_{c_{2}}}$ of the rotated Dicke model
$\op{H}_{\mathrm{RD}}(t)$, (iii) Fock state $\ket{n=0}\ket{j,m=-j}$
and (iv) the ground state $\ket{\mathrm{GS}}$ of the Dicke model.
In order to understand the influence of the
geometric phase on the non-equilibrium dynamics we compared the time
evolution for the rotationally driven Dicke model
$\op{H}_{\mathrm{RD}}(t)$ and the evolution of the Dicke model
$\op{H}_{\mathrm{D}}$ without driving.

Here is the summary of our findings for different physically relevant
initial states.
(i) For the stationary Dicke initial state (stationary state
of the usual Dicke model $\op{H}_{\mathrm{D}}$) 
we observe in the TDL a reentrant meta stable phase in the
super-radiant phase. 
This is clearly manifested in the time averaged scaled mean photon
number.
We observe that the system developed a dynamical critical coupling strength
$\lambda_{c}^{(\mathrm{dyn})}$ and dynamical critical driving
velocity $\delta_{\phi,c}^{(\mathrm{dyn})}$.
Dynamical quantum phase transition is identified as a sudden change in the dynamical
behavior of observables as a function of the driving parameters in
real time.
We suggest that this critical behavior results form a competition of the geometric
phase and the dynamical phase which can give rise to a
resonance phenomenon described in Ref.~\cite{TPG}.
By looking at the time averaged scaled mean photon number we constructed a
non-equilibrium phase diagram.
The time evolution of the parity operator reveals a similar phenomenon of appearance 
of a meta stable phase and provides a complementary information about a new metastable phase.
Further, we observe that this type of non-equilibrium driving allows
one to probe the equilibrium quantum criticalities ``from a
distance'' in parameter space, namely by encircling a quantum critical
surface in parameter space without actual crossing it.
This may provide a useful experimental hint since
the quantum state will not be destroyed by crossing the quantum
critical point.

(ii) For the stationary circle initial state
$\ket{\alpha_{c_{2}}}\ket{\zeta_{c_{2}}}$ (stationary coherent state
of the rotated Dicke model $\op{H}_{\mathrm{RD}}$) we found
a shift in the critical coupling by the amount given by the applied rotation velocity
$\lambda_{c}^{(\mathrm{rot})}=\sqrt{\omega(\omega_{0}+\delta_{\phi})}/2$.
This defines a critical rotation driving velocity
$\delta_{\phi,c}^{(\mathrm{rot})}=\frac{4\lambda^{2}}{\omega}-\omega_{0}$.
Which can otherwise be understood by making a transformation into a rotating frame and considering
the effective Hamiltonian $\op{H}_{\mathrm{ROT}}$ in this basis.
The expectation value of the parity operator shows a constant time evolution and goes to zero at the critical
coupling $\lambda_{c}^{(\mathrm{rot})}$.

(iii) For the Fock state with no photons and lowest spin projection
$\ket{n=0}\ket{j,-j}$ we obtain, for a finite number of atoms, a 
mean photon number which is different from zero.
We observe the same rotational critical coupling
$\lambda_{c}^{(\mathrm{rot})}=\sqrt{\omega(\omega_{0}+\delta_{\phi})}/2$ in this case
and therefore
$\delta_{\phi,c}^{(\mathrm{rot})}=\frac{4\lambda^{2}}{\omega}-\omega_{0}$,
the same as for the initial state $\ket{\alpha_{c_{2}}}\ket{\zeta_{c_{2}}}$.
The mean field solution does not evolve in time and produces trivial
results if we choose the parameters of the initial conditions as
$\alpha_{0}=\alpha_{0}^{\ast}=\zeta_{0}=\zeta_{0}^{\ast}=0$.
However, calculations for a very small initial value of the parameters (which can be interpreted as quantum fluctuations)
$\alpha_{0}=\alpha_{0}^{\ast}=\zeta_{0}=\zeta_{0}^{\ast}=10^{-\epsilon}$ reveal a non
zero mean photon number in the TDL using the mean field
equations.
For the time averaged quantities we also observe 
the same rotational critical coupling
$\lambda_{c}^{(\mathrm{rot})}=\sqrt{\omega(\omega_{0}+\delta_{\phi})}/2$
and 
$\delta_{\phi,c}^{(\mathrm{rot})}=\frac{4\lambda^{2}}{\omega}-\omega_{0}$, as before.
The parity is always constant and equals one for a finite number of two-level
atoms which also holds in the TDL.
On the other hand, for small non-zero initial conditions we
observe an oscillatory behavior of the parity expectation value.

(iv) For the initial state given by the ground state $\ket{\mathrm{GS}}$ (which is
computed numerically by diagonalizing the Dicke Hamiltonian
$\op{H}_{\mathrm{D}}$) we observe the same dynamical behavior as for the stationary Dicke
state $\ket{\alpha_{c_{\mathrm{st}}}}\ket{\zeta_{c_{\mathrm{st}}}}$.
This shows that these two states become equivalent in the TDL. 

Summarizing, non-equilibrium dynamics of the Dicke model shows several
universal properties independent on the precise nature of the initial
state and exhibits interesting non-equilibrium quantum critical
behavior which can be traced back to a competition of geometric and
dynamical phases. 

\section{Acknowledgments}

This work is supported by Swiss National Science Foundation.
V.G. is grateful to KITP for hospitality. 


\end{document}